\documentclass[traditabstract, twocolumns]{aa}

\usepackage{graphicx}
\usepackage{txfonts}
\usepackage{natbib}
\usepackage{xspace}
\usepackage{xcolor}
\usepackage{scalerel}
\usepackage[normalem]{ulem}
\usepackage{enumitem} 
\usepackage{multirow}
\usepackage{rotating}
\usepackage{adjustbox}
\usepackage{pbox}

\usepackage{soul}   

\newcommand{\pycs}{{\tt PyCS}\xspace}
\newcommand{\pycsmult}{{\tt PyCS-mult} \xspace}
\newcommand{\pycssum}{{\tt PyCS-sum} \xspace}
\newcommand{\dthreecs}{{\tt D3CS}\xspace}

\newcommand{\hc}{$H_0\ $}

\newcommand{\lcdm}{$\mathrm{\Lambda CDM}$\xspace}
\newcommand{\HEzerozero}{HE~0047$-$1756\xspace}
\newcommand{\HEzerodeux}{HE~0230$-$2130\xspace}
\newcommand{\HEzeroquatre}{HE~0435$-$1223\xspace}
\newcommand{\HEvingtetun}{HE~2149$-$2745\xspace}
\newcommand{\HSzerohuit}{HS~0818$+$1227\xspace}
\newcommand{\Jzeroun}{Q~J0158$-$4325\xspace}
\newcommand{\Jzerodeuxquatresix}{SDSS~J0246$-$0825\xspace}
\newcommand{\Jzerohuit}{SDSS~J0832$+$0404\xspace}
\newcommand{\Jzeroneuf}{SDSS~J0924$+$0219\xspace}
\newcommand{\Jdouze}{SDSS~J1226$-$0006\xspace}
\newcommand{\Jtreizevingt}{SDSS~J1320$+$1644\xspace}
\newcommand{\Jtreizevingtdeux}{SDSS~J1322$+$1052\xspace}
\newcommand{\Jtreizetrentecinq}{SDSS~J1335$+$0118\xspace}
\newcommand{\Jtreizequaranteneuf}{SDSS~J1349$+$1227\xspace}
\newcommand{\Jquatorzezerocinq}{SDSS~J1405$+$0959\xspace}
\newcommand{\Jquatorzecinquantecinq}{SDSS~J1455$+$1447\xspace}
\newcommand{\Jquinzequinze}{SDSS~J1515$+$1511\xspace}
\newcommand{\Jseizevingt}{SDSS~J1620$+$1203\xspace}
\newcommand{\Qtreize}{Q~1355$-$2257\xspace}
\newcommand{\Qvingtdeux}{Q~2237$+$0305\xspace}
\newcommand{\RXJonze}{RX~J1131$-$1231\xspace}
\newcommand{\UMsix}{UM~673}
\newcommand{\Qzeroun}{Q~0142$-$100\xspace}
\newcommand{\WFIvingtvingtsix}{WFI~J2026$-$4536\xspace}
\newcommand{\WFIvingttrentetrois}{WFI~J2033$-$4723\xspace}
\newcommand{\Jdouzerosix}{SDSS~J1206$+$4332\xspace}
\newcommand{\Bseize}{B~1608$+$656\xspace}
\newcommand{\Qzero}{Q~0957$+$561\xspace}
\newcommand{\PGonze}{PG~1115$+$080\xspace}
\newcommand{\be}{\begin{equation}}
\newcommand{\ee}{\end{equation}} 
\def\kmspc{${\textrm{km}\, \textrm{s}^{-1} \textrm{Mpc}^{-1}}$}

\begin{document}
\title{COSMOGRAIL XIX: Time delays in 18 strongly lensed quasars from 15 years of optical monitoring\thanks{All light curves presented in this paper are only available in electronic form
at the CDS via anonymous ftp to \url{cdsarc.u-strasbg.fr} (130.79.128.5)
or via \url{http://cdsarc.u-strasbg.fr/viz-bin/cat/J/A+A/640/A105}}.\\}

\author{
M.~Millon\inst{\ref{epfl}} \and
F.~Courbin\inst{\ref{epfl}} \and
V.~Bonvin\inst{\ref{epfl}} \and
E.~Paic \inst{\ref{epfl}} \and
G.~Meylan \inst{\ref{epfl}} \and
M.~Tewes \inst{\ref{Bonn}} \and 
D.~Sluse \inst{\ref{Liege}} \and
P.~Magain \inst{\ref{Liege}} \and
J.~H.~H.~Chan \inst{\ref{epfl}} \and
A.~Galan \inst{\ref{epfl}} \and
R.~Joseph \inst{\ref{epfl}, \ref{princeton}} \and
C.~Lemon \inst{\ref{epfl}} \and
O.~Tihhonova \inst{\ref{epfl}} \and
R.~I.~Anderson\inst{\ref{unige},\ref{eso}} \and 
M.~Marmier\inst{\ref{unige}} \and 
B.~Chazelas\inst{\ref{unige}} \and 
M.~Lendl\inst{\ref{unige}} \and 
A.~H.~M.~J.~Triaud \inst{\ref{unige}, \ref{unibirmigham}} \and 
A.~Wyttenbach\inst{\ref{unige}, \ref{unigrenoble}} 
}

\institute{
Institute of Physics, Laboratory of Astrophysics, Ecole Polytechnique 
F\'ed\'erale de Lausanne (EPFL), Observatoire de Sauverny, 1290 Versoix, 
Switzerland \label{epfl}\goodbreak \and
Argelander-Institut f\"ur Astronomie, Auf dem H\"ugel 71, 53121
Bonn, Germany \label{Bonn} \goodbreak \and
 STAR Institute, Quartier Agora - All\'ee du six Ao\^ut, 19c B-4000 Li\`ege, Belgium \label{Liege} \goodbreak \and
Department of Astrophysical Sciences, Princeton University, Princeton, NJ 08544, USA \label{princeton} \goodbreak \and
Observatoire astronomique de l'Universit\'e de Gen\`eve, 51 ch. des Maillettes, 1290 Versoix, Switzerland \label{unige} \goodbreak \and
European Southern Observatory, Karl-Schwarzschild-Str. 2, 85748 Garching b. M\"unchen, Germany \label{eso} \goodbreak \and 
School of Physics \& Astronomy, University of Birmingham, Edgbaston, Birmingham, B15 2TT, United Kingdom. \label{unibirmigham} \goodbreak \and
Universit\'e Grenoble Alpes, CNRS, IPAG, 38000 Grenoble, France \label{unigrenoble} \goodbreak 
}

\date{\today}
\abstract{We present the results of 15 years of monitoring lensed quasars, which was conducted by the COSMOGRAIL programme at the Leonhard Euler 1.2m Swiss Telescope. The decade-long light curves of 23 lensed systems are presented for the first time. We complement our data set with other monitoring data available in the literature to measure the time delays in 18 systems, among which nine reach a relative precision better than 15 \% for at least one time delay. To achieve this, we developed an automated version of the curve-shifting toolbox \pycs to ensure robust estimation of the time delay in the presence of microlensing, while accounting for the errors due to the imperfect representation of microlensing. We also re-analysed the previously published time delays of \RXJonze and \HEzeroquatre, by adding six and two new seasons of monitoring, respectively, and confirming the previous time-delay measurements. When the time delay measurement is possible, we corrected the light curves of the lensed images from their time delay and present the difference curves to highlight the microlensing signal contained in the data. To date, this is the largest sample of decade-long lens monitoring data, which is useful to measure $H_0$ and the size of quasar accretion discs with microlensing as well as to study quasar variability.}


\titlerunning{}
\maketitle

\section{Introduction}
In the \lcdm paradigm, the Universe is composed of cold dark matter (CDM) and includes a cosmological constant, $\Lambda$. For a flat topology, it is described by a set of six free parameters. The current expansion rate of the Universe, the Hubble constant $H_0$, can either be derived from these six parameters or measured directly. It is therefore playing a major role in verifying the agreement between theory and observations. 
One way of determining \hc is to measure distance parallaxes of Cepheid stars in the Milky Way and in the Magellanic Clouds and then use them to calibrate brighter standard candles, such as type Ia supernovae (SNIa), thus reaching larger distances. Using this method, the most precise estimate so far is $H_0 = 74.03 \pm 1.42$ \kmspc\,  \citep{Riess2019}. This method is known as the distance ladder method.

Conversely, it is also possible to measure \hc by using the physical size of the baryon acoustic oscillation (BAO) at the time of recombination in maps of the cosmic microwave background (CMB) at $z\sim1100$ \citep{Eisenstein2007} and then extrapolate it with a model fitted on the same CMB maps, down to $z=0$. This is known as the inverse distance ladder and it gives $H_0 = 67.4 \pm 0.5$  \kmspc\, in a flat cosmology \citep{Planck2018}. 
The two values of $H_0$ are currently in tension, indicating that there is either unknown sources of systematics in the measurements or that the \lcdm model needs further extensions \citep[e.g. see][for reviews]{Suyu2018, deGrijs2017}. Other independent methods probing the late Universe, such as gravitational waves \citep{Abbott2017, Feeney2019, Soares-Santos2019}, water megamasers \citep{Humphreys2013, Braatz2018, Pesce2020}, galaxy clustering \citep{DES2017, Philcox2020} or alternative calibrators of the SNIa such as Mira variables \citep{Huang2020} and the tip of the red giant branch (TRGB) method \citep{Freedman2020, Yuan2019} are extremely valuable to infer whether the apparent tension is real or not. In fact, the methods tend to support the existence of the tension between $H_0$ measurements from early and late Universe probes \citep[e.g.][]{Verde2019}.

A complementary and single-step technique to measure $H_0$ in the late Universe is called time-delay cosmography. The original idea was first proposed by \cite{Refsdal1964}. It uses the time delays between the images of strongly lensed quasars or supernovae to infer their so-called time delay distance $D_{\Delta t}$ \citep[see e.g.][]{Refsdal1964, Suyu2012}. When a background source is multiply imaged by a foreground galaxy, the travel time of the photons along each optical path is not exactly the same. This effect is due to the difference in length between the optical paths and to the gravitational delay caused by the potential well of the lensing galaxy as well as all mass contribution along the line of sight. As a consequence, if the unlensed source is photometrically variable, its variations are seen by the observer at different times in each lensed image. 
Quasars are ideal sources for time delay cosmography. Their high luminosity makes them visible over cosmological distances and their variability allows for the measurement of time delays. Since the discovery of the first multiply imaged quasar \citep[\Qzero]{Walsh1979}, the number of lensed quasars has drastically increased, especially in recent years  with the discovery of dozens of new systems in large sky surveys such as the Dark Energy Survey (DES) \citep{Agnello2015, Anguita2018, Agnello2018a} or the ESA \textit{Gaia} mission \citep{Lemon2018,Krone-Martins2018, Lemon2019}.
The first time-delay measurement, for \Qzero, was the subject of a ten-year controversy, with large uncertainties and even contradictory results from radio and optical observations \citep{Vanderriest1989, Schild1990, Roberts1991, Press1992}. This controversy was finally solved by \cite{Kundic1997}, but this reflects the difficulty of measuring time delays with poorly sampled light curves over short monitoring campaigns, mainly due to the microlensing by stars in the lens galaxy. As these pass in front of the quasar images, they introduce extrinsic variations atop the intrinsic variations of the quasar. Since these extrinsic variations are different in each quasar image, they distort the observed light curves and complicate the time-delay measurement. 

The COSMOGRAIL programme \citep{Courbin2005, Eigenbrod2005} is a long term lens monitoring programme whose goal is to provide the time delays for a large sample of strongly lensed quasars. The long-term follow-up of these objects was originally carried out by five 1-m class telescopes with, in particular, uninterrupted observations for 15 years at the Leonhard Euler 1.2m Swiss Telescope (hereafter Euler) at ESO La Silla, Chile. Some of the previous COSMOGRAIL results include precise time delays for \HEzeroquatre \citep{Courbin2011, Bonvin2017}, \RXJonze \citep{Tewes2013b}, \Jdouzerosix \citep{Eulaers2013}, \PGonze \citep{Bonvin2018a} and \WFIvingttrentetrois \citep{Bonvin2019}. 
Along with \Bseize, whose time delays have been measured in radio data by \cite{Fassnacht1999}, these systems were used by the H0LICOW collaboration \citep[H0 Lenses In COsmograil's Wellspring; see][for an overview of the programme]{Suyu2017} to measure $H_0 = 73.3 ^{+1.7}_{-1.8}$ \kmspc\, at 2.4\% precision \citep[][]{Wong2019}. This result is in good agreement with the local distance ladder method but in 3.1$\sigma$ tension with the \textit{Planck} observation. Improving the precision on $H_0$ and devising tests to check possible systematic errors will require more lenses with measured time delays, which is the contribution of the present work. 

The paper describes our observations in Sect.~\ref{Obs} as well as the reduction process followed by deconvolution-photometry to extract the light curves from the blended quasar images. Sect.~\ref{TD} describes how we measure the time delays in an automated and robust way, accounting for microlensing variations and providing uncertainties using simulated light curves. Sect.~\ref{results} presents the time-delay measurements for all our objects and Sect.~\ref{cfini} gives our conclusion and future plans for COSMOGRAIL. 

\section{Observation and data reduction}
\label{Obs}

\subsection{ECAM and C2 data reduction}
\begin{figure*}[htbp!]
    \begin{minipage}[c]{\textwidth}
     \centering
    \includegraphics[width=0.92\textwidth]{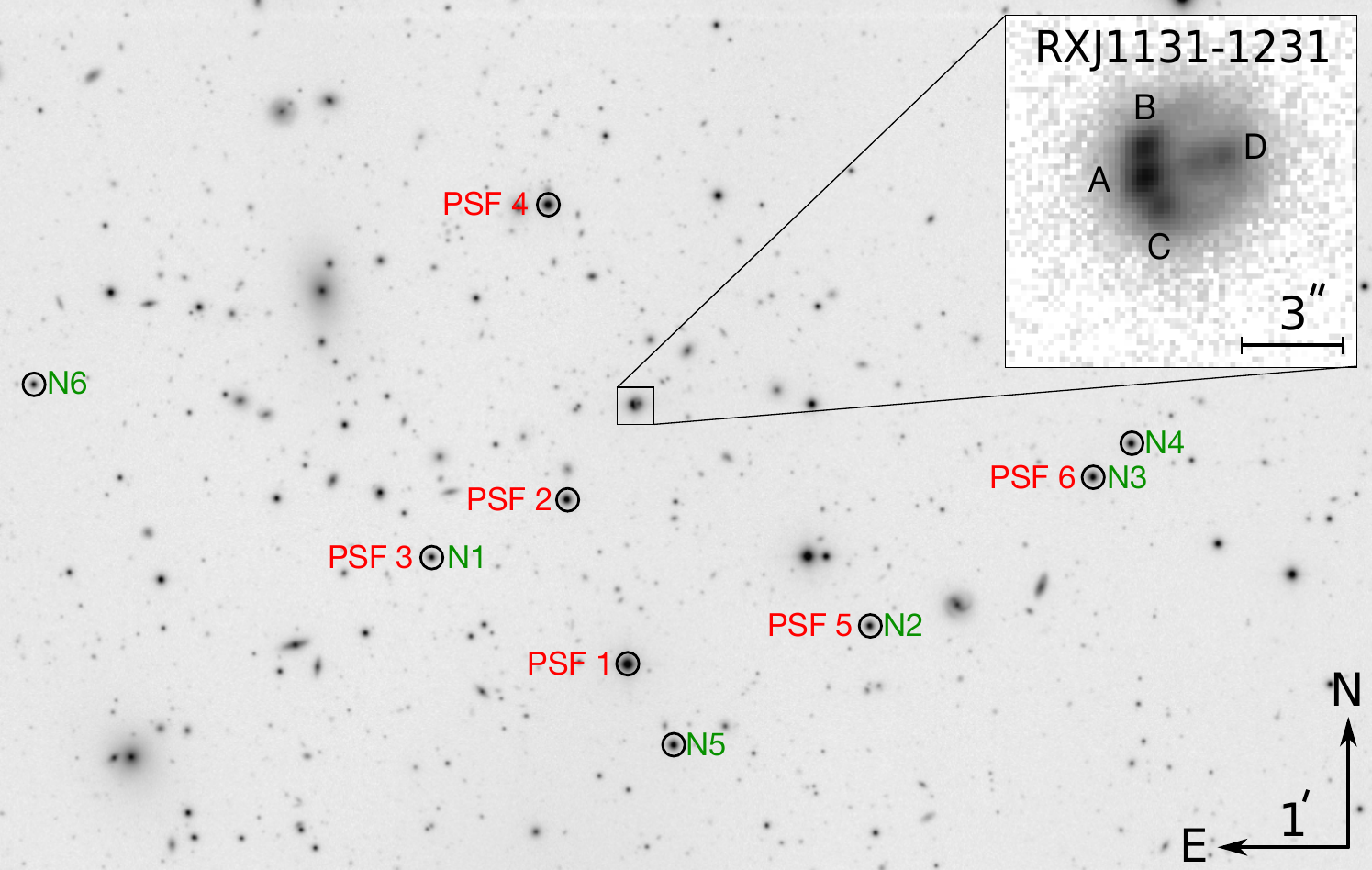}
    \end{minipage} 
    \begin{minipage}[c]{\textwidth}
     \centering
    \includegraphics[width=0.92\textwidth]{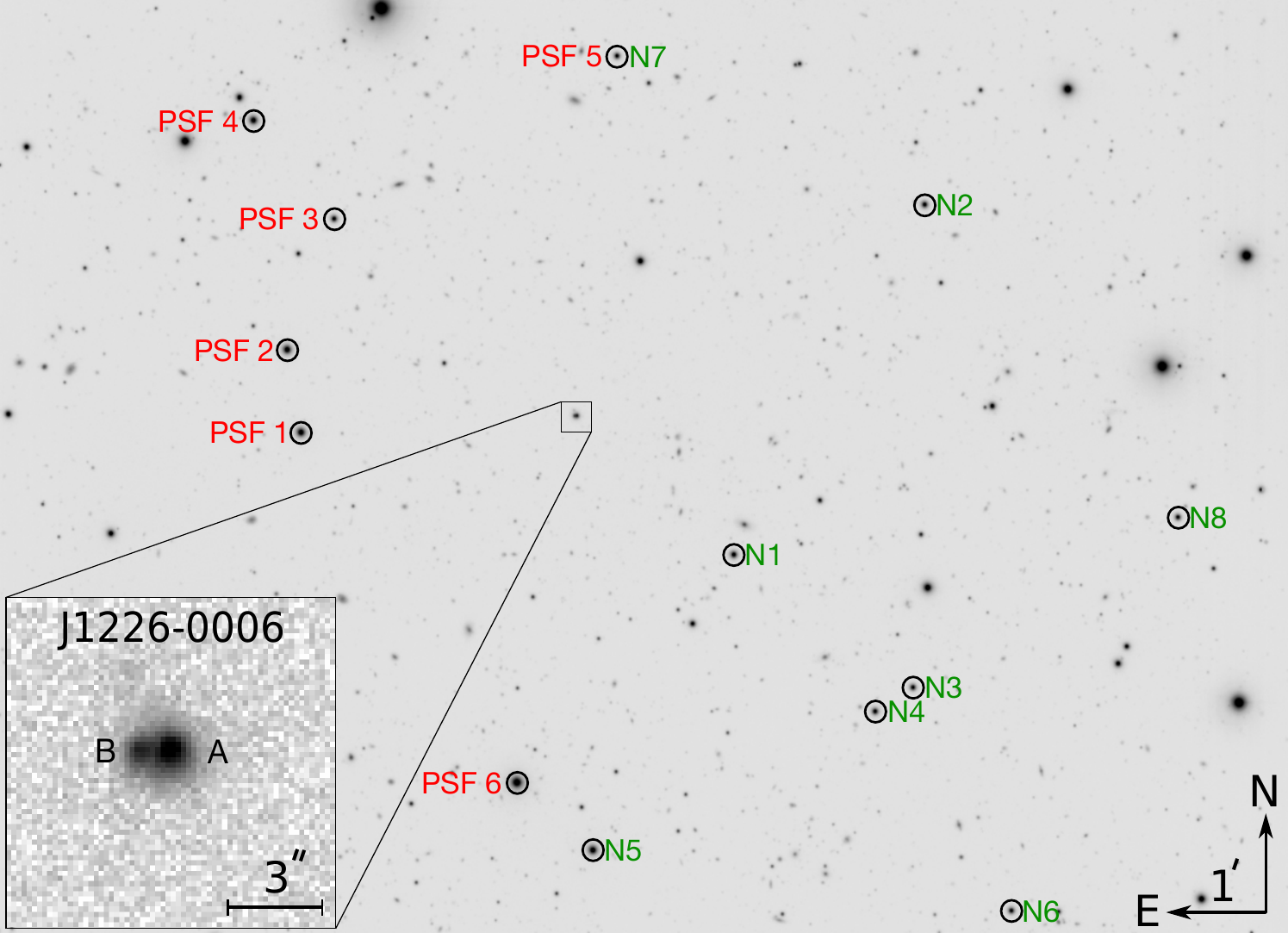}
    \end{minipage} 
        
    \caption{Part of the fields of view of \RXJonze (top) and \Jdouze (bottom) as seen by the Leonhard Euler 1.2m Swiss telescope at La Silla Observatory. A total of 195 (388) frames with seeing $<$1.25 (1.4) and PSF ellipticity $<$0.15 (0.18) are stacked to produce the deep field image for \RXJonze (\Jdouze). The insert shows a single exposure of 360 seconds in $R$ band. The stars used for the PSF construction are labelled PSF$_i$, in red, and the reference stars used for flux normalisation are labelled in green, $N_j$. The field of view for the other lensed quasars are presented in Appendix \ref{AppendixA}, Figs.~\ref{fig:nicefield_annex}, \ref{fig:nicefield_annex2} and \ref{fig:nicefield_annex3}}
    \label{fig:nicefield}
\end{figure*}
\begin{figure*}[htbp!]
    \centering
    \begin{minipage}[c]{0.95\textwidth}
    \includegraphics[width=\textwidth]{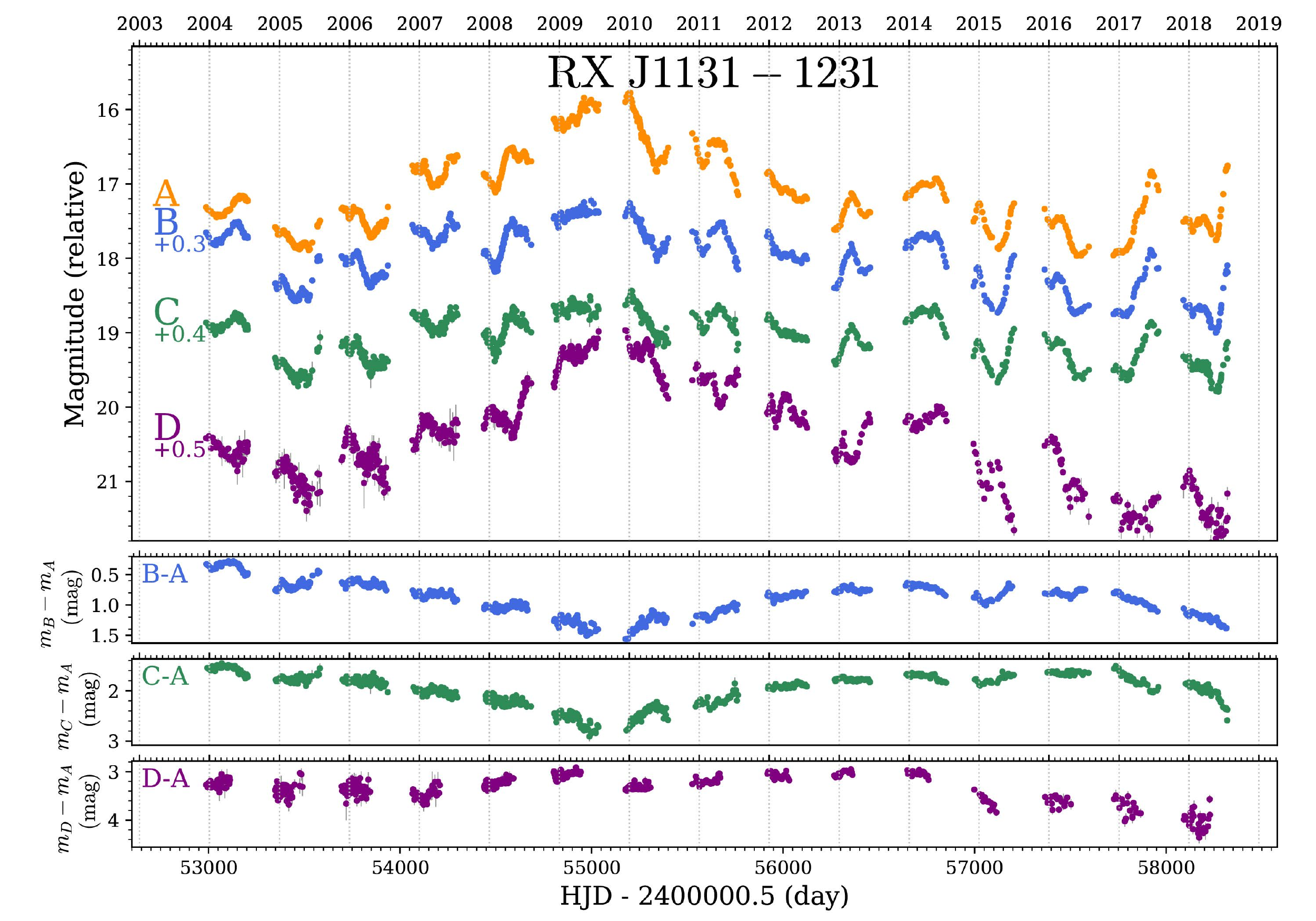}
    \end{minipage}
    \begin{minipage}[c]{0.95\textwidth}
    \includegraphics[width=\textwidth]{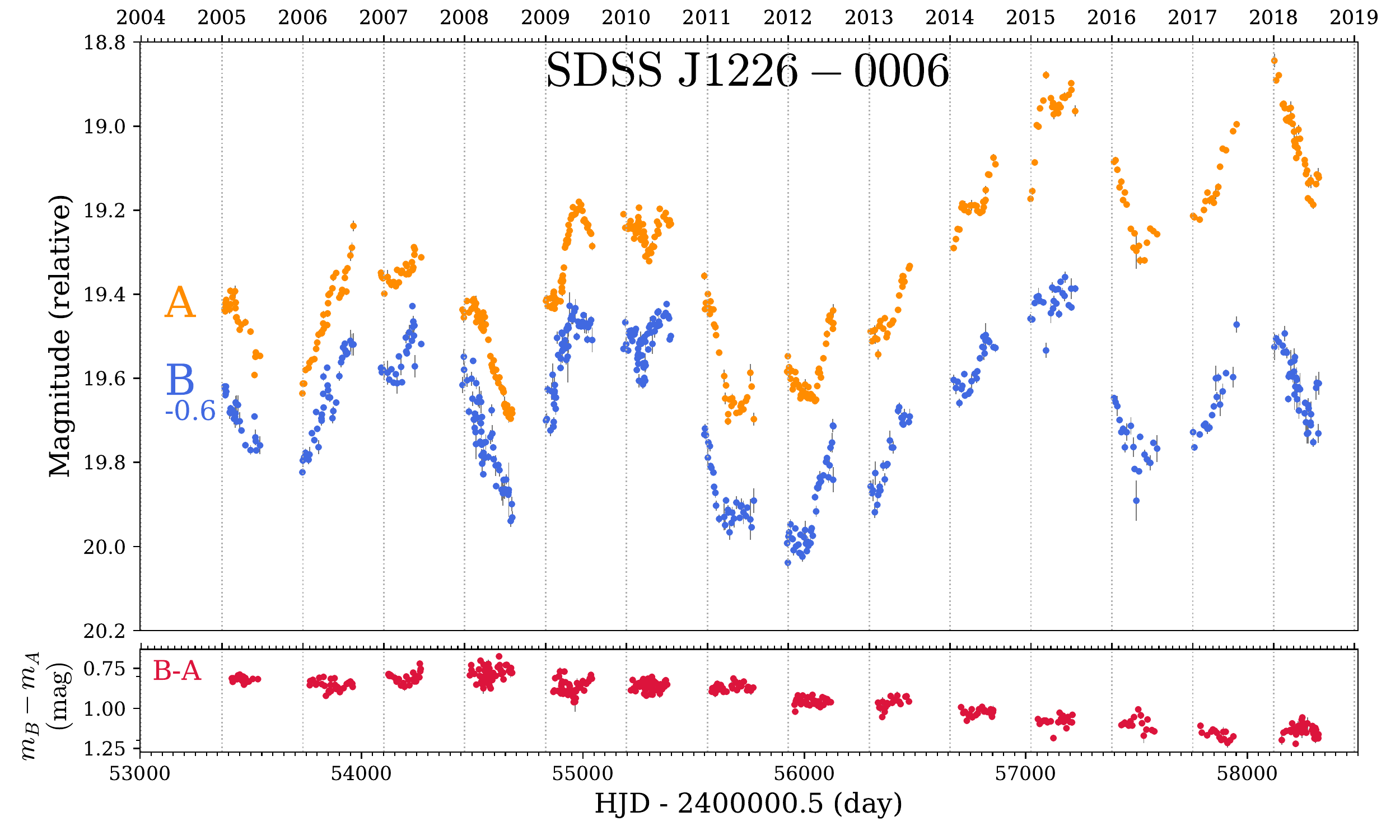}
    \end{minipage}    
    \caption{Full COSMOGRAIL $R$-band light curves for \RXJonze and \Jdouze, as also summarised in Table~\ref{tab:tabdata}. The bottom panels for each object show the difference curves between pairs of multiple images, shifted by the corresponding measured delay and interpolated with the intrinsic spline model fitted to the data (see Sect.~\ref{TD}). Light curves for all the other lensed quasars presented in this paper are shown in Appendix \ref{AppendixB} (Fig.~\ref{fig:annex_lcs1} to Fig.~\ref{fig:annex_lcs11})}
    \label{fig:lcs_1131_1226}
\end{figure*}

Most of our data were acquired with the Swiss 1.2m Euler telescope at the ESO La Silla observatory (Chile). The scheduled cadence of observation was originally of one observation every four days. The data presented here were acquired with two different instruments, namely C2 and EulerCAM (hereafter ECAM). The C2 camera was used from January 2004 to September 2010. It has a field of view of 11.4\arcmin $\times$ 11.4\arcmin and a pixel size of 0.344\arcsec. ECAM collected data from November 2010 to April 2018. It has a pixel size of 0.215\arcsec and a field of view of 14\arcmin $\times$ 14\arcmin. Table~\ref{tab:tabdata} summarises all data used in this paper. In regular operation mode, each epoch corresponds to five dithered exposures of 360 seconds each in $R$ band. Among all lens systems presented here, \RXJonze, \HEzeroquatre, \Jdouze and \HSzerohuit were monitored during the 15 years of the programme, from 2004 to spring 2018. We add respectively two and six new monitoring seasons to \HEzeroquatre and \RXJonze compared to their last appearance in a COSMOGRAIL publication \citep{Bonvin2017, Tewes2013b}.

The images were reduced in a consistent fashion using an updated version of the COSMOULINE pipeline\footnote{\url{https://github.com/COSMOGRAIL/COSMOULINE}}, in a very similar approach as \cite{Tewes2013a}. In this pipeline, one of the most crucial steps is the deblending of the light from the quasar images and the foreground lens galaxy; this has been achieved using the MCS deconvolution algorithm \citep{Magain1998, Cantale2016} according to the following procedure.

Firstly, the science exposures are corrected for the bias and the flat-field exposures to remove the additional bias level and correct for the CCD pixel efficiency variations. To do so, we follow standard but careful CCD reduction procedures with a pipeline described in \cite{Tewes2013a}. Secondly, the Point Spread Function (PSF) is measured on every individual exposure for a set of  stars (e.g. labelled PSF1 to PSF6 on Fig.~\ref{fig:nicefield}). The selected stars are chosen to be relatively close to the lens and with a magnitude comparable to the brightest quasar image. The PSF is composed of an analytical Moffat profile plus a pixel-grid correction. Thirdly, a deconvolution of reference stars (e.g. N1 to NX on Fig.~\ref{fig:nicefield}) is performed using the PSF build in the previous step. The most photometrically stable stars are selected to compute a median photometric normalisation coefficient for each exposure. This process also allows us to correct for image-to-image systematics that are introduced by PSF variations across the field. Finally, all images of the lensed quasar are then normalised using the coefficients computed previously and deconvolved simultaneously. The MCS algorithm outputs a model composed of a list of Gaussian point sources with improved resolution, representing the multiple images of the source and a `pixel channel', that includes all possible extended sources such as the lens galaxy, extended features of the source and companions of the lens galaxy. During the deconvolution process, the pixel channel is fixed for all exposures, as well as the relative astrometry of the quasar images, but the intensity of the latter is allowed to vary from exposure to exposure. As a result, the output of the process is the photometry of the quasar images for each exposure. The photometric data points are finally combined by observing nights to produce the light curves along with their photometric 1-$\sigma$ uncertainties. Depending on telescopes and sites there are between three and six individual frames per epoch to compute the photometric uncertainties. 

We illustrate the result of this process in Fig.~\ref{fig:lcs_1131_1226}, showing the COSMOGRAIL light curves\footnote{All light curves are publicly available at \url{www.cosmograil.org}.} for two objects with prominent variability: \RXJonze and \Jdouze. The light curves for 21 other lensed QSOs are shown in Appendix \ref{AppendixB}, from Fig.~\ref{fig:annex_lcs1} to Fig.~\ref{fig:annex_lcs11}. In some cases, the separation between two quasar images is too small and there can be flux leaking between images despite the deconvolution scheme. This typically happens when the image separation is below $0\arcsec75$. In such cases, we sum the fluxes of the two affected images. This is the case for \HEzerodeux where we show the total flux of A+B, for \Jzeroneuf, where we show A+D \cite[as in][]{Macleod2015}, and for \WFIvingtvingtsix, where we show A1+A2. These three objects are in a fold configuration each featuring a narrow pair of bright images. Since these images reside almost at the same place in the arrival-time surface, they are expected to be delayed by the same amount, thus having a roughly zero delay between them \citep{Bonvin2019}. Hence, our delays given for pairs rather than for resolved images are unlikely to be be biased at a level larger than a one day. We note, however, that such pairs with short or zero delay may turn out useful to test the microlensing time-delay hypothesis by \citet{Tie2017}: if a significant delay (days) is measured between narrow pairs, it may be entirely due to microlensing time delay. In at least one case where resolved photometry was possible \citep{Bonvin2019}, there is no measurable trace of such microlensing time delay, although its presence cannot be excluded.

\subsection{Additional data}
\label{sec:available_data}
In addition to the Euler light curves, we complement our data set with photometric monitoring data publicly available in the literature. For \Jquinzequinze, we added the data taken at the Liverpool Telescope published in \cite{Shalyapin2017}. We also include data taken at the SMARTS 1.3 m telescope with the ANDICAM optical and infrared camera for the lensed quasars \Jzeroneuf and \Jzeroun \citep{Macleod2015, Morgan2012}. In addition, we made use of the COSMOGRAIL data of \HEzeroquatre and \RXJonze taken at the 1.5m telescope at the Maidanak Observatory, Uzbekistan, at the Mercator Belgian telescope at Roque de Los Muchachos Observatory in La Palma, Canary Islands and at the SMARTS 1.3 m telescope at Cerro Tololo Inter-American Observatory, Chile, previously published in \cite{Courbin2011} and \cite{Tewes2013b}. When a lens system was monitored by several instruments, we either analyse the data sets separately or we merge the light curves and perform the analysis on the merged data set (see Sect.~\ref{sec:combining_datasets} for details). In the latest case and when the monitoring campaigns do overlap, we fitted a flux and magnitude correction to the data to compensate for slight photometric offsets caused by differences in the filter and detector responses, coupled with differences in the spectral energy distribution of the quasar and reference stars. When the monitoring campaigns do not overlap, we simply adjust the magnitude zero-point in order to obtain a reasonable match between the light curves.

\begin{table*}[htbp]
\caption{Summary of the COSMOGRAIL monitoring data. Measurements for each epoch consist in most cases of 5 dithered exposures in the $R$ filter. \label{tab:tabdata}}
\resizebox{\textwidth}{!}{%
\begin{tabular}{l|llllll}
\hline
Object                                 & $z_{\rm lens}$       & $z_{\rm source}$ & Telescope (Instrument) & \#Epochs     & \begin{tabular}[c]{@{}l@{}}Total exposure\\ time [hours]\end{tabular} & Duration of monitoring \\ \hline
\HEzerozero                            & 0.407       & 1.678  & Euler (C2)            & 237    & 124.1                                                                 & Aug. 2005 - Sep. 2010  \\
                                       &             &       & Euler (ECAM)          & 316    & 160.8                                                                 & Sep. 2010 - Oct. 2016  \\
\UMsix~(\Qzeroun) & 0.491       & 2.73  & Euler (C2)            & 175     & 92.2                                                                  & Aug. 2005 - Sep. 2010  \\
                                       &             &       & Euler (ECAM)          & 179     & 91.4                                                                  & Sep. 2010 - Oct. 2016  \\
\Jzeroun                               & 0.317       & 1.29  & Euler (C2)            & 222    & 113.9                                                                 & Oct. 2004 - Sep.2010   \\
                                       &             &       & Euler (ECAM)          & 305    & 154.1                                                                 & Sep. 2010 - Fev. 2018  \\
                                       &             &       & SMARTS(Andicam)       & 270     & 83.4                                                                  & Aug. 2003 - Dec. 2010  \\
\HEzerodeux                            & 0.523       & 2.162 & Euler (C2)            & 51      & 34.5                                                                  & Jul. 2004 - Oct. 2006  \\
                                       &             &       & Euler (ECAM)          & 122     & 63.2                                                                  & Aug. 2013 - Oct. 2016  \\
\Jzerodeuxquatresix                    & 0.723       & 1.689 & Euler (C2)            & 122     & 71.0                                                                  & Nov. 2006 - Sep. 2010  \\
                                       &             &       & Euler (ECAM)          & 249    & 130.0                                                                 & Sep. 2010 - Apr. 2018  \\
\HEzeroquatre                          & 0.454       & 1.693 & Euler (C2)            & 301    & 150.5                                                                 & Jan. 2004 - Apr. 2010  \\
                                       &             &       & Euler (ECAM)          & 419    & 221.2                                                                 & Sep. 2010 - Aug. 2018  \\
                                       &             &       & SMARTS(Andicam)       & 136     & 40.8                                                                  & Aug. 2003 - Apr. 2005  \\
                                       &             &       & Mercator (MEROPE)     & 104     & 52.0                                                                  & Sep 2004. - Dec 2008   \\
                                       &             &       & Maidanak (SITE)       & 26     & 26.0                                                                  & Oct. 2004  - Jul. 2006 \\
                                       &             &       & Maidanak (SI)         & 8       & 4.8                                                                   & Aug. 2006  - Jan. 2007 \\
\HSzerohuit                            & 0.39        & 3.113 & Euler (C2)            & 215   & 118.6                                                                 & Jan. 2005 - May 2010   \\
                                       &             &       & Euler (ECAM)          & 151   & 76.6                                                                  & Nov. 2010 - Apr. 2018  \\
\Jzerohuit                             & 0.659       & 1.116 & Euler (ECAM)          & 237   & 121.0                                                                 & Nov. 2010 - May. 2018  \\
\Jzeroneuf                             & 0.393       & 1.523 & Euler (C2)            & 25     & 15.8                                                                  & Jan. 2004 - Oct. 2005  \\
                                       &             &       & Euler (ECAM)          & 106    & 52.4                                                                  & Nov. 2010 - Dec. 2015  \\
                                       &             &       & SMARTS(Andicam)       & 158     & 61.9                                                                  & Nov. 2003 - May 2011   \\
\RXJonze                               & 0.295       & 0.657 & Euler (C2)            & 265   & 132.5                                                                 & Mar. 2004 - Jul. 2010  \\
                                       &             &       & Euler (ECAM)          & 311    & 162.0                                                                 & Nov. 2010 - Jul. 2018  \\
                                       &             &       & SMARTS(Andicam)       & 288     & 86.4                                                                  & Dec. 2003 - May 2011   \\
                                       &             &       & Mercator (MEROPE)     & 78      & 39.0                                                                  & Jan. 2005 - Jun. 2008  \\
\Jdouze                                & 0.517       & 1.123 & Euler (C2)            & 257   & 135.2                                                                 & Jan. 2005 - Jul. 2010  \\
                                       &             &       & Euler (ECAM)          & 226   & 118.9                                                                 & Dec. 2010 - Jul. 2018  \\
\Jtreizevingt                          & 0.899       & 1.502 & Euler (ECAM)          & 41     & 21.3                                                                  & Mar. 2013 - Jul. 2016  \\
\Jtreizevingtdeux                      & $\sim 0.55$ & 1.717 & Euler (ECAM)          & 115     & 61.5                                                                  & Jan. 2011 - Jul. 2016  \\
\Jtreizetrentecinq                     & 0.44        & 1.570  & Euler (C2)            & 214    & 114.2                                                                 & Jan 2005 - Aug. 2010   \\
                                       &             &       & Euler (ECAM)          & 161     & 80.9                                                                  & Jan. 2011 - Aug. 2017  \\
\Jtreizequaranteneuf                   & $\sim 0.65$ & 1.722 & Euler (ECAM)          & 144     & 72.6                                                                  & Jan. 2011 - Aug. 2017  \\
\Qtreize                               & 0.702       & 1.370  & Euler (C2)            & 311   & 167.2                                                                 & Jul. 2003 - Aug. 2010  \\
                                       &             &       & Euler (ECAM)          & 89      & 44.1                                                                  & Jan. 2011 - Jul. 2015  \\
\Jquatorzezerocinq                     & $\sim 0.66$ & 1.81  & Euler (ECAM)          & 58     & 30.5                                                                  & Feb. 2014 - Aug. 2017  \\
\Jquatorzecinquantecinq                & $\sim 0.42$ & 1.424 & Euler (ECAM)          & 130     & 65.6                                                                  & Feb. 2011 - Aug. 2017  \\
\Jquinzequinze                         & 0.742       & 2.054 & Euler (ECAM)          & 63      & 31.5                                                                  & Mar. 2014 - Aug. 2017  \\
                                       &             &       & Liverpool(IO:O)       & 150     & 30.0                                                                  & May. 2014 - Sep. 2016  \\
\Jseizevingt                           & 0.398       & 1.158 & Euler (C2)            & 12       & 7.1                                                                   & Mar. 2010 - Sep. 2010  \\
                                       &             &       & Euler (ECAM)          & 194    & 100.6                                                                 & Feb. 2011 - Sep. 2017  \\
\WFIvingtvingtsix                      & $\sim 1.04$ & 2.23  & Euler (C2)            & 338    & 130.9                                                                 & Avr. 2004 - Sep. 2010  \\
                                       &             &       & Euler (ECAM)          & 210    & 112.8                                                                 & Oct. 2010 - Nov. 2016  \\
\HEvingtetun                           & 0.603       & 2.033 & Euler (C2)            & 222    & 121.0                                                                 & Aug. 2006 - Sep. 2010  \\
                                       &             &       & Euler (ECAM)          & 263    & 135.0                                                                 & Oct. 2010 - Dec.2017   \\
\Qvingtdeux                            & 0.039       & 1.69  & Euler (ECAM)          & 62      & 33.0                                                                  & Nov. 2010 - Jul. 2015  \\ \hline
Total :                                &             &       &                       & 8336 & 4094.0                                                                &                       
\end{tabular}}
\end{table*}

\section{Time-delay measurements}
\label{TD}

A broad range of curve-shifting algorithms have been developed in the past to measure time delays between light curves \citep[e.g.][]{Press1992, Pelt1994, Kelly2009, Hirv2011, Hojjati2012, Hojjati2014, Aghamousa2015}. Efficient algorithms must be robust against photometric noise, coarse temporal sampling, season gaps and must also take into account the presence of microlensing caused by  moving stars in the lensing galaxy.  Microlensing variations affect the observed flux independently in each quasar image. For this reason they are often referred to as extrinsic variation, as opposed to the intrinsic quasar variations shared by all quasar images. 

To test the robustness of curve-shifting algorithms, realistic light curves containing microlensing were simulated and proposed for a blind analysis to the community \citep[Time-Delay Challenge; hereafter TDC]{Dobler2015}. The curve-shifting methods adopted in COSMOGRAIL make use of the `free-knot spline' estimator \citep[see][for an in-depth presentation]{Tewes2013a} and the regression-difference estimator as well as the uncertainty estimation scheme from the \pycs toolbox\footnote{PyCS can be downloaded from the COSMOGRAIL website \url{www.cosmograil.org}}. Both methods performed very well in term of precision and accuracy on the TDC data \citep{Liao2016, Bonvin2016}, however, with a better performance of the former over the latter. These two methods are the basis of most current time-delay cosmography work, for example, by the H0LiCOW team.

In this work, we used the two different time-delay estimators implemented in \pycs. As these estimators provide `point estimates' of the time delays, the uncertainties must be evaluated with realistic simulated light curves that mimic both the intrinsic and extrinsic variations in the lensed images, which we detail below. 

\subsection{The \pycs package}
\label{section:pycs}
\begin{figure*}[htbp]
    \centering
    \includegraphics[width=\textwidth]{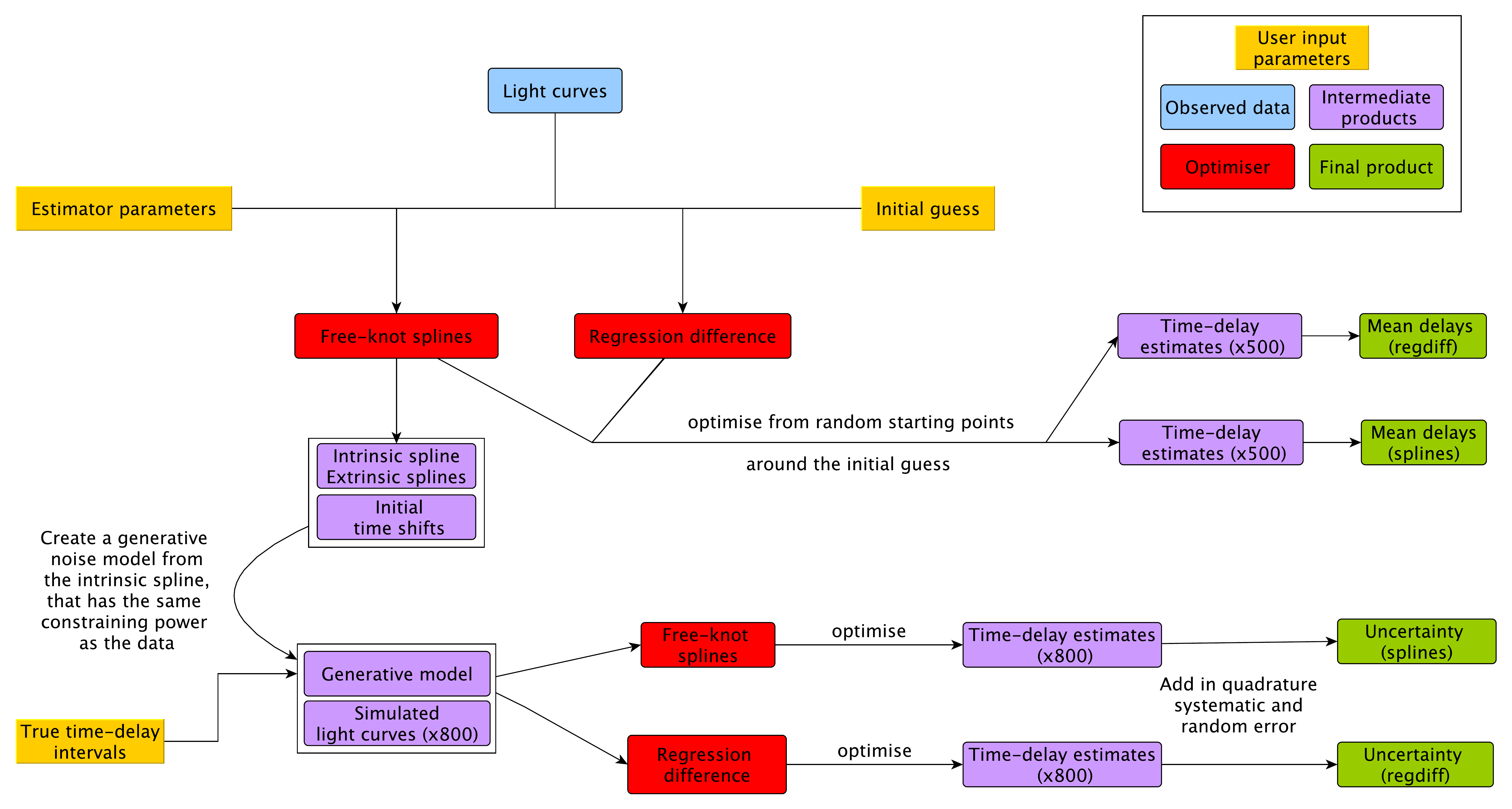}
    \caption{Sketch for the \pycs time-delay measurement pipeline. This procedure is repeated for each set of estimator parameters. The combination of the final products for each set of estimator parameters is discussed in Sect.~\ref{combiningtd}.}
    \label{fig:pycs_scheme}
\end{figure*}

The \pycs package proposes a fully data driven approach for time-delay measurements in the presence of microlensing. The variations of the quasar are directly derived from the data either using Gaussian processes or free-knot spline. Although there is no physical model behind the algorithms in \pycs, a number of  parameters are still required to model the quasar and microlensing variability. These cannot be explored in a fully Bayesian way due to the very large amount of computational time that such an approach would require. The same would hold true if we had chosen a physical model for microlensing, with the additional drawback that a wrong choice of a physical model may lead to biased measurements. We therefore propose a semi-Bayesian approach were we marginalise over a pre-selected grid of parameters to keep the computational time manageable on a small scale computing cluster. In previous COSMOGRAIL publications, external choices of parameter values were included as a robustness test. This procedure is now automatised to systematically explore broader parameter ranges. The full procedure is described in the following. 

In this work, we adopt the same terminology as in \cite{Bonvin2018a}. For clarity, we recall here several definitions:

\begin{itemize}

\item A \textit{curve-shifting technique} is a procedure that takes the monitoring data as inputs and results in a time-delay estimate along with associated uncertainties,

\item  An \textit{estimator} is an algorithm that returns the optimal time-delay between two light curves,

\item The \textit{estimator parameters} control the behaviour of the estimator (e.g. its convergence, number of degrees of freedom, etc.),

\item A \textit{generative model} is used to create simulated light curves that mimic the data and to evaluate the uncertainties on the time delays provided by the estimator.

\end{itemize}

We also recall the following terms to describe how we define and handle the different data sets, consistently with \cite{Bonvin2018a}: 
\begin{itemize}

\item A \textit{data set}, $\vec{D}$, usually corresponds to a monitoring campaign conducted with one instrument. In the present work this is ECAM, C2, SMARTS or Liverpool in one single filter. We also sometimes concatenate the light curves coming from several instruments in one single data set. For example, the joint ECAM+C2 light curves are referred to as the Euler data set.

\item A \textit {time-delay estimate}, $\vec{E} = \Delta t^{+\delta t_+}_{-\delta t_-}$, is composed of a point estimate with an uncertainty estimate of the time delay between two light curves. It corresponds to one particular choice of estimator and associated set of parameters.

\item A \textit {group} of time-delay estimates, $\vec{G}=[\vec{E_{AB}}, \vec{E_{AC}}, \vec{E_{BC}}]$, is a set of time-delay estimates between all pairs of light curves for a particular lensed system, obtained using the same curve-shifting technique. A group corresponds to the output of the pipeline presented in Sect.~\ref{sec:measurement pipeline} below for a given estimator, set of estimator parameters and generative model. 

\item A \textit{series} of time-delay estimates, $\vec{S}=[\vec{G_1},...\vec{G_i},...\vec{G_N}]$, for $i \in N$ is an ensemble of groups of time-delay estimates. They typically share the same data set and estimator but make use of different set of {\it estimator parameters}.

\end{itemize}

The two estimators used in this work are introduced in the next two sections.  

\subsubsection{Free-knot spline estimator}

The free-knot spline estimator models light curves as analytical spline functions. More precisely, we consider one unique free-knot spline for the intrinsic variations of the quasar, which is fitted simultaneously on all light curves, plus independent splines fitted individually on each light curve to model the extrinsic (microlensing) variations. The position of the knots of both the intrinsic and extrinsic splines as well as the time delays between light curves are optimised simultaneously. 

The parameters of the free-knot spline estimator are the initial spacing between the knots of the intrinsic spline, $\eta$, and of the extrinsic splines, $\eta_{ml}$. These parameters control the flexibility, that is, the number of degrees of freedom, of the intrinsic and extrinsic splines. If the initial spacing is too large, fast variations are missed and the precision and accuracy of the method is affected. A too small initial spacing between knots leads to an over-fitting of the data, also affecting the results.  The choice of $\eta$ and $\eta_{ml}$ must therefore be adapted to the data quality, which mainly depends on the cadence and photometric noise. The range of parameters used in this work for the different data sets are described in Table \ref{tab:regdiff_param}. We choose $\eta_{ml}$ being larger than $\eta$ as the microlensing variation typically occurs on longer time-scales than the intrinsic quasar variation. This is supported by the data themselves: taking the flux ratio between pairs of images, after correction by the time delay, leads to fairly smooth and long-term variations over several years as seen, for example, in the lower panels of Fig.~\ref{fig:lcs_1131_1226}, while intrinsic quasar variations are of the order of a few weeks to months as seen in the upper panels of the same figures.

\subsubsection{Regression difference estimator}
In the regression difference method, a Gaussian process regression is performed on each quasar image light curve independently. The regressions are then shifted in time and subtracted pair-wise, resulting in one difference curve for each pair of light curves and its associated uncertainties envelope. The algorithm minimises the variability of the difference curves by varying the time shift.

The estimator parameters are the covariance function of the Gaussian process,  its smoothness degree, $\nu$, its amplitude, A, its characteristic time scale, {\tt scale}, and an additional scaling factor for the photometric uncertainties, {\tt errscale} \citep[see][for more details]{Tewes2013a}. We use several combinations of these parameters which are summarised in Table \ref{tab:regdiff_param}. 

\begin{figure*}[t!]
    \begin{minipage}[r]{\textwidth}
    \centering
    \includegraphics[width=0.83\textwidth]{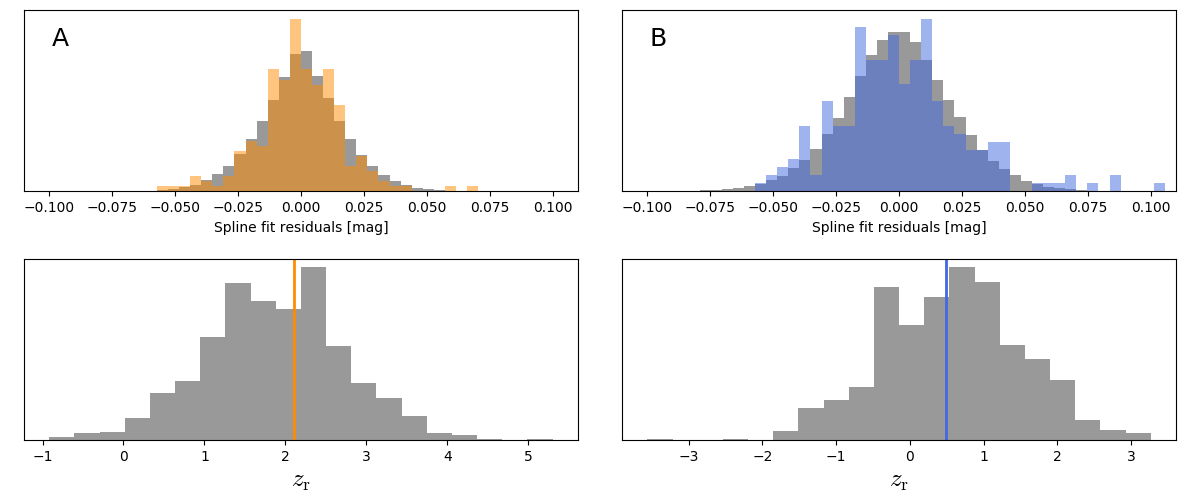}
    \hspace*{-1cm}
    \includegraphics[width=0.87\textwidth]{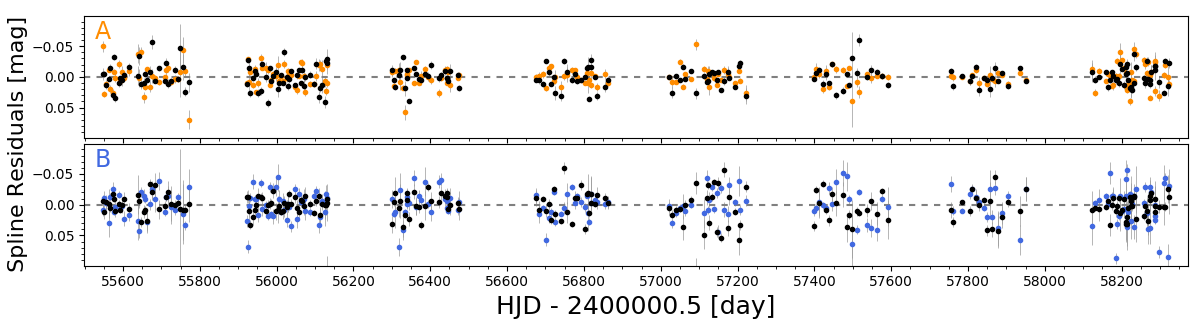}
    \end{minipage} 
    \caption{\textit{Top:} Distribution of the free-knot spline fit residuals for model parameter $\eta$ = 45 days, $\eta_{ml} = 300$ days in the ECAM light curves of \Jdouze. The orange and blue histograms represent the distributions of data residuals whereas the grey histogram corresponds to the distribution of the residuals for the 800 synthetic light curves created using the generative model. \textit{Middle panels:} Normalised number of runs $z_r$ for the synthetic curves (grey histogram) and for the data (orange and blue vertical lines). \textit{Bottom panels:} Residuals for the free-knot spline fit to the data, shown in blue and orange for image A and B, respectively. One example of the typical residuals obtained on applying the free-knot spline curve shifting technique on a simulated light curve is shown in black. Residuals are statistically similar to the observed data in terms of dispersion of the residuals, $\sigma$, and normalised number of runs, $z_r$ (Eq. ~\ref{eq:zr}). }
    \label{fig:resihist}
\end{figure*}

\begin{figure*}[h!]
    \centering
    \begin{minipage}[c]{\textwidth}
    \centering
    \includegraphics[width=0.9\textwidth]{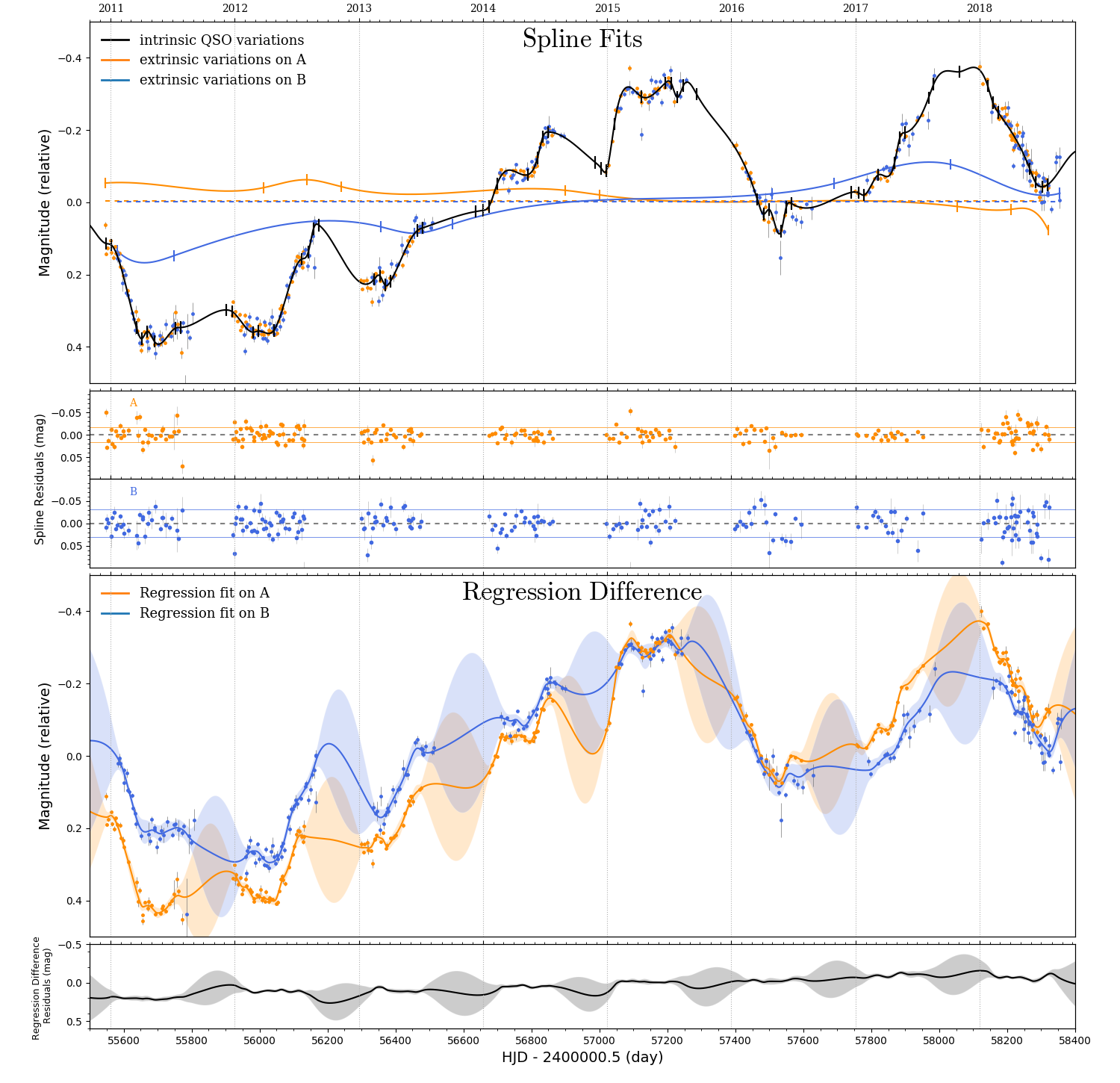}
    \end{minipage} 
    \caption{\textit{Top panel: } Spline fit of the ECAM data of \Jdouze. The black line represents the intrinsic variation shared by both curves ($\eta = 45 $ days) whereas the blue and orange lines model the extrinsic variations of A and B, respectively ($\eta_{ml} = 300 $ days). The curves are shifted in time by the optimal time delay, and corrected from their modelled extrinsic variations. \textit{Middle panel}: Residuals of the spline fit. \textit{Bottom panel: } Regression fits using Gaussian processes and their 1-$\sigma$ envelopes. The difference curve between the regression curves fitted on image A and image B is also shown in black. The algorithm is optimising the time shift in order to minimise the variability in the difference curve.}
    \label{fig:curve_fit_1226}
\end{figure*}

\subsection{Measurement pipeline}
\label{sec:measurement pipeline}
In the present work, we develop an extension of \pycs to handle a large number of different data sets in a homogeneous and automated way. We extend the procedure described in \citet{Bonvin2018a} to check the robustness of the two estimators against the choice of parameters. A schematic description of the procedure is presented in Fig.~\ref{fig:pycs_scheme}, leading  to a time-delay measurement along with its associated 1$\sigma$ uncertainties. We perform the following steps: 

\begin{enumerate}[label=(\roman*)]

\item We visually inspect the light curves in order to estimate a first guess for the time delays. To do this, we use the publicly available web application \dthreecs \footnote{All COSMOGRAIL light curves can be inspected with this interactive tool at \url{https://obswww.unige.ch/~millon/d3cs/COSMOGRAIL_public/}} \citep{Bonvin2016}.

\item We use this first guess as a starting point for the free-knot spline estimator. For a given set of parameters, we then obtain the optimised intrinsic and extrinsic splines. This also provides us point estimates of the time delays and with residuals of the fit.

\item We build a generative model using the spline fit resulting from the previous step. We then generate simulated light curves from the optimised intrinsic splines by randomizing the `true' time delay applied to each individual curve, following the procedure described by \citet{Tewes2013a} and we add noise. In previous works, this step was achieved by adding correlated and Gaussian noise in order to account, at the same time, for the residual extrinsic variability that is not included in the `microlensing' spline and for the shot-noise respectively. This required manually fine-tuning the noise power spectrum. We now use an automated way of generating simulated light curves by estimating the power spectrum of the residuals. In this new approach, we directly generate the noise of the simulated light curves from the measured power spectrum with randomised phases in Fourier space. This procedure results in simulated light curves with the same statistical noise properties, as presented in (vi). 

\item We run the estimators from 500 different random starting points around the guess delay to estimate the intrinsic error of the estimator, which is an indication of its robustness \citep[see][]{Tewes2013a}. The median time delays for each pair of light curves is our final point estimates for each estimator, that is, either the free-knot spline or the regression difference.

\item To estimate the uncertainties associated for the two estimators, we sample 800 simulated light curves from the generative model computed at step (iii) with different time delays and apply the estimator on them. These {\it true} time delays are chosen in the range of $\pm20$ days around the median delay obtained in (iv) and allows us to test whether our results depend on our choice of true time delays. For some noisy light curves, we extend this range not to be limited by this prior. We obtain the uncertainty of our estimator in an empirical way by combining in quadrature the random and systematic part of the uncertainties.

\item Finally, we ensure  that  the  statistics  of  the  residuals  on  the  simulated curves fit matches the one from the data. To do so, we compute two statistics on the residuals resulting from the free-knot spline fit, following the procedure described in \cite{Tewes2013a}. We expect that the residuals exhibit the same standard deviation $\sigma$ and the same normalised number of runs, $z_r$ (as defined below). A {\it run} corresponds to a sequence of adjacent residuals that are all either positive or negative. For truly independent residuals, we expect that the distribution of the runs (either positive or negative) is normally distributed with a mean and variance following:
\be
\mu_r = \frac{2N_{+}N_{-}}{N} + 1 \mathrm{\     and\     } \sigma_{r}^2 =\frac{(\mu_{r}-1)(\mu_{r}-2)}{N-1},
\ee
where $N_{+}$ and $N_{-}$ are the number of positive and negative residuals, and $N$ is the total number of data points. The normalised number of runs $z_r$ is given by: 
\be 
z_r = \frac{r - \mu_r}{\sigma_r}, 
\label{eq:zr}
\ee
where $r$ is the measured number of runs. 
Deviations of $z_r$ from 0 indicates that the residuals are correlated (negative $z_r$) or anti-correlated (positive $z_r$). We assess that this statistic computed for the simulated light curves distributes well around the value computed on the real data. This ensures that the simulated light curves and the real data have the same constraining power.
In Fig.~\ref{fig:resihist}, the distribution of the two relevant statistics, $\sigma$ and $z_r$, are shown for \Jdouze, with the same set of parameters as used for the fit in Fig.~\ref{fig:curve_fit_1226}. If the two statistics computed on the real and simulated data do not agree within $0.5\sigma$, we iteratively adapt the spectral window used for the generative model at step (iii).

\end{enumerate}
\begin{table*}[htpb]
\centering
\caption{Set of parameters used for the free-knot spline estimator and the regression difference estimator, for different data sets. These parameters are described in Section \ref{section:pycs}.\label{tab:regdiff_param}}
\begin{tabular}{l|cc|cccccc}
\multicolumn{1}{c|}{}                           & \multicolumn{2}{c|}{free-knot splines}                                    & \multicolumn{6}{c}{regression difference}                                                  \\ 
\hline\hline
                                                & \multirow{3}{*}{ $\eta$ }              & \multirow{3}{*}{25, 35, 45, 55}     & \multicolumn{1}{l}{} & Set 1~  & Set 2   & Set 3   & Set 4   & Set 5                       \\ 
\cline{5-9}
\multirow{5}{*}{ECAM}                           &                                        &                                  & $\nu$                & 1.7     & 2.2     & 1.5     & 1.3     & 1.7                         \\
                                                &                                        &                                  & A                    & 0.5     & 0.4     & 0.4     & 0.2     & 0.3                         \\
                                                & \multirow{3}{*}{$\eta_{\mathrm{ml}}$ } & \multirow{3}{*}{150, 300, 450, 600} & scale                & 200     & 200     & 200     & 200     & 200                         \\
                                                &                                        &                                  & errscale             & 20      & 25      & 20      & 5       & 5                           \\
                                                &                                        &                                  & kernel               & Mat\'ern  & Mat\'ern & Mat\'ern & Mat\'ern & Pow. Exp.                   \\ 
\hline
                                                & \multirow{3}{*}{ $\eta$ }              & \multirow{3}{*}{25, 35, 45, 55}     & \multicolumn{1}{l}{} & Set 1~  & Set 2   & Set 3   & Set 4   & Set 5                       \\ 
\cline{5-9}
\multirow{5}{*}{C2 and SMARTS } &             &     & $\nu$                & 2.2     & 1.8     & 1.9     & 1.3     & 1.7                         \\
                                                &                                        &                                  & A                    & 0.5     & 0.7     & 0.6     & 0.3     & 0.7                         \\
                                                & \multirow{3}{*}{$\eta_{\mathrm{ml}}$ } & \multirow{3}{*}{150, 300, 450, 600} & scale                & 200     & 200     & 200     & 150     & 300                         \\
                                                &                                        &                                  & errscale             & 25      & 25      & 20      & 10      & 25                          \\
                                                &                                        &                                  & kernel               & Mat\'ern & Mat\'ern & Mat\'ern & Mat\'ern & Pow. Exp.  \\
\hline
\end{tabular}
\end{table*}

We systematically apply this procedure for the sets of estimator parameters described in Table \ref{tab:regdiff_param}. For the free-knot spline technique we adopt a mean spacing between the knots $\eta$ of 25, 35, 45 and 55 days for the intrinsic spline and of $\eta_{ml} =$ 150, 300, 450 and 600 days for the extrinsic splines. Combining all possible combinations of $\eta$ and $\eta_{ml}$, we obtain 16 different sets of parameters. We slightly adapt this grid of parameters for shorter light curves since the mean knot separation cannot exceed the duration of the monitoring campaign. For the regression difference technique, we explored five different sets of parameters to test the robustness of the estimator against changes in the properties of the Gaussian process used to fit the data.\\

\subsection{Combining the time-delay estimates}
\label{combiningtd}

The pipeline presented here aims at systematically testing the robustness of the measurement against different assumptions for the variability of the quasar and of the extrinsic components, which are controlled by the estimator parameters. We decide not to optimise on the estimator parameters since we aim to marginalise over the various plausible microlensing models in our final time-delay uncertainties. On the other hand, marginalising over all estimator parameters is not optimal either because estimator parameters that are not necessarily well-suited to represent the data would be accounted for in the final estimate and this would degrade its precision.


Instead, we adopt an hybrid approach between marginalisation and optimisation, as described in \cite{Bonvin2018a} and \cite{Bonvin2019}. Among all the groups of time-delay estimates computed for various estimator parameters, we first select the most precise one as our reference group. We then compute the `tension' between the reference group and all the other groups. The tension between two time-delay estimates $\mathbf{E_A} = A^{+a_{+}}_{-a_{-}}$ and $\mathbf{E_B} = B^{+b_{+}}_{-b_{-}}$ with $A > B$, in units of $\sigma$ is defined as \citep{Bonvin2018a}: 
\begin{equation}
    \tau(\mathbf{E_{A}}, \mathbf{E_{B}}) = (A -B)/\sqrt{a_{-}^2 + b_{+}^2}.
\end{equation}
We also define the tension between two groups $\mathbf{G_1}$ and $\mathbf{G_2}$ as the maximum tension between the time-delay estimates from corresponding pairs of light curves : 
\begin{equation}
    \tau(\mathbf{G_1}, \mathbf{G_2}) = \max_{j}(\tau(\mathbf{E_{1,j}}, \mathbf{E_{2,j}})).
\end{equation}
We marginalise only if the tension $\tau$ between the reference group and the other groups, $\sigma$, exceeds a threshold of $\tau_{\rm thresh} = 0.5$. If this is the case, we select the most precise group among those whose tension exceeds $\tau_{\rm thresh}$ and combine it with the reference. This combination becomes the new reference group and we repeat this process until no more groups are in a tension that exceeds $\tau_{\rm thresh}$ with the reference. We note that in this approach, choosing $\tau_{\rm thresh} = 0$ corresponds to a proper marginalisation over all the groups whereas choosing $\tau_{\rm thresh} \sim \infty$ is equivalent to selecting only the most precise group. 

\begin{figure*}[htbp!]
    \begin{minipage}[c]{\textwidth}
    \includegraphics[width=0.49\textwidth]{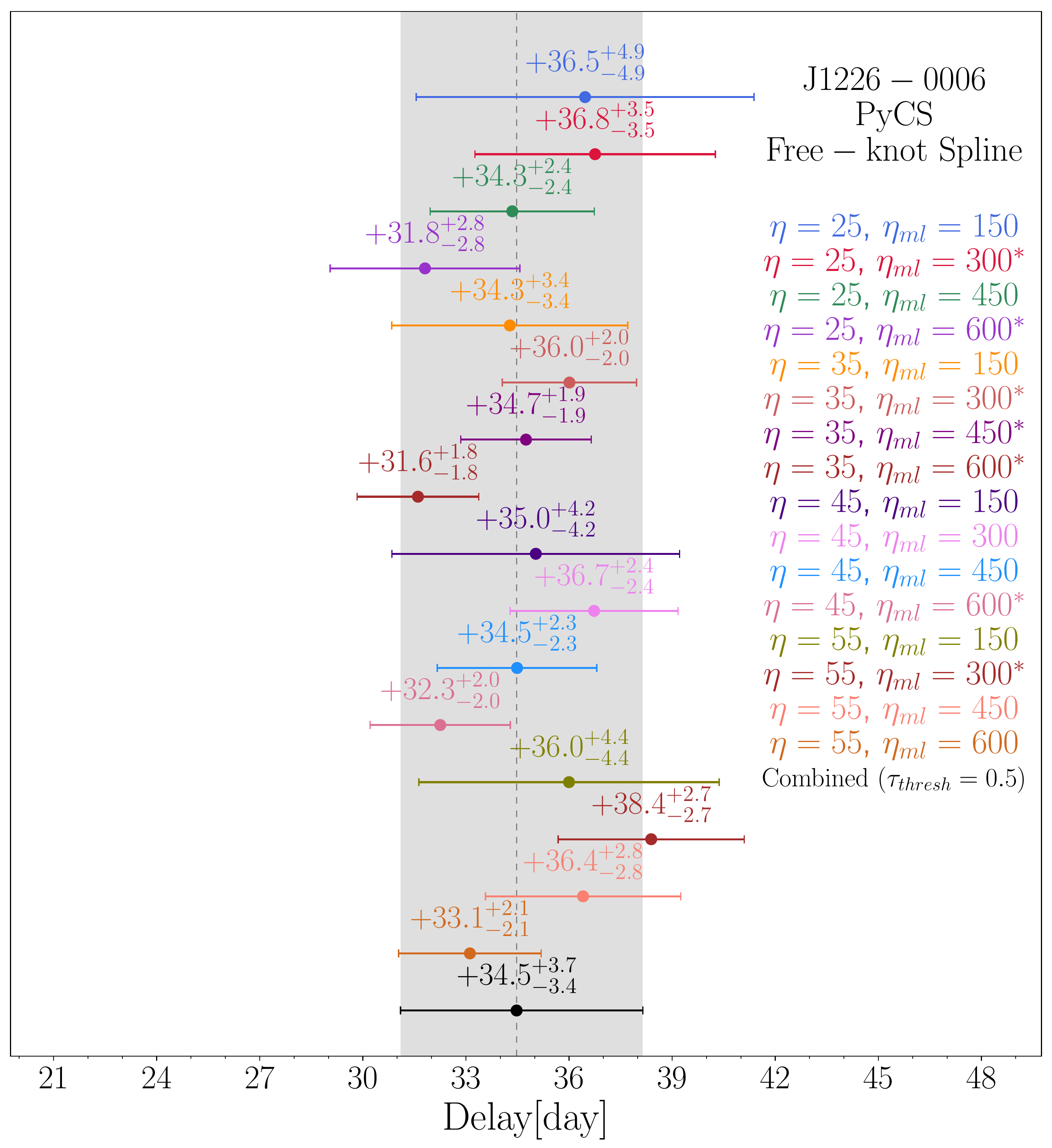}
    \includegraphics[width=0.49\textwidth]{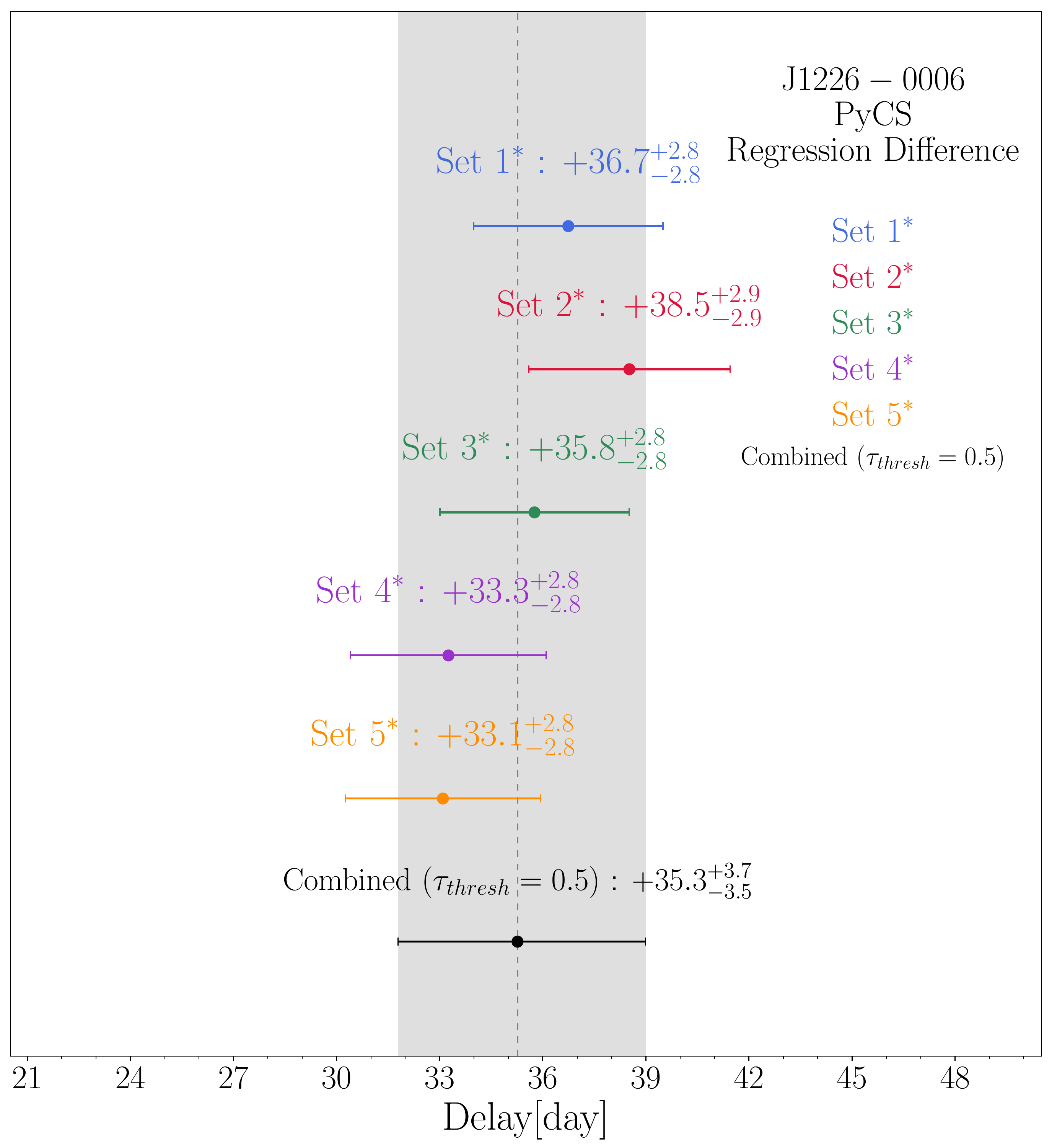}
    \end{minipage}
    \caption{Illustration of how we combine groups of time-delay estimates in a series on the ECAM data set for the lensed quasar \Jdouze. \textit{Left panel: } Series of time-delay estimates obtained for the free-knot spline technique. Each time-delay estimate presented on the plot corresponds to a particular choice of estimator parameters for the mean spacing between the knots of the intrinsic spline $\eta$, and of the extrinsic splines $\eta_{ml}$. The combined estimate is shown both in black at the bottom of the panel and in a grey shaded band, to visually ease the comparison with individual estimates and it corresponds to the combination described in section \ref{combiningtd} with $\tau_{\rm thresh} = 0.5$. The estimates used to produce this combination are marked with $^*$. \textit{Right panel: } Series of time-delay estimates obtained with the regression difference technique. The sets of parameter are detailed in Table \ref{tab:regdiff_param}.
    }
    \label{fig:delay_1226}
\end{figure*}

For each estimator, we obtain a series of time-delay estimates, which is composed of several groups that share the same data set but contain different combinations of the estimator parameters. We combine the groups within a series according to the procedure described above. Fig.~\ref{fig:delay_1226} illustrates each group's result, and compares them with the combined estimate for the free-knot spline and regression difference estimator in the case of \Jdouze. 

\begin{figure}
    \centering
    \includegraphics[width=0.49\textwidth]{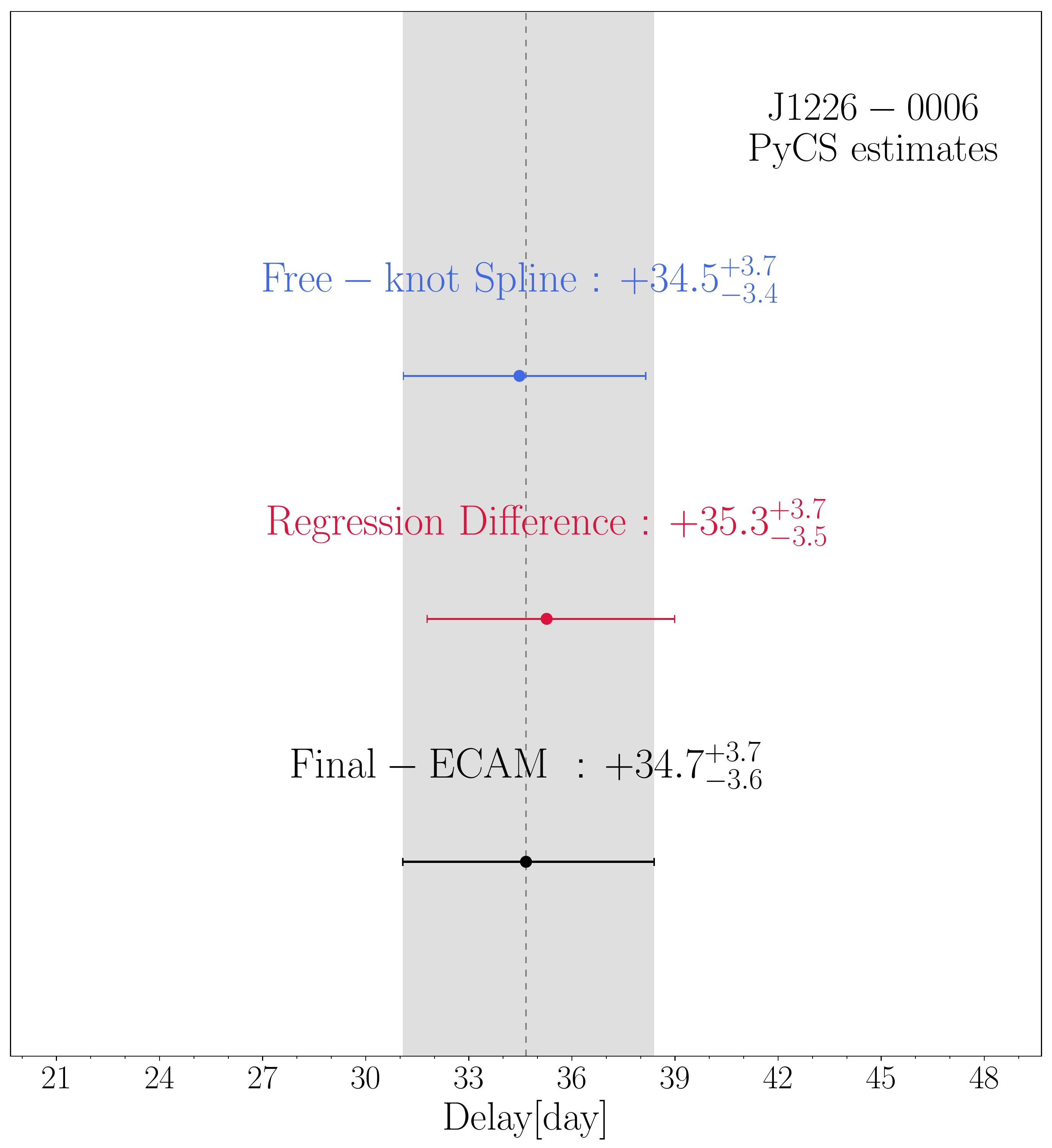}
    \caption{Final time-delay estimate for the regression difference and free-knot spline estimator. The marginalisation over these two estimates yields our final time-delay estimate for the ECAM data set, shown in a shaded grey band.}
    \label{fig:delay_1226b}
\end{figure}

After combining the groups in both series, we typically obtain two groups of time-delay estimates, one for each estimators. As the two estimators are applied on the same data set, they cannot be considered as independent. We therefore marginalise over the free-knot spline and regression difference techniques, resulting in our final time-delay estimates for a particular data set. Fig.~\ref{fig:delay_1226b} illustrates this process in the case of \Jdouze. 

\subsection{Combining data sets}
\label{sec:combining_datasets}
The last step consists in combining the data sets when several instruments were used to monitor the same target. When the data quality is not sufficient to obtain reliable time-delay estimates with one single data set, we prefer to merge the light curves taken from different instruments and run the full analysis on the joint data set. This typically occurs when one of the monitoring campaign does not contain enough well-defined sharp features in the light curves to constrain the time delay. Many factors can explain these failures, for example when the time delay is long compared with the length of the observing season  and limits the overlap between the useful parts of the light curves, or due to low variability of the quasar, photometric noise or limited sampling. 

\begin{figure}
    \begin{minipage}[c]{0.49\textwidth}
    \includegraphics[width=\textwidth]{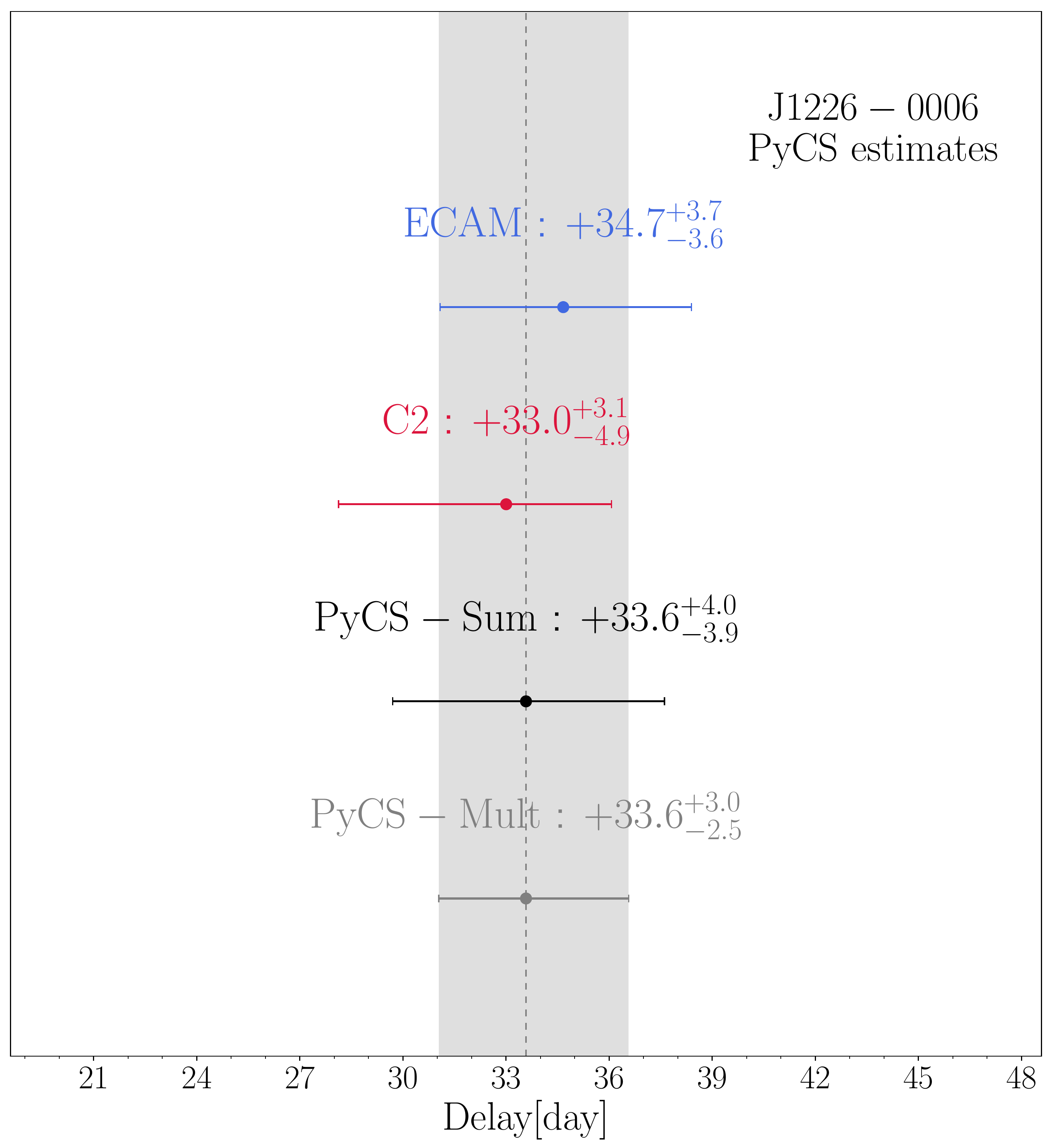}
    \end{minipage}
    \caption{Example of combination of ECAM and C2 data sets for \Jdouze. The \pycssum estimate corresponds to the marginalisation over the two data sets. The \pycsmult assumes that the two data sets are independent and corresponds to the multiplication of the  probability distributions for the two methods.}
    \label{fig:dataset_combination_1226}
\end{figure}

If the data quality is sufficient, that is, if there are enough well-defined inflection points that can be identified in all the light curves, we apply the measurement pipeline on each individual data set and then combine the resulting time-delay measurements. The most conservative way to combine different data sets consists in marginalizing over the different data sets, which is a valid approach in the case of overlapping monitoring campaigns. In this case, the measurements can be affected by the same systematic bias due to microlensing or to microlensing time delay and are therefore not independent. We call the final time-delay estimate obtained after marginalisation over the data sets \pycssum, as it effectively results from the sum of the individual groups. The second option consists in multiplying the groups of the individual data set, making the assumption that they do not share unaccounted systematics and that they are therefore independent measurements of the same quantity, which is the case when considering non-overlapping monitoring campaigns like the C2 and ECAM data sets. We call this the \pycsmult result. As an example, we show in Fig.~\ref{fig:dataset_combination_1226} the combination of the ECAM and C2 time-delay estimates for \Jdouze, for which both the \pycsmult and \pycssum results are shown.

As the microlensing time delay \citep{Tie2017} could potentially affect our measurements, making the observed time delay variable over a time scale similar to the microlensing time scale (i.e. from several months to several years) one should explicitly account for this effect as it was done in \cite{Bonvin2018a}. This is beyond the scope of this paper as including microlensing time delay is usually done at the same time the measured time delays are incorporated in the mass model of the lens \citep[e.g.][]{Rusu2019, Chen2018, Chen2019}. However, there is so far no evidence for our time-delay measurements being significantly time variable in the data. For example, the time-delay values we obtain when splitting our data in several chunks do not change more than the expected uncertainties \citep[e.g.][]{Tewes2013b, Bonvin2017} and we also do not measure any significant time delay between resolved narrow blends of images that have zero predicted cosmological (i.e. from macro model of the lens) time delay \citep[e.g.][]{Bonvin2019}. Future work may show or not if the microlensing time delay becomes significant when the precision of the data improve, for example, with high cadence monitoring \citep{Courbin2018}.

\section{Notes on individual objects}
\label{results}

\begin{figure*}[htbp!]
  \begin{adjustbox}{addcode={\begin{minipage}{\width}}{\caption{%
Relative uncertainties on the measured time delays for objects presented in this work (\textit{left}) and for other time delays available 
in the literature (\textit{right}). In the case of quadruply imaged systems and when several independent time delays are measured, we multiply the independent time-delay estimates to obtain the relative uncertainty that is achievable on $H_0$. The light blue region corresponds to a precision of less than 15\%. Systems marked with a $^*$ have uncertain or multi-modal time delays. The time delays from the literature can be found in (1) \cite{Biggs2018}, (2) \cite{Courbin2018}, (3) \cite{Hainline2013}, (4) \cite{Jakobson2005}, (5) \cite{Rathna2013}, (6) \cite{Poindexter2007}, (7) \cite{Bonvin2018a}, (8) \cite{Eulaers2013}, (9) \cite{Goicoechea2016}, (10) \cite{Akhunov2017}, (11) \cite{Patnaik2001}, (12) \cite{Shalyapin2019}, (13) \cite{Burud2002b}, (14) \cite{Burud2000}, (15) \cite{Fassnacht2002}, (16) \cite{Vuissoz2007}, (17) \cite{Lovell1998}, (18) \cite{Bonvin2019}, (19) \cite{Eulaers2013}. We note that these time delays from the literature were not measured in an homogeneous way, sometimes not taking into account the microlensing in the light curves. The uncertainties for some of these systems are therefore likely to be largely underestimated.\label{fig:sumarry}
      }\end{minipage}},rotate=90,center}
    \includegraphics[scale =0.44]{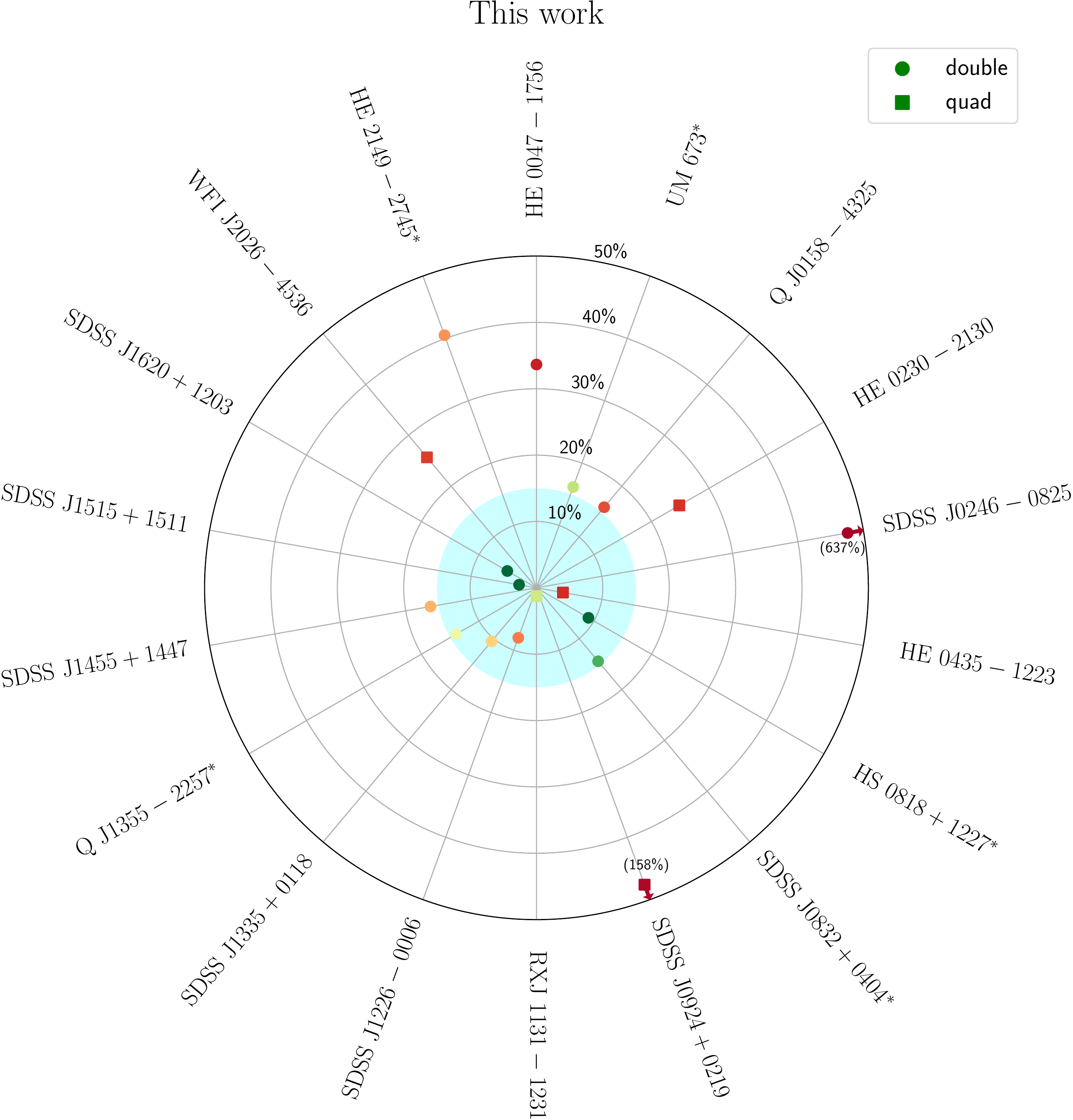}%
    \hspace{0.1cm}
    \includegraphics[scale =0.47]{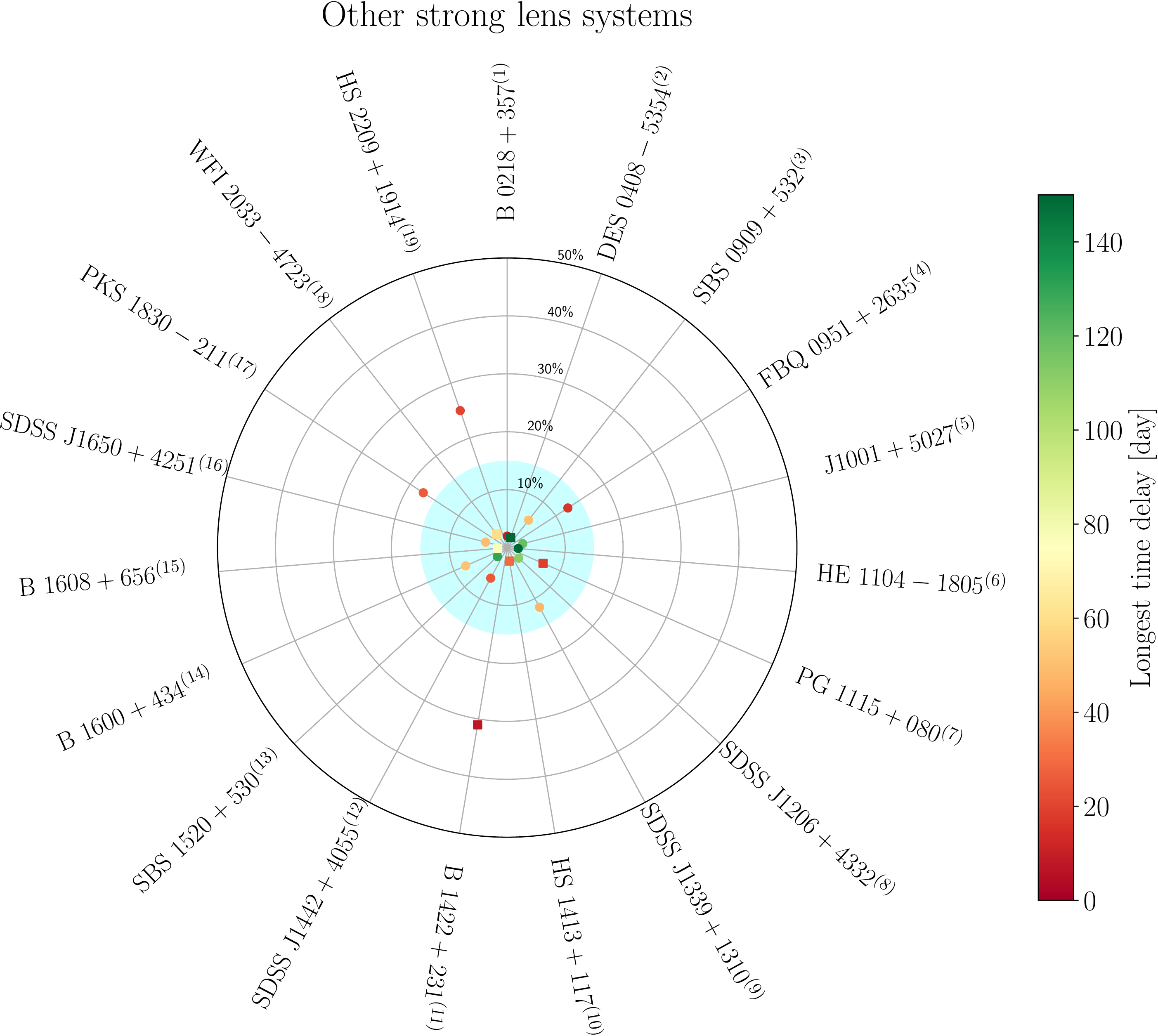}
  \end{adjustbox}
\end{figure*}

The new automated \pycs pipeline was applied to 18 lensed quasars, for which the data quality was sufficient to obtain robust time delays. As each object has its own specificities such as the duration of the monitoring campaigns, the mean sampling of the light curves, the instrument used, etc., we detail below the results for individual objects and give our best time-delay values along with the associated 1-$\sigma$ uncertainties. In giving our results for a pair of images A and B, a negative value of $\Delta t_{AB}$ means that the information is visible first in image A. Conversely, a positive time delay $\Delta t_{AB}$ means that B is the leading image. All time delays measured in this paper, along with previously published time delays available in the literature are summarised in Table \ref{tab:delay} and Fig.~\ref{fig:sumarry}. We list here only the system where the background quasar is lensed by a single galaxy and not by a cluster or a group of galaxies. When comparing the uncertainties measured in this work with previous time delays estimates published in the literature, we recall that the modelling of the microlensing is extensively tested here, which was not always the case in previous works. For quadruply imaged systems and when several independent time delays are measured, the relative precision that we quote on Fig.~\ref{fig:sumarry} corresponds to the combination of the independent time-delay estimates. This corresponds to the minimum relative uncertainty that is achievable on $H_0$.

\subsection{\HEzerozero}
\HEzerozero is a doubly imaged quasar at redshift $z_{\rm source}=1.678$ with the lens galaxy at redshift $z_{\rm lens}=0.407$ \citep{Wisotzki2004, Eigenbrod2006, Sluse2012}. Time delays were previously measured by \cite{Giannini2017} who found $\Delta t_{AB} = -7.6\pm 1.8$ days with five seasons of monitoring at the 1.54 m Danish telescope at the ESO La Silla observatory. With the ECAM and C2 data at Euler, we find $\Delta t_{AB} = -9.5^{+3.9}_{-4.0}$ days and $\Delta t_{AB} = -12.8^{+8.3}_{-8.3}$ days respectively. Combining our two Euler measurements, our final \pycsmult estimate is $\Delta t_{AB} = -10.4^{+3.5}_{-3.5}$ days.

\subsection{\UMsix~ (\Qzeroun)}
\UMsix~(\Qzeroun) is a doubly imaged quasar at redshift $z_{\rm source}=2.73$ discovered by \cite{Surdej1987} with a lens galaxy at redshift $z_{\rm lens}$ = 0.491 \citep{Eigenbrod2007}.
The time delay was first measured by \cite{Koptelova2012} from ten seasons of monitoring at the 1.5m telescope of Maidanak observatory. The authors obtained  $\Delta t_{AB} = -95^{+5 +14}_{-16 -29}$ days (68 and 95 \% confidence interval). \cite{Oscoz2013} also found $\Delta t_{AB} = -72 \pm 22$ days from 12 years of observation at the Teide Observatory. Our measurement is compatible with previous estimate but yield larger 1-$\sigma$ uncertainties than the value of \cite{Koptelova2012} as our analysis explicitly includes several microlensing models, which were not accounted for in previous studies. We find $\Delta t_{AB} = -97.7 ^{+16.1}_{-15.5}$ days, but we note that there is weak evidence for another value around $\Delta t_{AB} = -200$ days. 

\subsection{\Jzeroun}
The doubly imaged QSO, \Jzeroun, at redshift $z_{\rm source}=1.29$, also known as CTQ 414, was discovered by \cite{Morgan1999} during the Cal\'an-Tololo Quasar survey.  \cite{Faure2009} measured a tentative lens redshift of $z_{\rm lens} = 0.317$. A model of the lens was also proposed in this work and  a truncated pseudo-isothermal elliptical mass distribution (TPIEMD) was favoured by the data. Under these assumptions, the predicted time delay between image A and B was around $\Delta t_{AB}=-14 $ days. In their microlensing analysis, \cite{Morgan2008} failed to measure the time delay but successfully estimated the accretion disc size. \cite{Morgan2012} added four new seasons to the previous analysis and predicted a delay $\Delta t_{AB} = -12.4$ days using a singular isothermal sphere (SIS) as a lens model and flat-\lcdm cosmology with $h$=0.7, $\Omega_M=0.3$ and $\Omega_{\Lambda} = 0.7$. The simultaneous microlensing and time-delay analysis also favoured a negative AB delay (i.e. A leads B). 

In this work we add seven seasons of monitoring to the data set published in \cite{Morgan2012}. Significant variations of the quasar after 2011 allows us to constrain the time delay to $\Delta t_{AB} = -22.5 ^{+3.5}_{-4.1}$ days from the joint ECAM and C2 data set (Euler set). We attempt to measure the time delay independently on the ECAM and C2 data set, but the C2 light curve quality is insufficient. We also re-analyse the SMARTS data from \cite{Morgan2012} and obtain $\Delta t_{AB} = -22.3 ^{+9.5}_{-8.0}$ days, in excellent agreement with the Euler estimate. Combining these two estimates, we find $\Delta t_{AB} = -22.7 ^{+3.6}_{-3.6}$ days. 

This object is significantly affected by microlensing, with microlensing variations of about one magnitude over the 13 years of monitoring, as can be seen on the top panel of Fig.~\ref{fig:annex_lcs2}. The difference light curve exhibits strong short-timescale microlensing variations in the five first seasons of monitoring in addition to the long term trend. 

\subsection{\HEzerodeux}
\HEzerodeux is a quadruply imaged quasar at $z_{\rm source}=2.162$ discovered by  \cite{Wisotzki1999}. The redshift of the lens $z_{\rm lens}$=0.523, was measured independently by \cite{Eigenbrod2006} and \cite{Ofek2006}. Images A and B are in a close fold configuration and separated by only $0\arcsec74$. This two images are not resolved in the Euler exposures even after deconvolution. Their fluxes are therefore summed in one single virtual image $A'=A+B$. Both the C2 and ECAM data sets have relatively strong intrinsic features that allow us to measure the time delay between image A$'$ and C. Image D is however too faint to obtain a robust measurement. We measure $\Delta t_{A'C} = +13.2 ^{+5.8}_{-5.3}$ days from the ECAM data set, $\Delta t_{A'C} = +16.4 ^{+14.3}_{-16.6}$ days from the C2 data set and $\Delta t_{A'C} = +15.7 ^{+4.2}_{-3.6}$ days from the joint C2-ECAM light curves. We adopt the later as our final time-delay estimate. 

\subsection{\Jzerodeuxquatresix}
\Jzerodeuxquatresix is a doubly imaged quasar discovered by \cite{Inada2005} ($z_{\rm source} = 1.689$, $z_{\rm lens} = 0.723$). We present here the results of the monitoring campaign from  November 2006 to April 2018. The time-delay estimate from the ECAM data set ($\Delta t_{AB} = -1.9 ^{+6.0}_{-7.2}$ days) is compatible with the time delay measured from the C2 data set ($\Delta t_{AB} = +2.1 ^{+10.5}_{-9.7}$ days). The \pycsmult estimate yields $\Delta t_{AB} = -1.3 ^{+5.8}_{-5.2}$ days, also in good agreement with the measurement performed on the joint ECAM-C2 light curve $\Delta t_{AB} = +0.8 ^{+5.0}_{-5.2}$ days, which is our final value. Although this object displays prominent variations, its time delay is compatible with zero, making it of little use for cosmological purposes.

\subsection{\HEzeroquatre}
\HEzeroquatre is quadruply imaged quasar in a cross configuration \citep[$z_{\rm source} = 1.693$, $z_{\rm lens} = 0.454$,][]{Wisotzki2002, Sluse2012}. The time delays were first
measured in \cite{Kochanek2006} with two seasons of optical monitoring, and later confirmed in \cite{Courbin2011} and \cite{Bonvin2017} with respectively seven and thirteen seasons of monitoring. We complement the data presented in \cite{Bonvin2017} with two additional seasons and measure the time delays with the automated pipeline presented in Sect.~\ref{sec:measurement pipeline}. We analyse separately the data presented in \cite{Courbin2011} corresponding to the C2, SMARTS, Maidanak and Mercator data (hereafter the C11 data set) and the ECAM data set. Our group of time-delay estimates on the C11 data set is $\vec{G_{C11}} = [\Delta t_{AB} = -9.8^{+1.0}_{-1.0}, \Delta t_{AC} = -1.6^{+0.9}_{-1.0}, \Delta t_{AD} = -14.2^{+1.1}_{-1.1}]$ days compatible with less than 1.4$\sigma$ tension with our group of estimates on the ECAM data set $\vec{G_{ECAM}} = [\Delta t_{AB} = -7.6^{+1.3}_{-1.3}, \Delta t_{AC} = +0.4^{+1.1}_{-1.1}, \Delta t_{AD} = -14.2^{+1.1}_{-1.1}]$ days. Our final \pycsmult group of estimates combining the C11 and ECAM gives $\vec{G_{\pycsmult}} = [\Delta t_{AB} = -9.0^{+0.8}_{-0.8}, \Delta t_{AC} = -0.8^{+0.8}_{-0.7}, \Delta t_{AD} = -13.8^{+0.8}_{-0.8}]$ days in excellent agreement with the estimates that we obtain on the joint C11+ECAM light curves $\vec{G_{all}} = [\Delta t_{AB} = -8.3^{+0.9}_{-0.8}, \Delta t_{AC} = -0.4^{+0.8}_{-0.8}, \Delta t_{AD} = -13.5^{+0.9}_{-0.8}]$ days. 

\subsection{\HSzerohuit}
\HSzerohuit is a doubly imaged quasar discovered in the Hamburg Quasar Survey. Its redshift is $z_{\rm source}=3.115$ \citep{Hagen2000} and the redshift of the lensing galaxy is $z_{\rm lens} = 0.39$. Using the joint ECAM-C2 light curves we propose a tentative delay of $\Delta t_{AB} = -153.8^{+13.2}_{-14.6}$ days. We also find evidence for a multi-modal delay around -53 days. 

\subsection{\Jzerohuit}
\Jzerohuit is doubly imaged quasar discovered by \cite{Oguri2008} at $z_{\rm source}=1.115$ and whose lens it at redshift $z_{\rm lens} = 0.659$. It was monitored by the Euler Swiss telescope shortly after its discovery, from November 2010 to May 2018. The measurement is challenging as the time delay favoured by the data is close to that of the season length. Our best estimate is $\Delta t_{AB} = -125.3^{+12.8}_{-23.4}$ days from the ECAM data set.

\subsection{\Jzeroneuf}
\Jzeroneuf is a quadruply imaged quasar at redshift $z_{\rm source}=1.523$ ($z_{\rm lens}=0.393$) discovered by \cite{Inada2003}. As image D is faint and very close angularly to image A ($0\arcsec69$), we sum the flux of these two images into one single virtual image A$'$. We are able to measure the time delay in both the SMARTS and Euler data set but only between the two brighter images, A$'$ and B. The light curve of image C is not of sufficient quality to distinguish features, essential for a reliable time-delay measurement. We find $\Delta t_{A'B} = +2.5 ^{+4.3}_{-4.3}$ days using the Euler light curves and $\Delta t_{A'B} = +3.0 ^{+8.2}_{-7.9}$ days with the SMARTS light curve. Combining these two measurements, we obtain the final \pycsmult estimate : $\Delta t_{A'B} = +2.4 ^{+3.8}_{-3.8}$ days.

\subsection{\RXJonze}
\RXJonze is a quadruply imaged quasar at redshift $z_{\rm source} = 0.657$ in a cusp configuration with a lens galaxy at redshift $z_{\rm lens} = 0.295$ \citep{Sluse2003, Sluse2007}. The first attempt to measure the time delays is reported in \cite{Morgan2006b} but appeared to be incorrect due to short light curves, insufficient data quality and possible microlensing variability that imitates the quasar variation \citep{Tewes2013b}. These time-delay estimates were corrected by \cite{Tewes2013b} with nine seasons of monitoring, demonstrating the need of long monitoring campaigns to obtain robust time-delay estimates unless high-cadence and high-signal-to-noise observations are available. We complement the data presented in \cite{Tewes2013b} with six new seasons of monitoring. In our analysis, we separate the data into two different sets; the data acquired between December 2003 and July 2010 at the SMARTS, Mercator, and Euler (C2) telescopes (hereafter the T13 data set) and the data acquired by the ECAM camera at the Euler Telescope (ECAM data set). We obtain a group of time-delay estimates on the T13 data set $\vec{G_{T13}} = [\Delta t_{AB} = 0.0^{+1.2}_{-1.0}, \Delta t_{AC} = -0.7^{+1.3}_{-1.4}, \Delta t_{AD} = -92.9^{+2.3}_{-2.5}]$ days. When applying our method on the ECAM data set we obtain $\vec{G_{ECAM}} = [\Delta t_{AB} = +2.6^{+0.8}_{-0.9}, \Delta t_{AC} = -1.4^{+2.3}_{-2.5}, \Delta t_{AD} = -91.6^{+2.7}_{-3.0}]$ days. Our \pycsmult final group of time-delay estimates is $\vec{G_{\pycsmult}} = [\Delta t_{AB} = +1.6^{+0.7}_{-0.7}, \Delta t_{AC} = -1.0^{+1.2}_{-1.2}, \Delta t_{AD} = -92.5^{+1.9}_{-1.8}]$. We obtain very similar time delays and uncertainties when running our measurement pipeline on the joint T13+ECAM light curves $\vec{G_{all}} = [\Delta t_{AB} = +1.3^{+1.0}_{-0.9}, \Delta t_{AC} = -1.0^{+1.6}_{-1.4}, \Delta t_{AD} = -92.7^{+2.3}_{-2.3}]$ days. 

\subsection{\Jdouze}
\Jdouze is a doubly imaged quasar at redshift $z_{\rm source} = 1.123$ and a lensing galaxy at $z_{\rm lens} = 0.517$ \citep{Inada2008}.  We use this object to illustrate the time-delay measurement pipeline developed for this paper. As shown on Fig.~\ref{fig:delay_1226}, we obtain $\Delta t_{AB} = +34.7 ^{+3.7}_{-3.6}$ days using the ECAM data set and $\Delta t_{AB} = +33.0 ^{+3.1}_{-4.9}$ days on the C2 data set. Combining the two measurements (\pycsmult) gives  $\Delta t_{AB} = +33.7 ^{+2.7}_{-2.7}$ days, which is our final measurement, in good agreement with the joint ECAM-C2 estimate $\Delta t_{AB} = +31.4 ^{+4.2}_{-5.8}$ days. 

\subsection{\Jtreizetrentecinq}
\Jtreizetrentecinq is a doubly imaged quasar  at $z_{\rm source} = 1.571$ and a lensing galaxy at $z_{\rm lens} = 0.440$, \citep{Oguri2004, Eigenbrod2006}. A tentative predicted delay $\Delta t_{\rm AB, pred} = -43$ days was proposed in \cite{Oguri2004}, assuming $H_0 = 70 $ \kmspc. We measure the time delay between images A an B on both the ECAM and C2 data sets. We obtain respectively $\Delta t_{AB} = -61.3 ^{+9.9}_{-10.3}$ days using the C2 data and $\Delta t_{AB} = -52.5 ^{+7.5}_{-7.2}$ days with ECAM data. As the data quality of both data sets allows a reliable time-delay estimate and since the two measurements are compatible within their 1-$\sigma$ uncertainties we can combine them. We obtain a final \pycsmult estimate of $\Delta t_{AB} = -56.0 ^{+5.7}_{-6.1}$ days, which is well compatible with the estimate from the joint ECAM-C2 light curves $\Delta t_{AB} = -54.8 ^{+7.3}_{-7.3}$ days. 

\subsection{\Qtreize}
\Qtreize is a doubly imaged quasar at redshift $z_{\rm source}=1.370$ \citep{Morgan2003}. The redshift of the lens is $z_{\rm lens} = 0.701$ \citep{Eigenbrod2006}. A time delay of $\Delta t_{AB} = -89 ^{+28}_{-39}$ days was predicted by \cite{Saha2006} assuming $H_0 = 70$ \kmspc and ensembles of non-parametric models. \Qtreize was monitored for seven seasons with the C2 camera and for additional three seasons with ECAM. We apply our measurement pipeline on the joint ECAM-C2 data set as the ECAM light curves alone does not contain enough structures to measure the time delay. We propose a tentative delay of $\Delta t_{AB} = -81.5 ^{+10.8}_{-12.0}$ days. Note, however, that while the light curve of the brightest quasar image, A, displays very well defined variations, the much fainter image B is very noisy and is the limiting factor in this time-delay measurement. 

\subsection{\Jquatorzecinquantecinq}
The doubly imaged quasar \Jquatorzecinquantecinq was discovered by \cite{Kayo2010} at redshift $z_{\rm source} = 1.424$. This object was monitored during seven seasons with the ECAM camera. We measured a time delay $\Delta t_{AB} = -47.2 ^{+7.5}_{-7.8}$ days using only the ECAM data set. 

\subsection{\Jquinzequinze}
\Jquinzequinze is a well separated and bright doubly imaged quasar at redshift $z_{\rm source} = 2.054$, $z_{\rm lens} = 0.742$ \citep{Inada2014}.  \cite{Shalyapin2017} first measured the time delay between images A and B and found $\Delta t_{AB} = -211 \pm 5$ days. In this work, we re-analyse the photometric data taken at the Liverpool Telescope between May 2014 and September 2016 with \pycs, testing the impact of several microlensing models on the time-delay measurement. We complement this data set with three seasons of monitoring taken with the ECAM instrument at the Euler Swiss Telescope between March 2014 and August 2017.
The light curves of images A and B only have a very small overlap due to the time delay being of roughly the same duration as the season gap. We therefore obtain a poor precision using the ECAM data set alone $\Delta t_{AB} = -199.6 \pm 16.0 $ days. The only intrinsic variation that is clearly visible in both the A and B light curve at the end of season 2015 in A and at the beginning of season 2016 is insufficiently sampled in the ECAM data set to obtain a precise estimate. Using the Liverpool data set alone, the better monitoring cadence and the longer monitoring season allows us to  measure $\Delta t_{AB} = -214.6 ^{+4.6}_{-7.6}$ days. This is in good agreement with previous estimates of \cite{Shalyapin2017} on the same data set but using a different methodology. The slightly larger uncertainties of our analysis can be explained by the fact that we extensively test against a broad variety of different microlensing models and that we marginalise over two different curve shifting techniques. 

By combining the two data sets into one single light curve, we significantly improve the sampling on the critical part of the curves that contain distinctive features in both A and B images. Our final estimate on the joint Euler+Liverpool light curves is $\Delta t_{AB} = -210.5 ^{+5.5}_{-5.7}$ days. 
 
\subsection{\Jseizevingt}
The doubly imaged quasar \Jseizevingt  ($z_{\rm source} = 1.158$, $z_{\rm lens}=0.398$) was discovered in the Sloan Sky Digital Survey (SDSS) by \cite{Kayo2010}. 
This object was mainly monitored with the ECAM instrument during seven seasons and with the C2 camera for one season. As the C2 data alone are largely insufficient to measure a time delay, we merge the two data sets in one single light curve. We apply our method on the joint ECAM-C2 light curves and obtain $\Delta t_{AB} = -171.5 ^{+8.7}_{-8.7}$ days. Similarly to \Jquinzequinze, this lensed quasar has a time delay of the same duration of the season gap which limits the achievable precision.

\subsection{\WFIvingtvingtsix}
\WFIvingtvingtsix is a quadruply imaged quasar in a fold configuration  ($z_{\rm source} = 2.23$, and uncertain lens redshift $z_{\rm lens} = 1.04$) discovered by \cite{Morgan2004}. \cite{Saha2006} predicted an AB time delay of $15^{+2}_{-6}$ days, from non-parametric modelling of point-like image positions. As images A1 and A2 are still not resolved after deconvolution because of their low image separation ($0\arcsec33$), we add the flux of these two images into one virtual image, A. Only the time delay between image A and B is measurable with our data. The light curve of image C is too noisy to identify intrinsic variation of the quasar and reliably estimate a time delay. In \cite{Cornachione2019} only the AB time delay is presented, whereas the AC delay is predicted by modelling the lens and assuming a fiducial cosmology. This paper is focusing on determining the accretion disc size of this object using the microlensing. We present here the details of the measurement of the AB time delay.  

Most of the constraining features of the light curves are concentrated in the ECAM data, especially in the 2014 and 2015 seasons. We use separately the ECAM data and the joint C2-ECAM data and obtain respectively $\Delta t_{AB} = +18.7 ^{+4.9}_{-4.7}$ days and $\Delta t_{AB} = +18.4 ^{+6.7}_{-9.6}$ days. The C2 data are particularly noisy and do not contain clear variations of the quasar visible in either of images A and B. Adding them to the ECAM data results in a loss of precision of the measurement without changing the median value of the estimate. For any follow-up work, we therefore recommend to use only the ECAM estimate. 

\subsection{\HEvingtetun}
The doubly imaged Broad Absorption Line (BAL) quasar \HEvingtetun was discovered in the Hamburg/ESO survey by \cite{Wisotzki1996}. The redshift of the quasar is $z_{\rm source}=2.033$ and the redshift of the lens is $z_{\rm lens} = 0.603$ \citep{Eigenbrod2006}. The time delay between image A and B was previously published by \cite{Burud2002}, who found $\Delta t_{AB} = -103 \pm 12 $ days but a reanalysis of the same data by \cite{Eulaers2011} concluded that the previous measurement was not reliable. 

We propose a tentative time-delay estimate of $\Delta t_{AB} = -39.0^{+14.9}_{-16.7}$ days (\pycsmult C2-ECAM), significantly different from the previous measurement although our data do not allow us to completely exclude the previous time-delay estimate. Our measurement on the joint C2-ECAM light curves gives $\Delta t_{AB} = -32.4^{+18.3}_{-9.3}$, in agreement with the \pycsmult estimate.  We note that this is a BAL quasar with therefore a complex internal structure. This quasar may be a good candidate to detect the microlensing time delay as described in \cite{Tie2017}, although in this case the source size effect may be due to the BAL structure around the quasar rather than to the time propagation of light in the accretion disc. 

\begin{table}[htbp!]
\caption{Summary of the quasar lensed by a single galaxy with known time delays. The upper part of the table gives our new measurements while the lower part gives a summary of the published time delays in the literature.\label{tab:delay}}
\renewcommand{\arraystretch}{1.35}
\resizebox{0.5\textwidth}{!}{%
\begin{tabular}{l l l l}
\hline
Lenses                             & Time-Delays [Day]                              & Comments                                           &  \\ \hline
\HEzerozero                        & $\Delta t_{AB} = -10.4^{+3.5}_{-3.5}$    & \pbox[t]{5cm}{\pycsmult C2-ECAM}                                &  \\
\UMsix~ (\Qzeroun)                             & $\Delta t_{AB} = -97.7^{+16.1}_{-15.5}$  & \pbox[t]{5cm}{from joint C2+ECAM light curves (uncertain)}        &  \\
\Jzeroun                           & $\Delta t_{AB} = -22.7^{+3.6}_{-3.6}$    & \pbox[t]{5cm}{\pycsmult C2-SMARTS-ECAM }                           &  \\
\HEzerodeux                        & $\Delta t_{A'C} = 15.7^{+4.2}_{-3.6}$     & \pbox[t]{5cm}{\pycsmult C2-ECAM    }                              &  \\
\Jzerodeuxquatresix                & $\Delta t_{AB} = 0.8^{+5.0}_{-5.2}$      & \pbox[t]{5cm}{from joint C2-ECAM light curves }                   &  \\
\HEzeroquatre                      & $\Delta t_{AB} = -9.0^{+0.8}_{-0.8}$     & \pbox[t]{5cm}{\pycsmult C11-ECAM} &  \\
                                   & $\Delta t_{AC} = -0.8^{+0.8}_{-0.7}$     &                                                    &  \\
                                   & $\Delta t_{AD} = -13.8^{+0.8}_{-0.8}$    &                                                    &  \\
                                   & $\Delta t_{BC} = 7.8^{+0.9}_{-0.9}$      &                                                    &  \\
                                   & $\Delta t_{BD} = -5.4^{+0.9}_{-0.8}$     &                                                    &  \\
                                   & $\Delta t_{CD} = -13.2^{+0.8}_{-0.8}$    &                                                    &  \\
\HSzerohuit                        & $\Delta t_{AB} = -153.8^{+13.2}_{-14.6}$ & \pbox[t]{5cm}{from joint C2-ECAM light curves (uncertain)}          &  \\
\Jzerohuit                         & $\Delta t_{AB} = -125.3^{+12.8}_{-23.4}$ & \pbox[t]{5cm}{from ECAM light curves (uncertain)}                  &  \\
\Jzeroneuf                         & $\Delta t_{A'B} = 2.4^{+3.8}_{-3.8}$      & \pbox[t]{5cm}{\pycsmult Euler-SMARTS }                            &  \\
\RXJonze                           & $\Delta t_{AB} = 1.6^{+0.7}_{-0.7}$      & \pbox[t]{5cm}{\pycsmult T13-ECAM}  &  \\
                                   & $\Delta t_{AC} = -1.0^{+1.2}_{-1.2}$     &                                                    &  \\
                                   & $\Delta t_{AD} = -92.5^{+1.9}_{-1.8}$    &                                                    &  \\
                                   & $\Delta t_{BC} = -2.0^{+1.4}_{-1.3}$     &                                                    &  \\
                                   & $\Delta t_{BD} = -93.7^{+2.0}_{-2.0}$    &                                                    &  \\
                                   & $\Delta t_{CD} = -91.8^{+2.2}_{-2.3}$    &                                                    &  \\
\Jdouze                            & $\Delta t_{AB} = 33.7^{+2.7}_{-2.7}$     & \pbox[t]{5cm}{\pycsmult C2-ECAM}                                  &  \\
\Jtreizetrentecinq                 & $\Delta t_{AB} = -56.0^{+5.7}_{-6.1}$    & \pbox[t]{5cm}{\pycsmult C2-ECAM}                                 &  \\
\Qtreize                           & $\Delta t_{AB} = -81.5^{+10.8}_{-12.0}$  & \pbox[t]{5cm}{from joint C2+ECAM light curves (uncertain)}        &  \\
\Jquatorzecinquantecinq            & $\Delta t_{AB} = -47.2^{+7.5}_{-7.8}$    & \pbox[t]{5cm}{from ECAM light-curve }                             &  \\
\Jquinzequinze                     & $\Delta t_{AB} = -210.2^{+5.5}_{-5.7}$   & \pbox[t]{5cm}{from joint ECAM+Liverpool light curves  }           &  \\
\Jseizevingt                       & $\Delta t_{AB} = -171.5^{+8.7}_{-8.7}$   & \pbox[t]{5cm}{from joint C2+ECAM light curves }                   &  \\
\WFIvingtvingtsix                  & $\Delta t_{AB} = 18.7^{+4.7}_{-4.9}$     & \pbox[t]{5cm}{from ECAM light curves  }                           &  \\
\HEvingtetun                       & $\Delta t_{AB} = -39.0^{+14.9}_{-16.7}$   & \pbox[t]{5cm}{\pycsmult C2-ECAM (uncertain)}                                        &  \\ \hline
B~0218$+$357\xspace      & $\Delta t_{AB} = -11.3^{+0.2}_{-0.2}$    & from \cite{Biggs2018}                              &  \\
DES~0408$-$5354\xspace   & $\Delta t_{AB} = -112.1^{+2.1}_{-2.1}$   & from \cite{Courbin2018}                            &  \\
                                   & $\Delta t_{AD} = -155.5^{+12.8}_{-12.8}$ &                                                    &  \\
                                   & $\Delta t_{BD} = -42.4^{+17.6}_{-17.6}$  &                                                    &  \\
SBS~0909$+$532\xspace    & $\Delta t_{AB} = -50.0^{+2.0}_{-4.0}$    & from \cite{Hainline2013}                           &  \\
FBQ~0951$+$2635\xspace   & $\Delta t_{AB} = 16.0^{+2.0}_{-2.0}$     & from \cite{Jakobson2005}                           &  \\
J1001$+$5027\xspace       & $\Delta t_{AB} = -119.3^{+3.3}_{-3.3}$   & from \cite{Rathna2013}                             &  \\
HE~1104$-$1805\xspace    & $\Delta t_{AB} = 152.2^{+2.8}_{-3.0}$    & from \cite{Poindexter2007}                         &  \\
PG~1115$+$080\xspace     & $\Delta t_{AB} = -8.3^{+1.5}_{-1.6}$     & from \cite{Bonvin2018a}                            &  \\
                                   & $\Delta t_{AC} = 9.9^{+1.1}_{-1.1}$      &                                                    &  \\
                                   & $\Delta t_{BC} = 18.8^{+1.6}_{-1.6}$     &                                                    &  \\
SDSS~J1206$+$4332\xspace & $\Delta t_{AB} = -111.3^{+3.0}_{-3.0}$   & from \cite{Eulaers2013}                            &  \\
SDSS~J1339$+$1310\xspace & $\Delta t_{AB} = 47.0^{+5.0}_{-6.0}$     & from \cite{Goicoechea2016}                         &  \\
HS~1413$+$117\xspace     & $\Delta t_{AB} = -17.4^{+2.1}_{-2.1}$    & from \cite{Akhunov2017}                            &  \\
                                   & $\Delta t_{AC} = -18.9^{+2.8}_{-2.8}$    &                                                    &  \\
                                   & $\Delta t_{AD} = 28.8^{+0.7}_{-0.7}$     &                                                    &  \\
B~1422$+$231\xspace      & $\Delta t_{AB} = -1.5^{+1.4}_{-1.4}$     & from \cite{Patnaik2001}                            &  \\
                                   & $\Delta t_{AC} = 7.6^{+2.5}_{-2.5}$      &                                                    &  \\
                                   & $\Delta t_{BC} = 8.2^{+2.0}_{-2.0}$      &                                                    &  \\
SDSS~J1442$+$4055\xspace & $\Delta t_{AB} = -25.0^{+1.5}_{-1.5}$    & from \cite{Shalyapin2019}                          &  \\
SBS~1520$+$530\xspace    & $\Delta t_{AB} = -130.0^{+3.0}_{-3.0}$   & from \cite{Burud2002b}                             &  \\
B~1600$+$434\xspace      & $\Delta t_{AB} = -51.0^{+4.0}_{-4.0}$    & from \cite{Burud2000}                              &  \\
B~1608$+$656\xspace      & $\Delta t_{AB} = 31.5^{+2.0}_{-1.0}$     & from \cite{Fassnacht2002}                          &  \\
                                   & $\Delta t_{CB} = 36.0^{+1.5}_{-1.5}$     &                                                    &  \\
                                   & $\Delta t_{DB} = 77.0^{+2.0}_{-1.0}$     &                                                    &  \\
SDSS~J1650$+$4251\xspace & $\Delta t_{AB} = -49.5^{+1.9}_{-1.9}$    & from \cite{Vuissoz2007}                            &  \\
PKS~1830$-$211\xspace    & $\Delta t_{AB} = -26.0^{+4.0}_{-5.0}$    & from \cite{Lovell1998}                             &  \\
WFI~2033$-$4723\xspace   & $\Delta t_{A1B} = 36.2^{+2.3}_{-1.6}$    & from \cite{Bonvin2019}                             &  \\
                                   & $\Delta t_{A2B} = 37.3^{+3.0}_{-2.6}$    &                                                    &  \\
                                   & $\Delta t_{BC} = -59.4^{+1.3}_{-3.4}$    &                                                    &  \\
HS~2209$+$1914\xspace    & $\Delta t_{AB} = 20.0^{+5.0}_{-5.0}$     & from \cite{Eulaers2013}                            & 
\end{tabular}
}
\end{table}

\subsection{Inconclusive attempts}
Some of our monitoring data do not provide robust time delays. We nevertheless release the data for possible combination with future data sets. For each we provide the discovery paper as well as the source and lens redshifts when available. 
These unfortunate cases are listed below. 
\paragraph{\Qvingtdeux :} This system \citep[$z_{\rm source} = 1.69$, $z_{\rm lens} = 0.039$,][]{Huchra1985} is a quadruply imaged quasar where the multiple images are formed in the bulge of the lensing spiral galaxy. The variability of this system is  dominated by the microlensing variability due to the very high stellar density at the position of multiple images. For this reason, the peak that can be observed in image C at the end of 2012 in Fig.~\ref{fig:annex_lcs11} cannot be matched in the other curves and is attributed to a microlensing caustic crossing event \citep{Goicoechea2020}. As a result of this large extrinsic variability, the measurement of time delays is not possible with this data set. 

\paragraph{\Jtreizevingt :} This doubly imaged system \citep[$z_{\rm source} = 1.502$, $z_{\rm lens} = 0.899$,][]{Rusu2013} do not yield any time-delay measurements because of the poor sampling of the light curves and the short monitoring season. With an average of $\sim$10 epochs per monitoring season, the observing cadence was not sufficient to measure a time delay despite the large amplitude variation ($\sim$0.4 mag) seen in image B. In addition, this target is located at DEC=+16, which is at the northern limit of what can be observed from La Silla Observatory with a sufficiently long visibility window. The time delay of this system is probably of the order of several years \citep[]{Rusu2013}, due to the very large image separation (8$\arcsec$57). 

\paragraph{\Jtreizevingtdeux :} We could not measure a time delay for this system \citep[$z_{\rm source} =1.717$, $z_{\rm lens} = \sim 0.55$,][]{Oguri2008} because of the faint B image coupled with the absence of intrinsic variation with an amplitude larger than 0.1 mag. The relatively noisy B light curve do not allowed us to clearly identify intrinsic variations that are also visible in image A.

\paragraph{\Jtreizequaranteneuf \ :} Despite the seven years of monitoring, we could not determine the time delay of this lens system \citep[$z_{\rm source} = 1.722$, $z_{\rm lens} = \sim 0.65$,][]{Kayo2010} because of the absence of intrinsic variation larger than 0.1 mag in image B. Image A shows a variation of the order of $\sim$0.15 mag at the end of 2016 and beginning of 2017 but this variation cannot be matched in image B. Since the time delay is predicted to be of the order of one year \citep{Kayo2010}, this variation might had appeared in the B light curve after the end of our monitoring campaign.

\paragraph{\Jquatorzezerocinq \ :}This doubly imaged lens quasar \citep[$z_{\rm source} = 1.81 $, $z_{\rm lens} = \sim 0.66$,][]{Jackson2012} do not yield any time-delay measurements due to the absence of a clear intrinsic signal above the noise level. This is due to the combination of the poor sampling of the light curves, the relatively faint B component that is also contaminated by light of the lens galaxy, the short monitoring seasons and the low variability of the quasar. 

\section{Conclusions}
\label{cfini}
%
We present long-term lens monitoring data acquired at the Euler Swiss telescope since the very beginning of the COSMOGRAIL programme in September 2003 up to April 2018. We show for the first time the decade-long light curves for a sample of 23 lensed quasars. 

A new automated procedure has been developed in order to process all light curves within a homogeneous framework. The curve-shifting package \pycs has been adapted to ensure that measurements are robust against the choice of parameters for the intrinsic variability of the quasar as well as for the microlensing variations. 

We successfully measure 18 time delays, among which nine have uncertainties below 15 \%. These objects are therefore possible good candidates for time-delay cosmography. Turning these time delays into cosmological constraints will require the modelling of the lenses and the analysis of the line of sight as done by the H0LiCOW \citep[e.g.][]{Suyu2017, Bonvin2017, Birrer2019, Wong2019}, STRIDES \citep[e.g.][]{Shajib2019} and SHARP collaborations \citep[e.g.][]{Chen2019}.  

We show that long term monitoring campaigns can provide precise time-delay estimates and overcome the degeneracy between the intrinsic variations of the quasar and the microlensing by collecting over the years several inflection points that can be seen in the light curves of the multiple images. However, we report that several systems do not exhibit sharp variations with amplitude that can be detected at the photometric precision of our data. As a result, the monitoring strategy should now be adapted to higher cadence and higher photometric precision. Recent monitoring campaigns conducted with 2m-class telescopes achieve a high level of precision on the time delay in only one monitoring season by successfully observing small variation of quasars, happening on shorter time-scales than typical microlensing variation \citep{Courbin2018, Bonvin2018a}.

Furture high-cadence lens monitoring data, along with their cosmological interpretation will, from now on, be presented in the new series of TDCOSMO (Time-Delay COSMOgraphy) papers \citep[e.g.][]{Millon2020}, which includes members of the present COSMOGRAIL collaboration in addition to the H0LiCOW, STRIDES and SHARP collaborations.

\begin{acknowledgements}
The authors would like to thank all the Euler observers as well as the technical staff of the Euler Swiss telescope who made the uninterrupted observation over 15 years possible. COSMOGRAIL is supported by the Swiss National Science Foundation (SNSF) and by the European Research Council (ERC) under the European Union’s Horizon 2020 research and innovation programme (COSMICLENS: grant agreement No 787886). This research made use of Astropy, a community-developed core Python package for Astronomy \citep{Astropy2013, Astropy2018} and the 2D graphics environment Matplotlib \citep{Hunter2007}. 

\end{acknowledgements}

\bibliographystyle{aa}
\bibliography{biblio}

\appendix
\section{Field of view of all lensed quasars}
\label{AppendixA}

We present below the field of view for each of our lensed quasars, showing the reference stars used for the relative photometry and the PSF stars used for the deconvolution photometry. All images are stacks of the best seeing frames taken in the $R$-band and often include dozens of hours of total integration. 

\begin{figure*}[htbp!]
    \centering
    \begin{minipage}[c]{0.49\textwidth}
    \includegraphics[width=\textwidth]{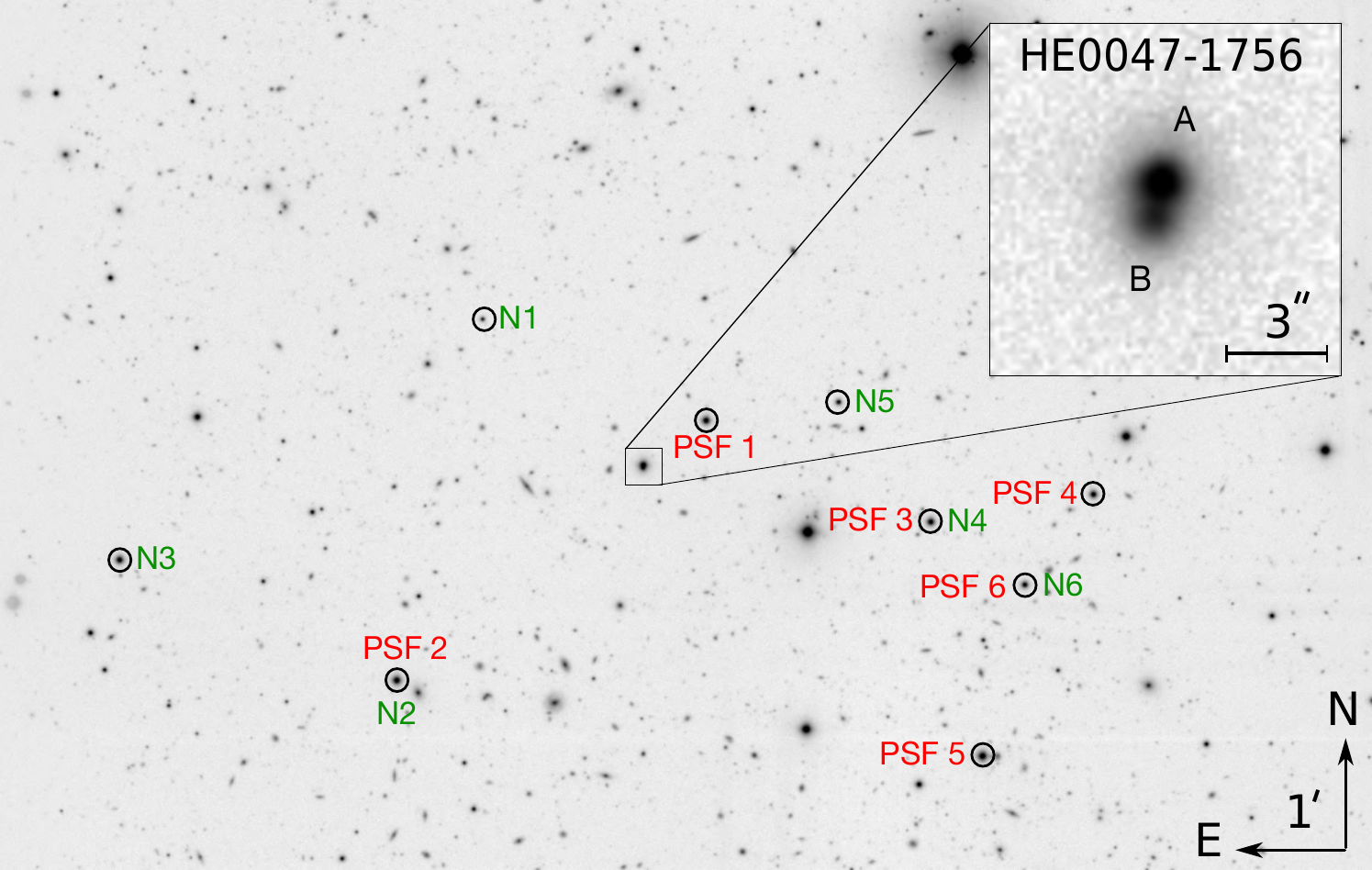}
    \end{minipage} 
    \begin{minipage}[c]{0.49\textwidth}
    \includegraphics[width=\textwidth]{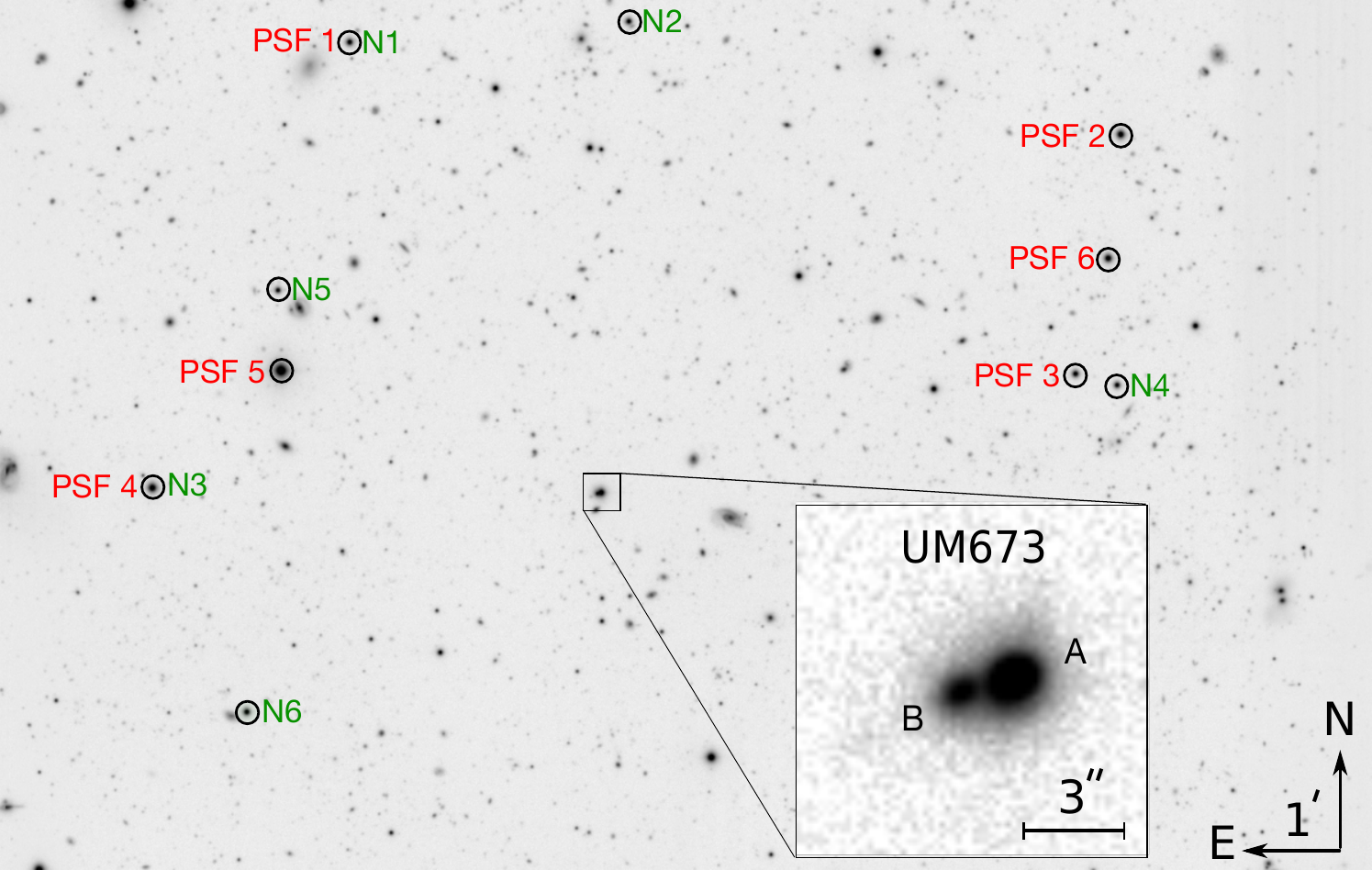}
    \end{minipage}

    \begin{minipage}[c]{0.49\textwidth}
    \includegraphics[width=\textwidth]{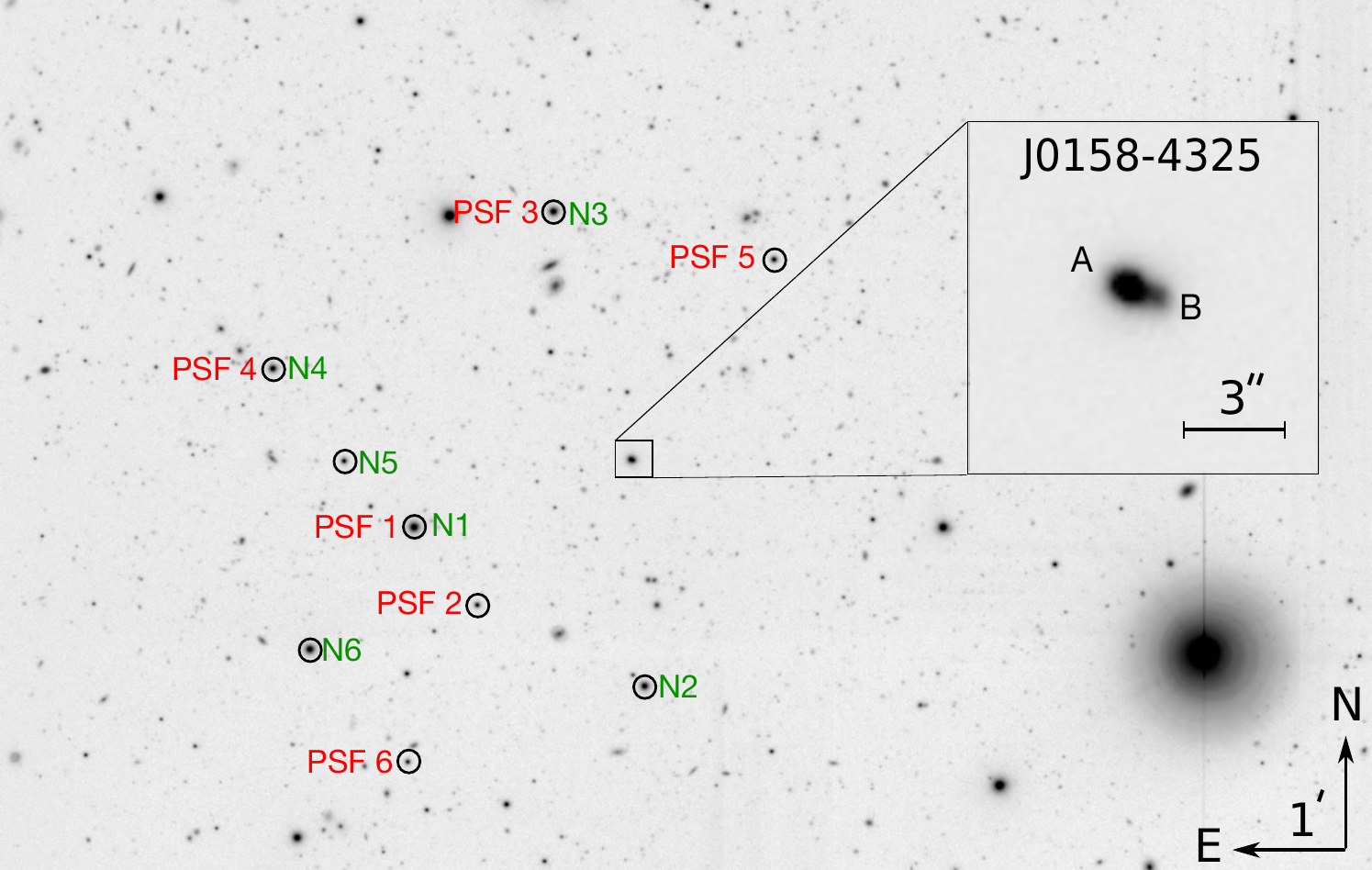}
    \end{minipage}
    \begin{minipage}[c]{0.49\textwidth}
    \includegraphics[width=\textwidth]{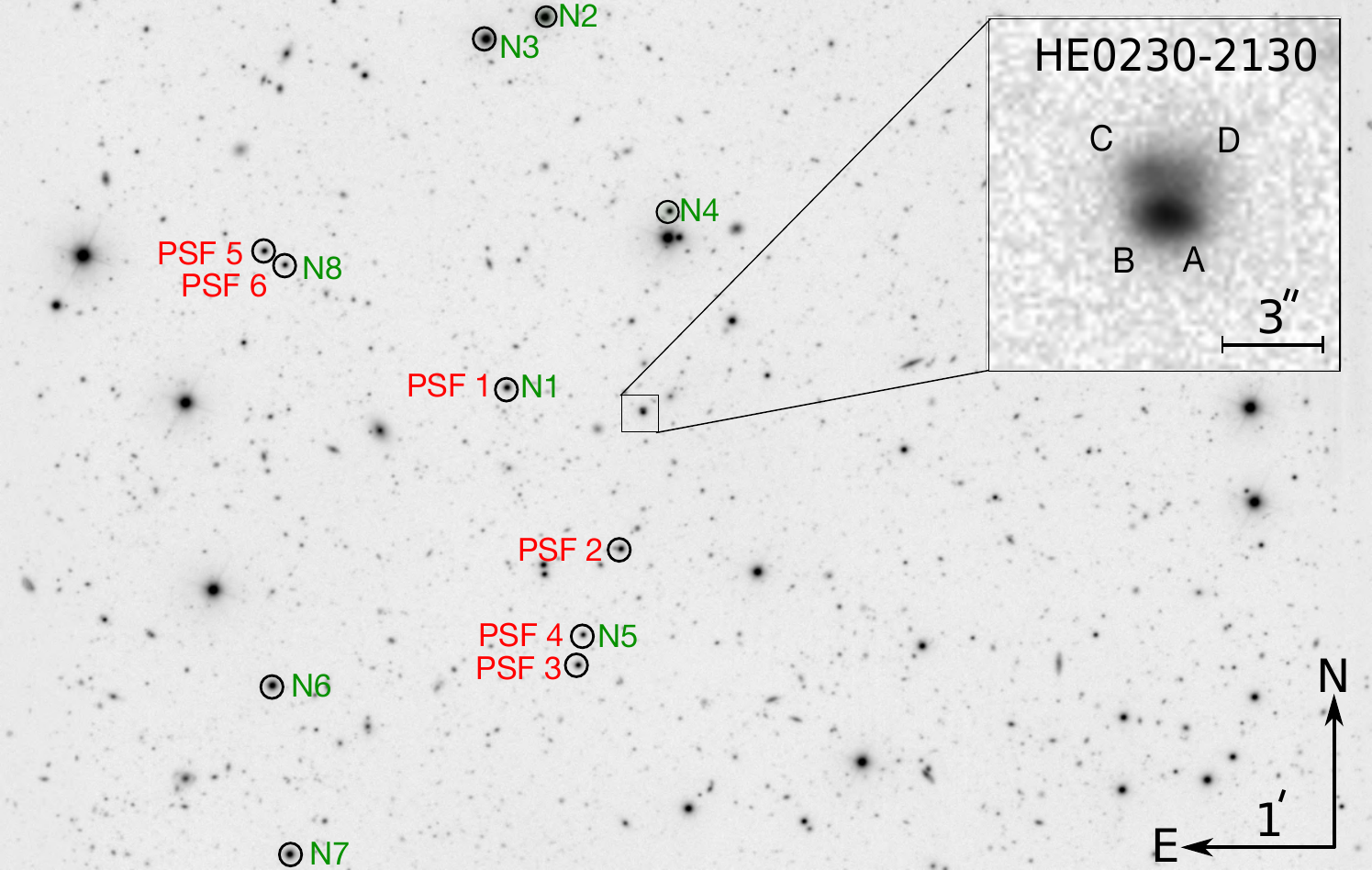}
    \end{minipage} 
    
    \begin{minipage}[c]{0.49\textwidth}
    \includegraphics[width=\textwidth]{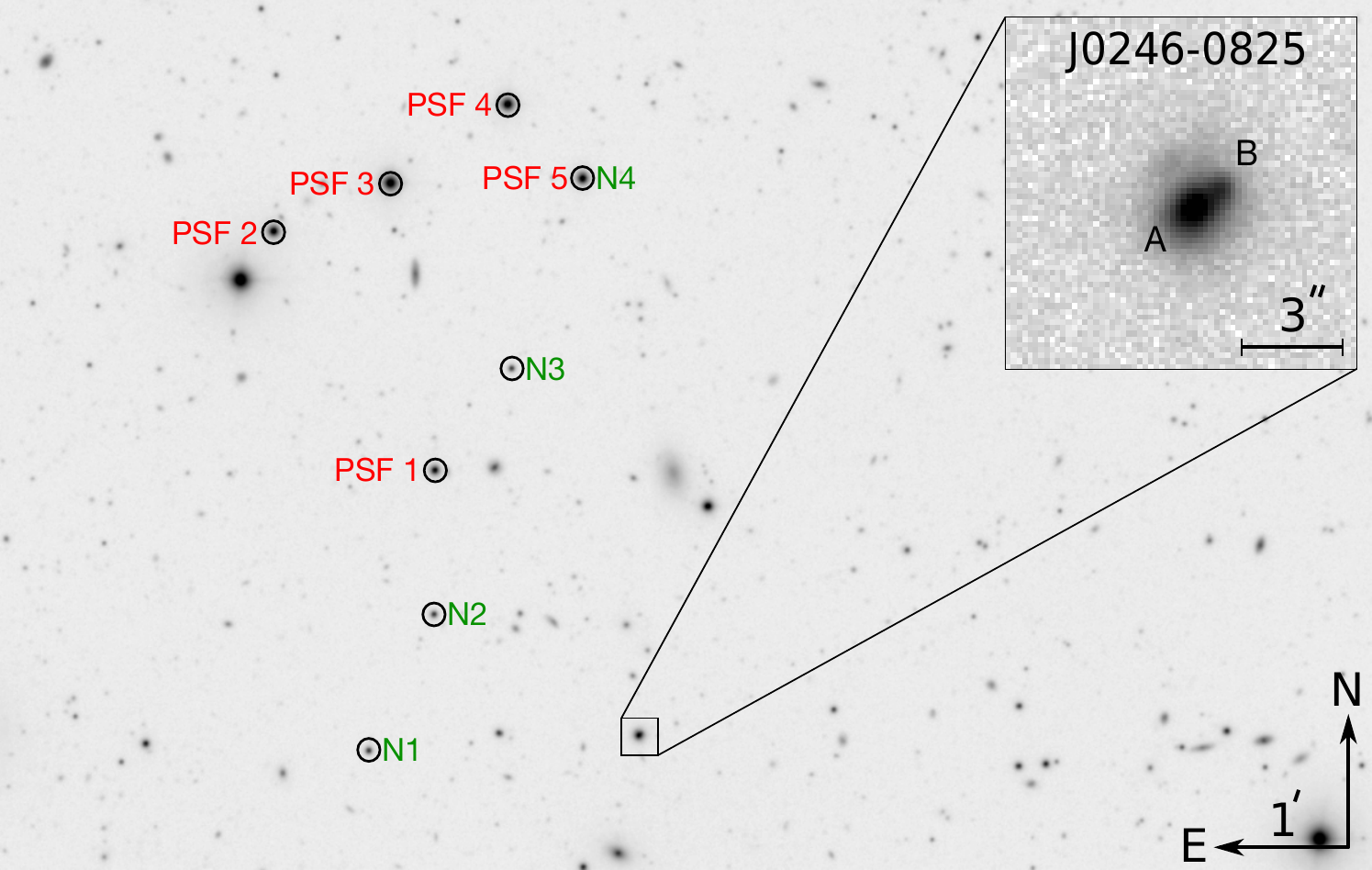}
    \end{minipage}   
    \begin{minipage}[c]{0.49\textwidth}
    \includegraphics[width=\textwidth]{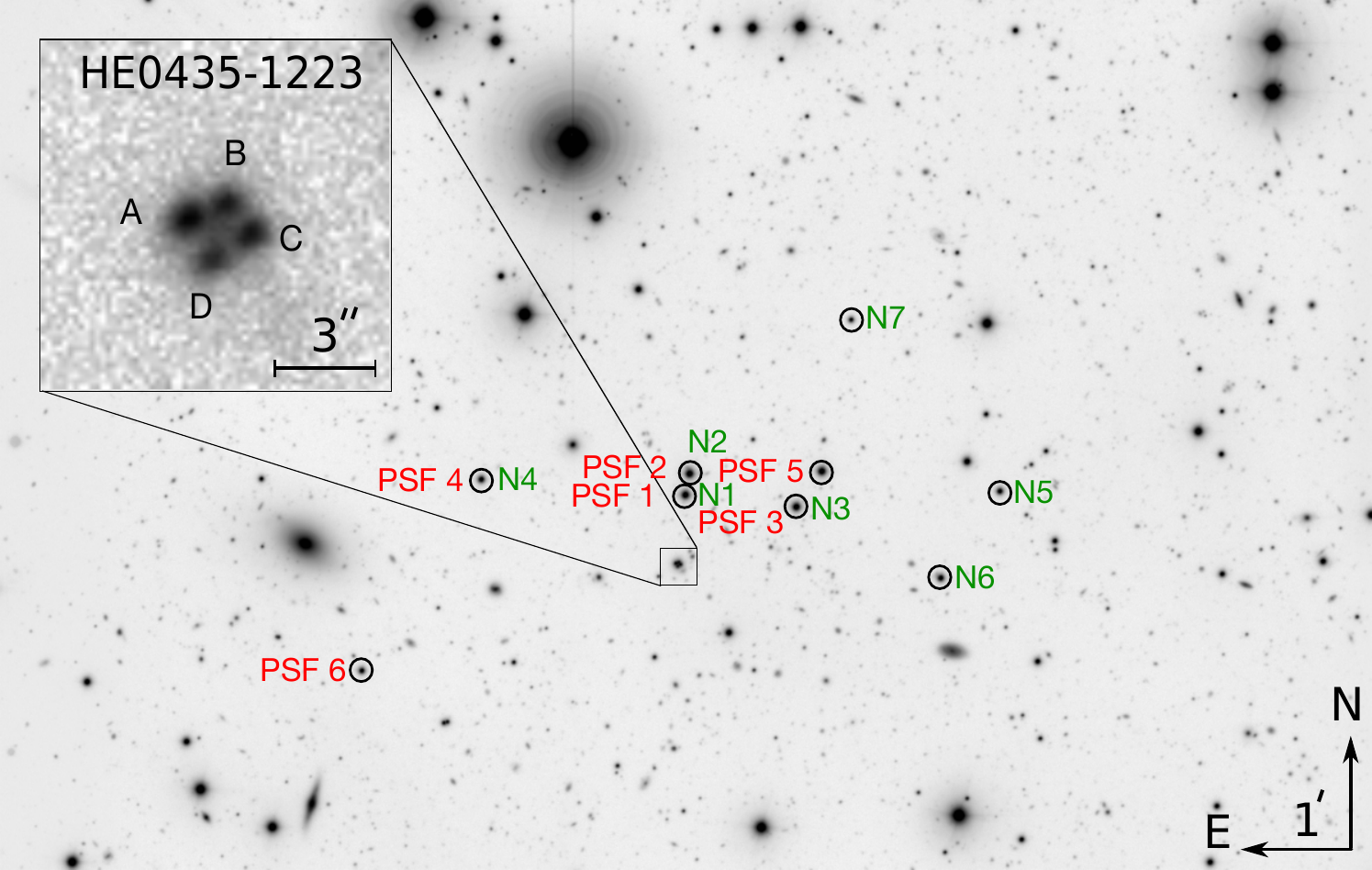}
    \end{minipage}

    \begin{minipage}[c]{0.490\textwidth}
    \includegraphics[width=\textwidth]{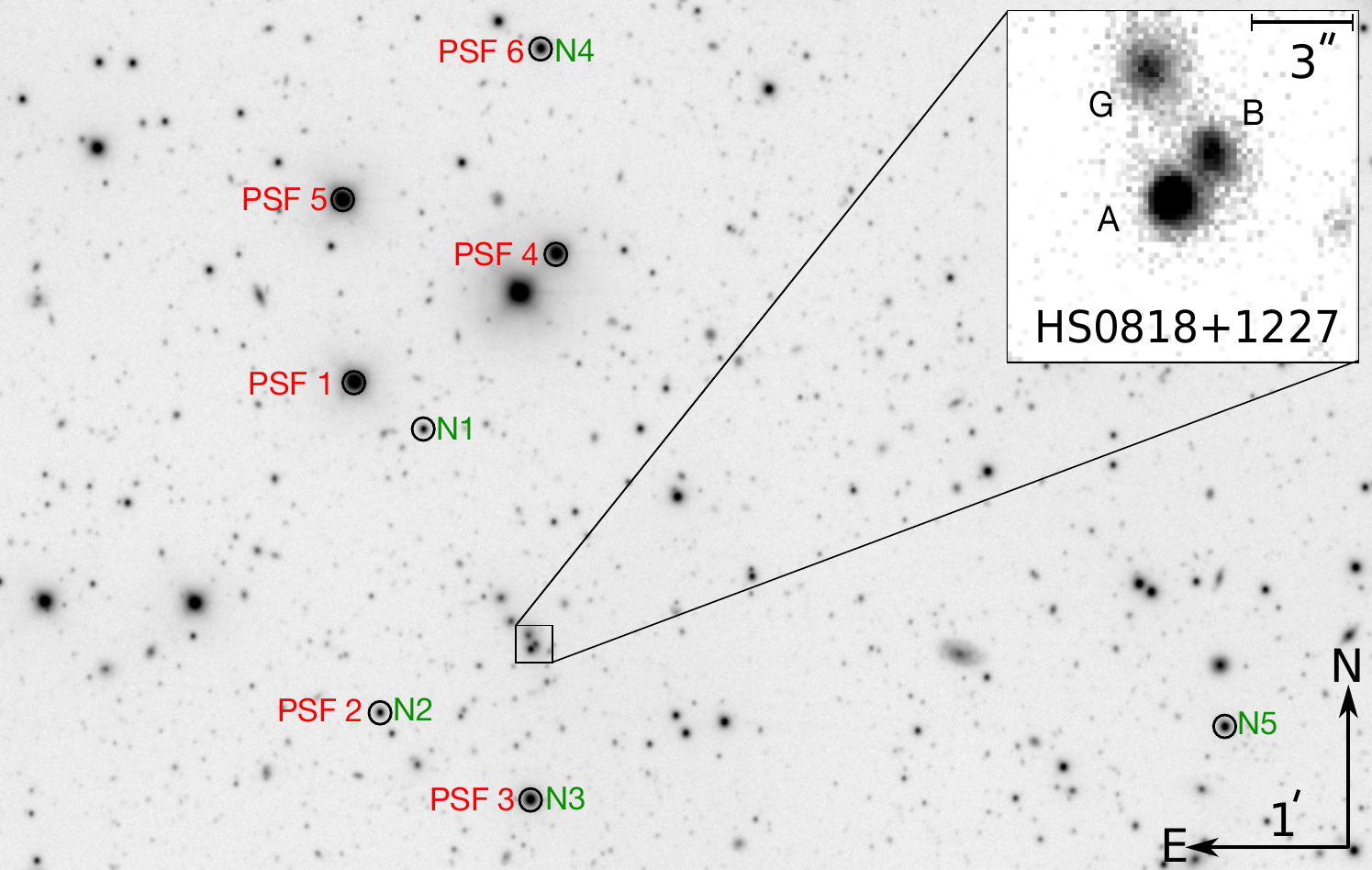}
    \end{minipage} 
    \begin{minipage}[c]{0.490\textwidth}
    \includegraphics[width=\textwidth]{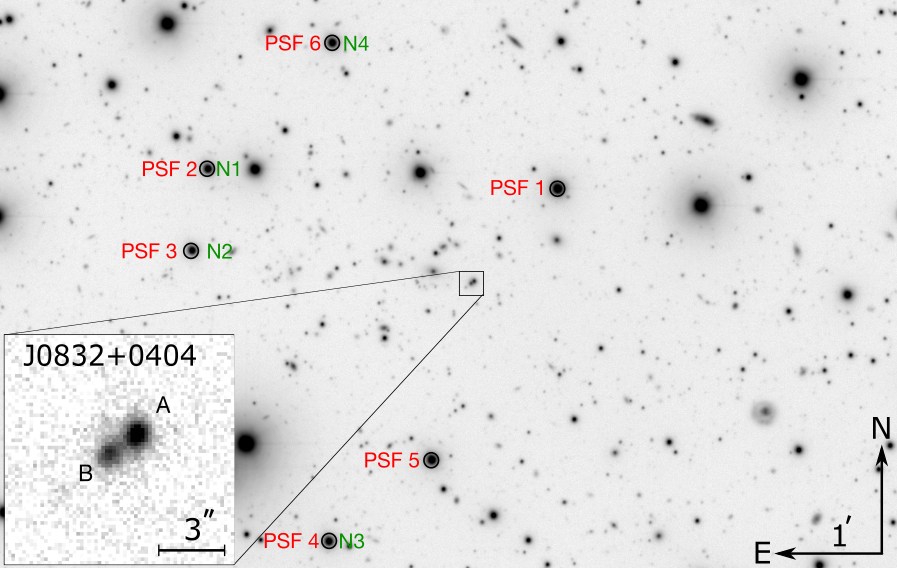}
    \end{minipage}

    \caption{Deep field images, similar to Fig.~\ref{fig:nicefield}, for the rest of the COSMOGRAIL lensed quasar sample.}
    \label{fig:nicefield_annex}
\end{figure*}

\begin{figure*}[htbp!]
    \centering
    \begin{minipage}[c]{0.490\textwidth}
    \includegraphics[width=\textwidth]{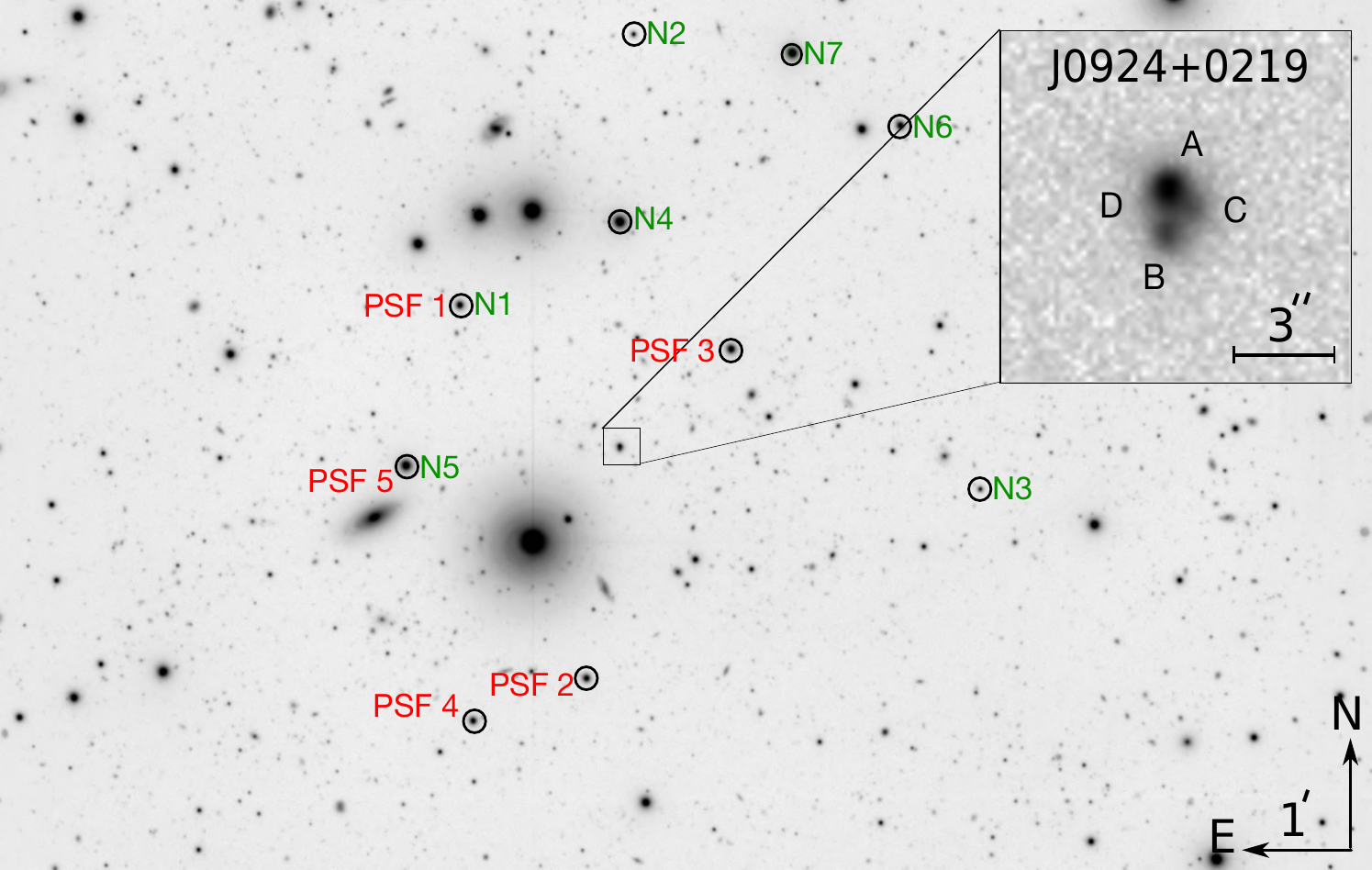}
    \end{minipage} 
    \begin{minipage}[c]{0.490\textwidth}
    \includegraphics[width=\textwidth]{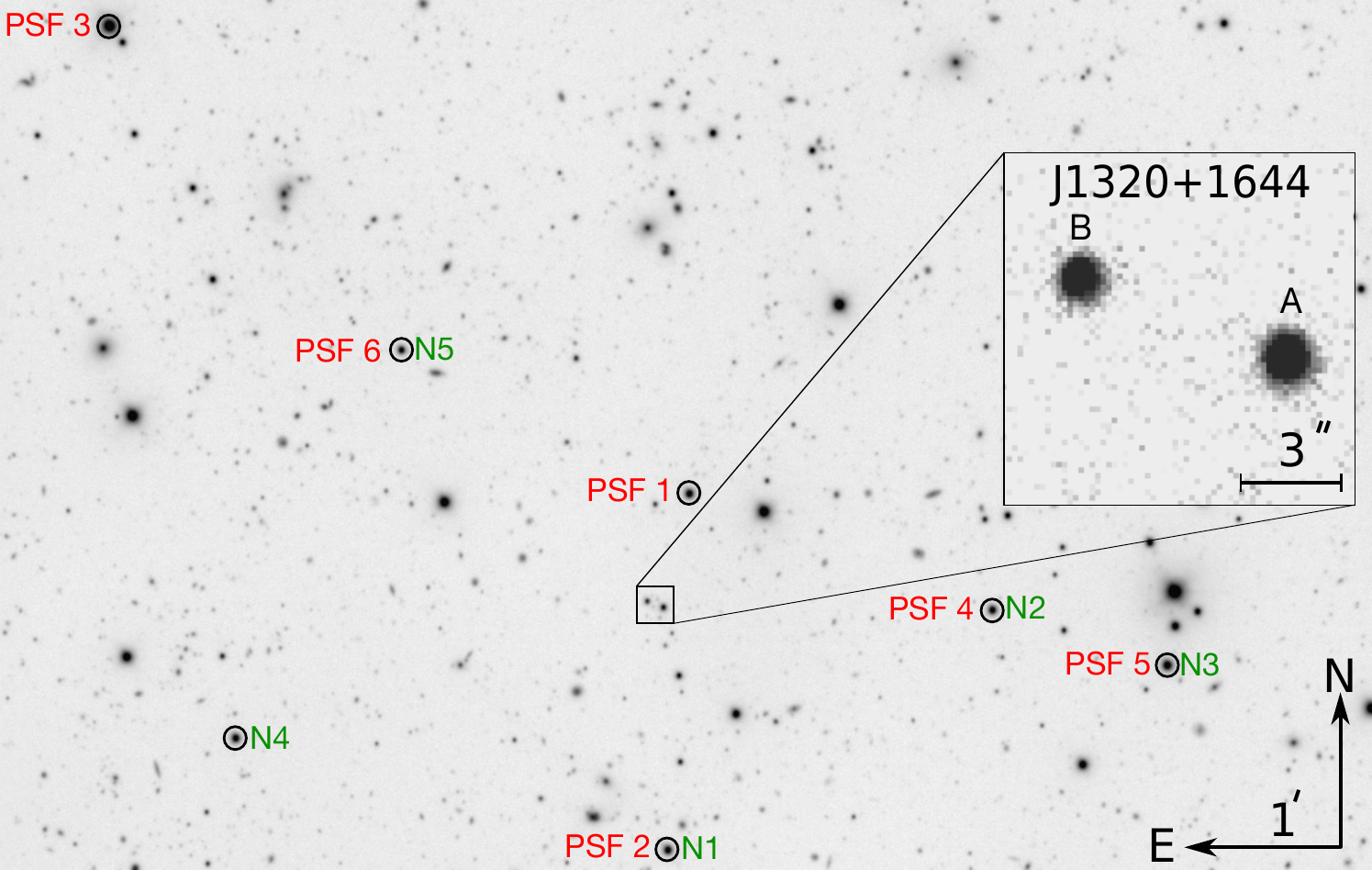}
    \end{minipage} 
    
    \begin{minipage}[c]{0.490\textwidth}
    \includegraphics[width=\textwidth]{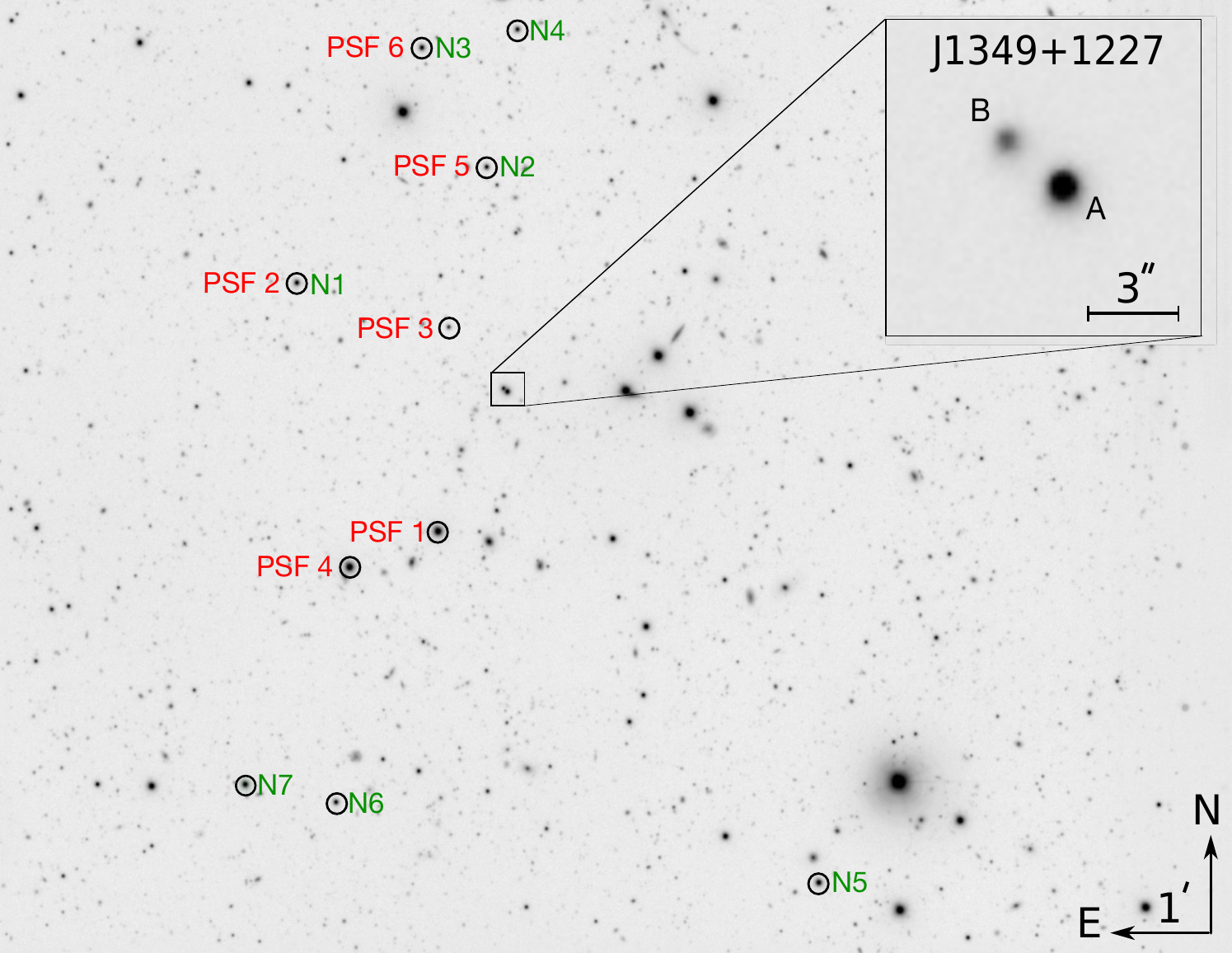}
    \end{minipage} 
    \begin{minipage}[c]{0.490\textwidth}
    \includegraphics[width=\textwidth]{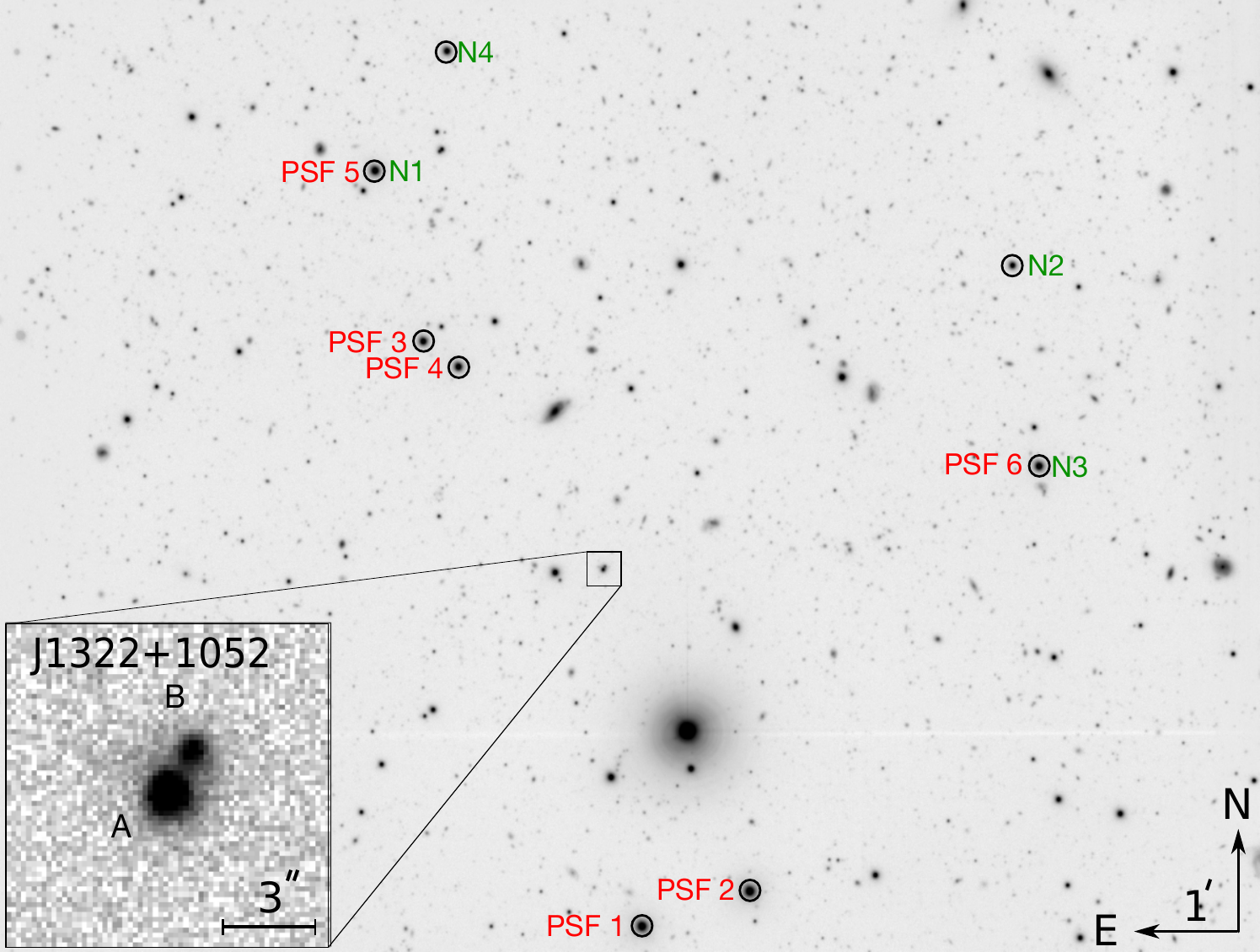}
    \end{minipage} 
    
    \begin{minipage}[c]{0.490\textwidth}
    \includegraphics[width=\textwidth]{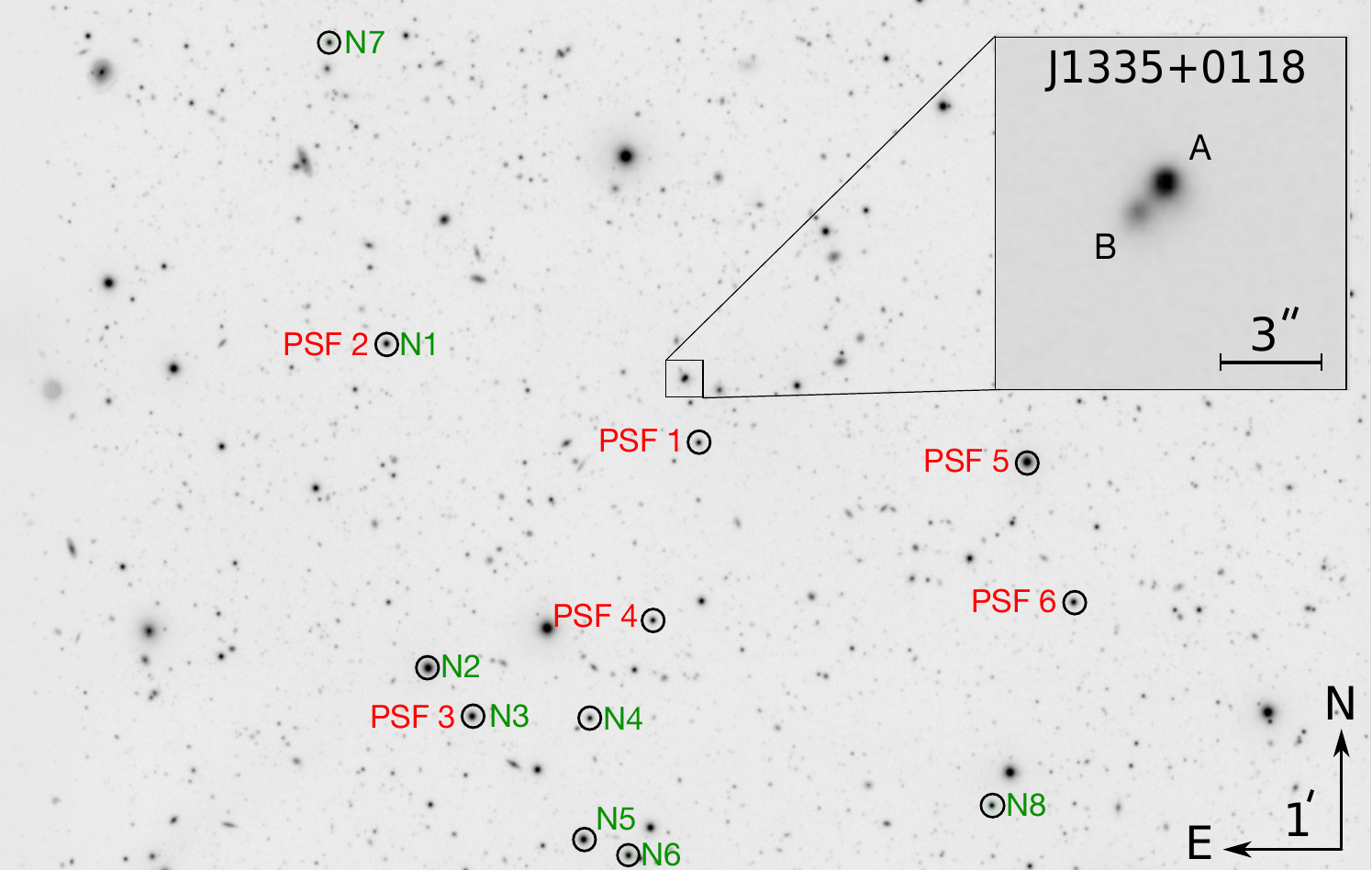}
    \end{minipage} 
    \begin{minipage}[c]{0.490\textwidth}
    \includegraphics[width=\textwidth]{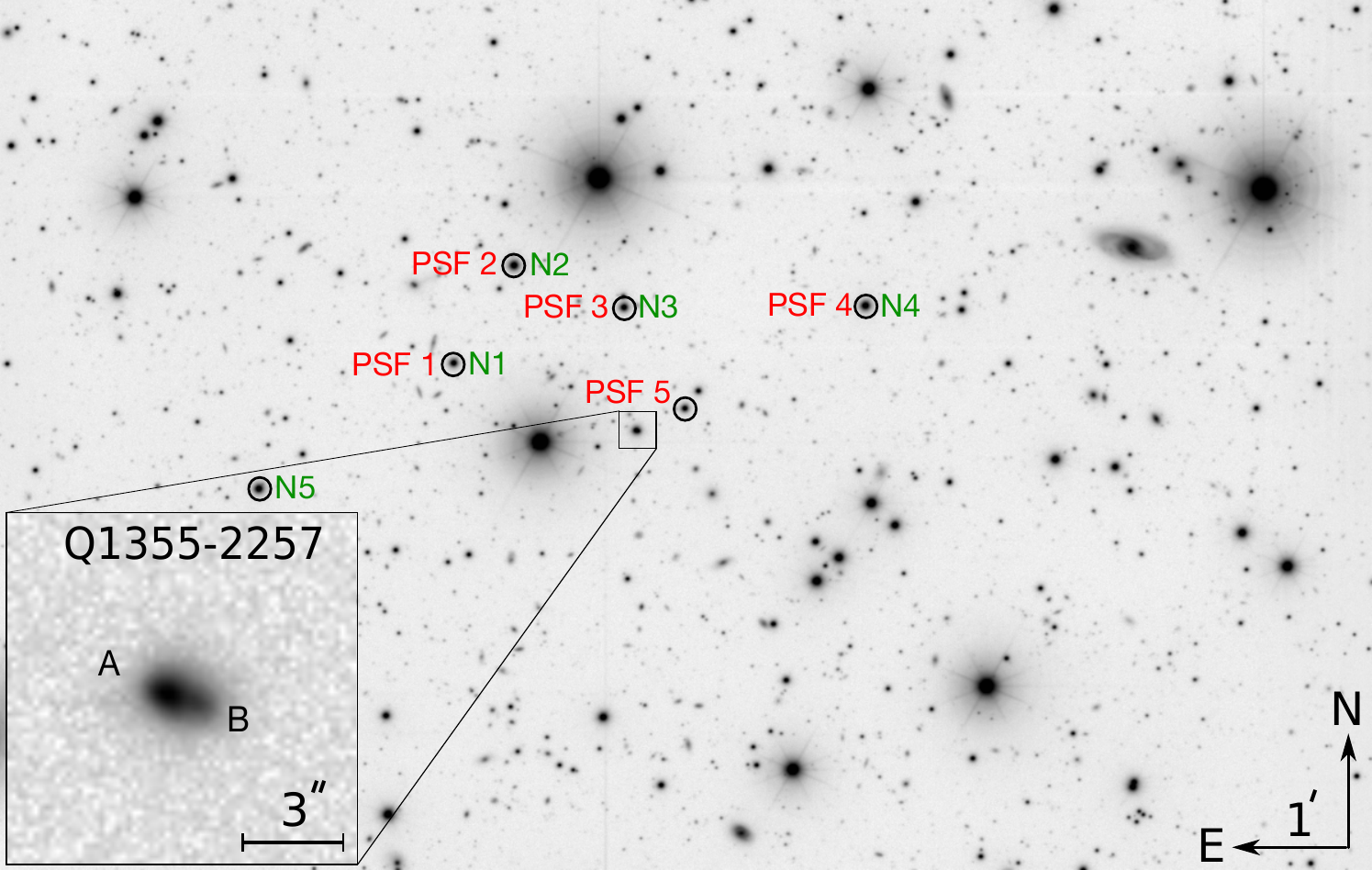}
    \end{minipage} 
    
    \begin{minipage}[c]{0.490\textwidth}
    \includegraphics[width=\textwidth]{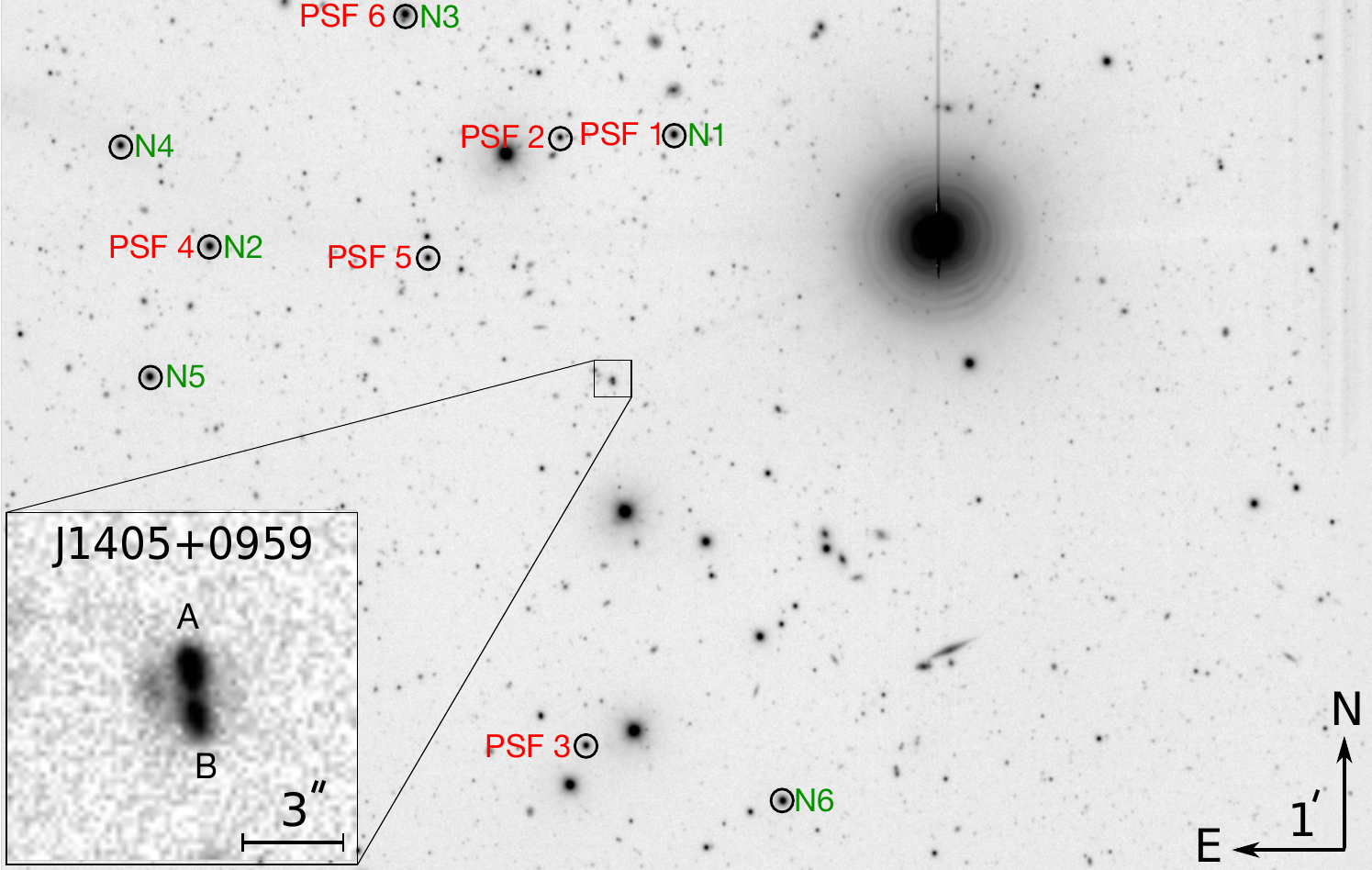}
    \end{minipage} 
    \begin{minipage}[c]{0.490\textwidth}
    \includegraphics[width=\textwidth]{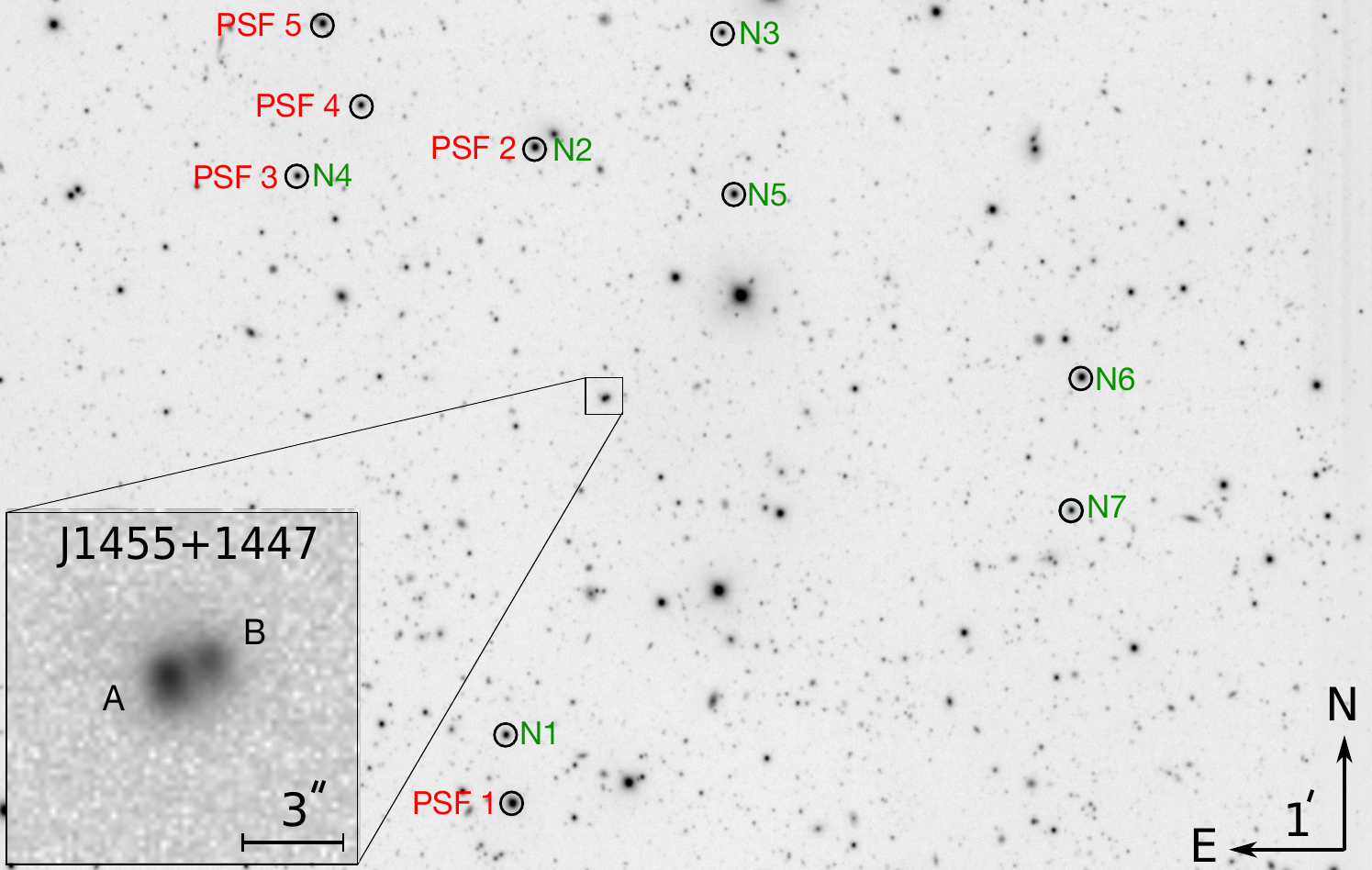}
    \end{minipage} 
    
    \caption{Continuation of Fig.~\ref{fig:nicefield_annex}}
    \label{fig:nicefield_annex2}
\end{figure*}
    
\begin{figure*}[htbp!]
    \centering   
    \begin{minipage}[c]{0.490\textwidth}
    \includegraphics[width=\textwidth]{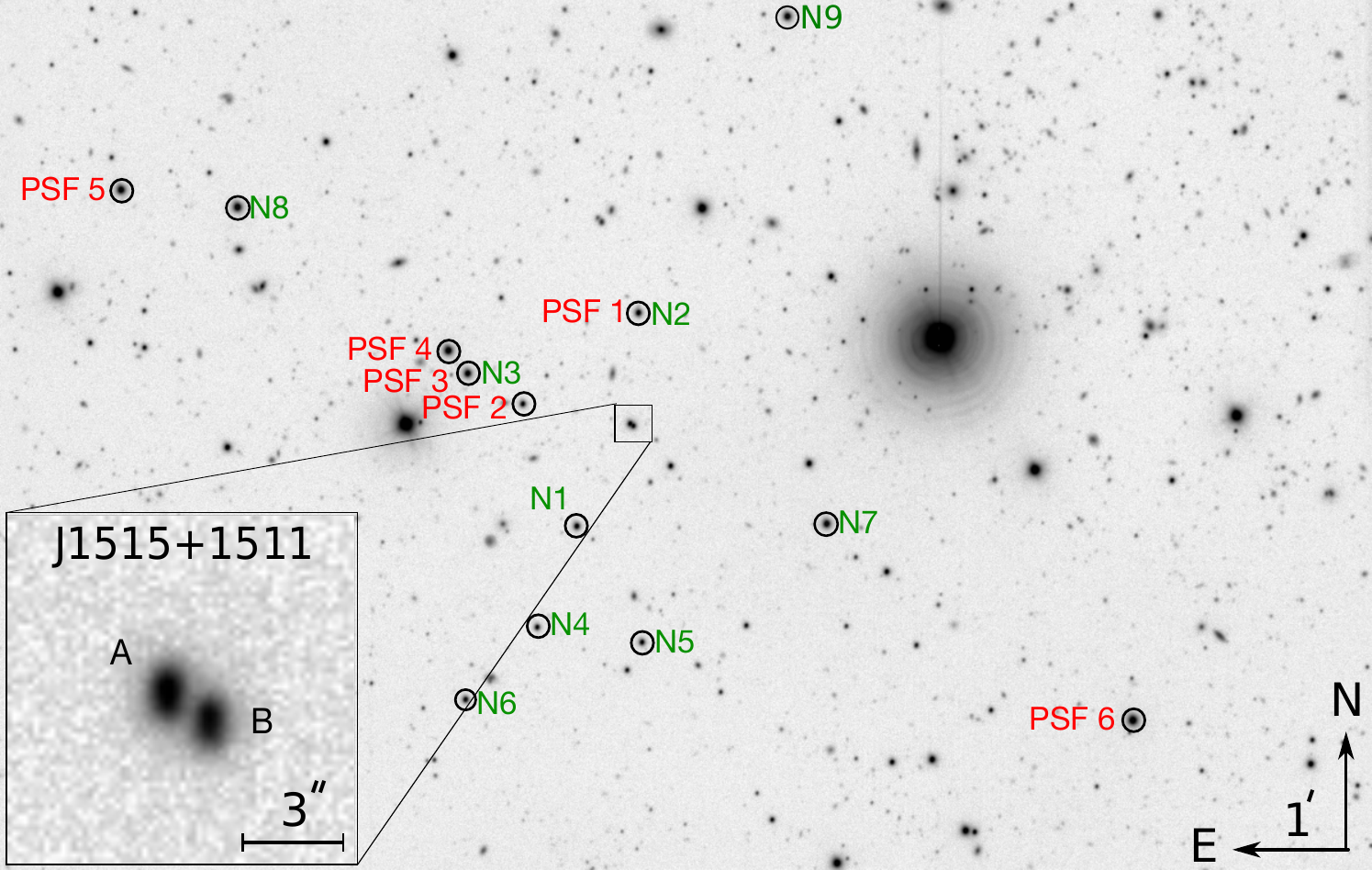}
    \end{minipage} 
    \begin{minipage}[c]{0.49\textwidth}
    \includegraphics[width=\textwidth]{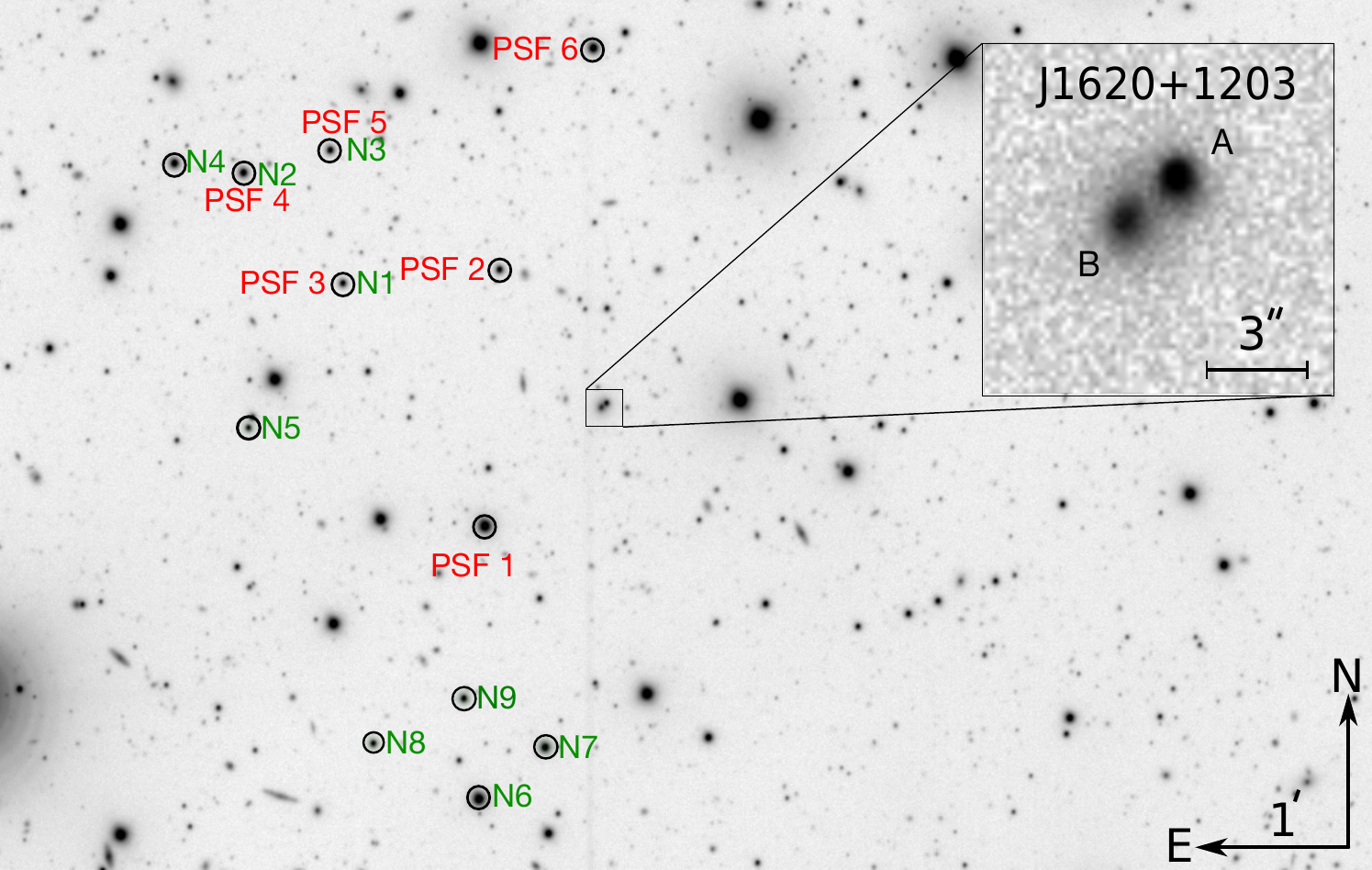}
    \end{minipage}
    \begin{minipage}[c]{0.49\textwidth}
    \includegraphics[width=\textwidth]{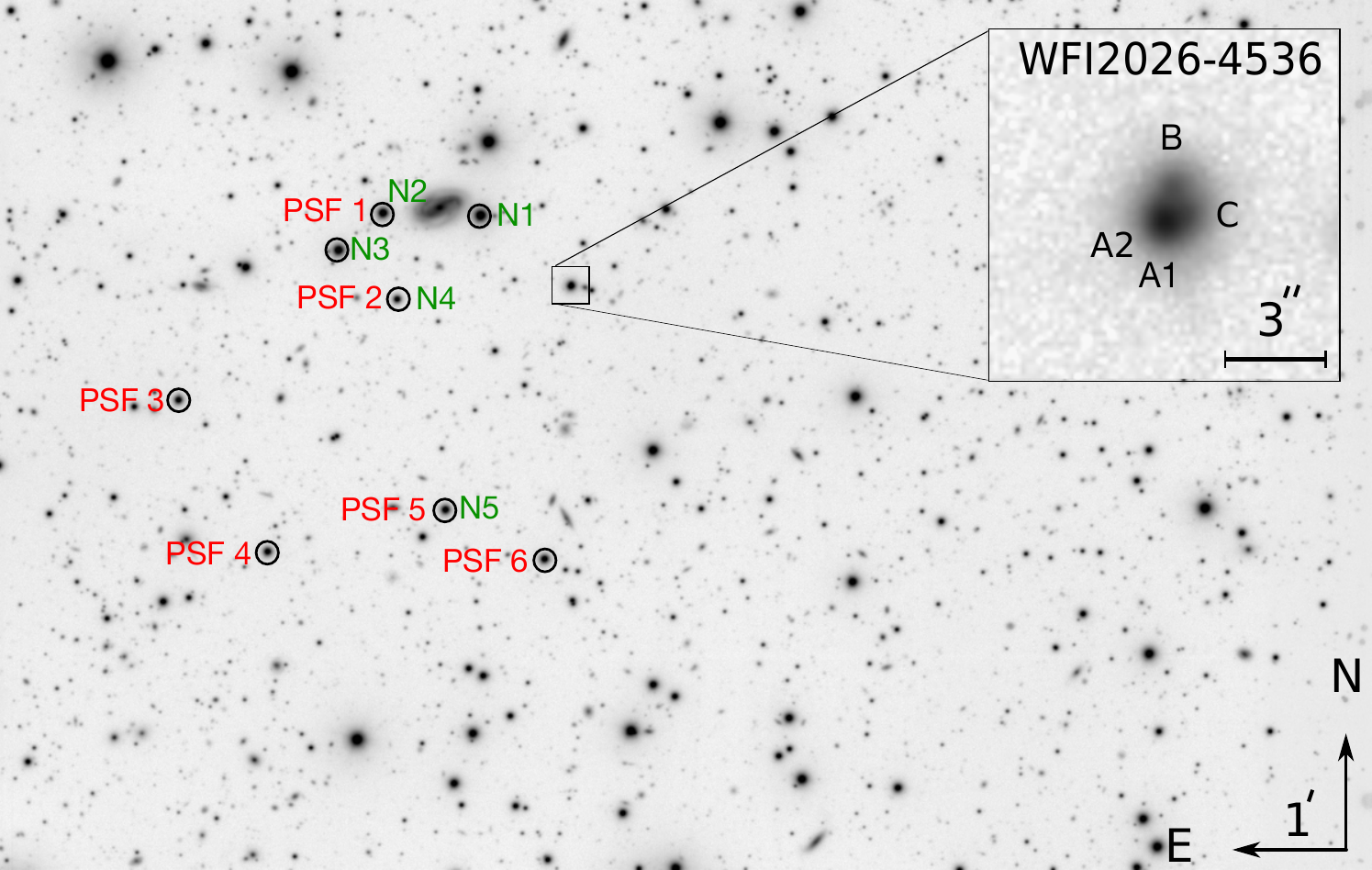}
    \end{minipage}  
    \begin{minipage}[c]{0.49\textwidth}
    \includegraphics[width=\textwidth]{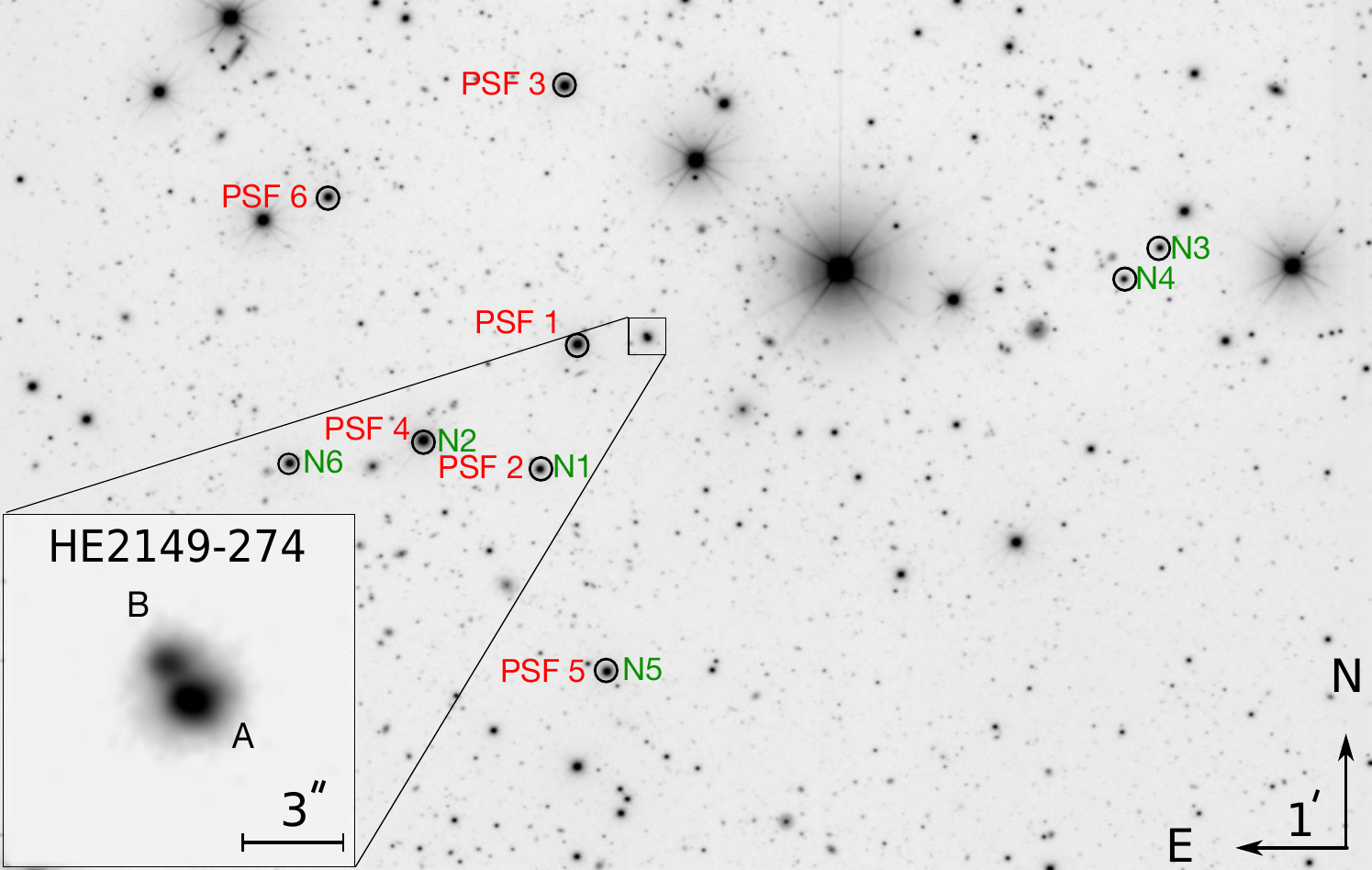}
    \end{minipage} 
    \begin{minipage}[c]{0.6\textwidth}
    \includegraphics[width=\textwidth]{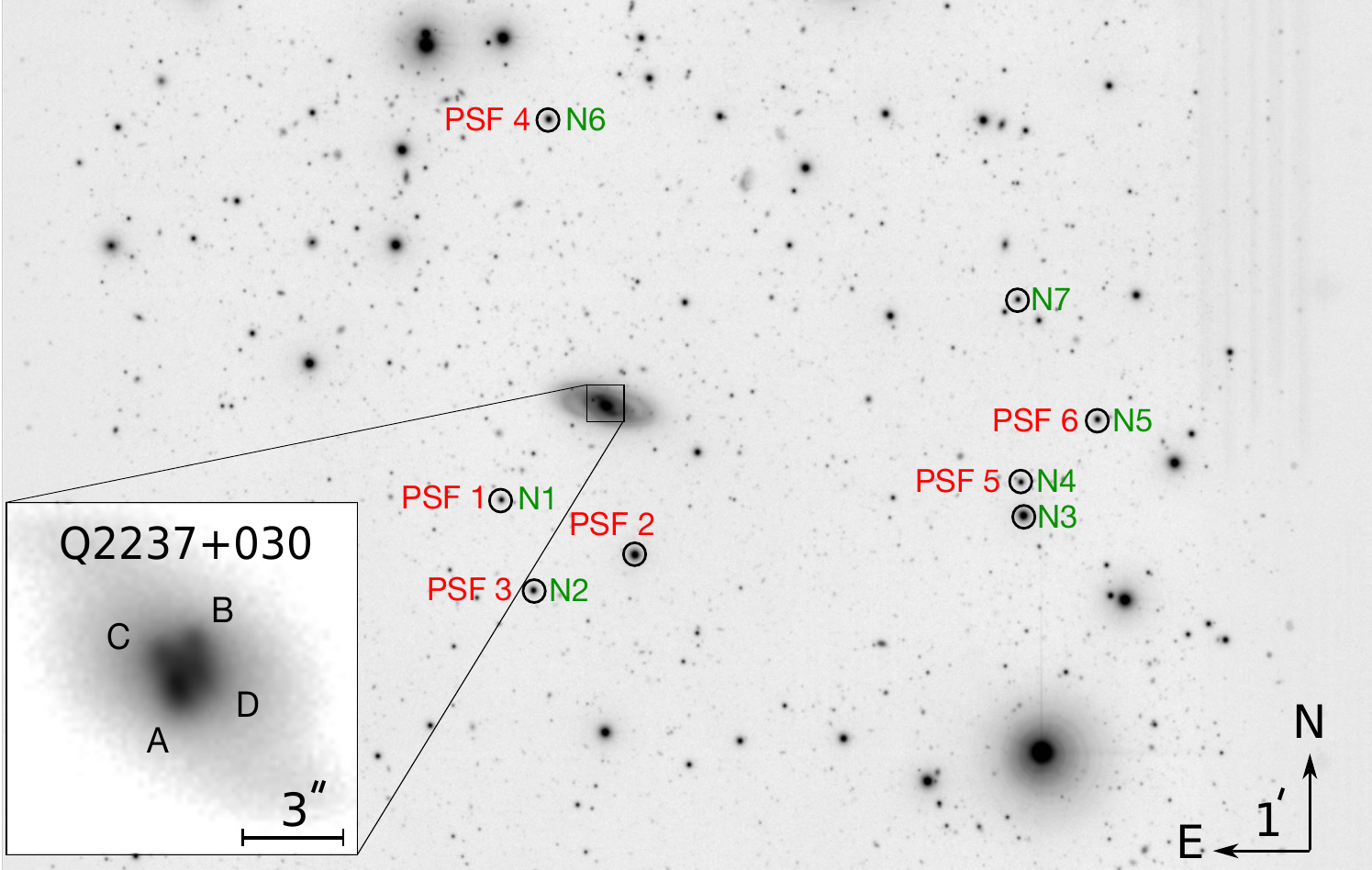}
    \end{minipage} 
    
    \caption{Continuation of Fig.~\ref{fig:nicefield_annex2}}
    \label{fig:nicefield_annex3}
\end{figure*}

\newpage
\section{Light curves for all lensed quasars}
\label{AppendixB}

In this section, we show the $R$-band light curves for all our objects, obtained using deconvolution photometry. Many span almost 15 years in length with a typical temporal sampling of one point every 3-5 days. The light curves can be visualised and downloaded from the \dthreecs interactive tool and the CDS data base. When additional photometric data are available in the literature, we add these data points onto our light curves (see Sect. \ref{sec:available_data}). For each curve, we also show the flux ratio between the quasar images after correction for the time delay measured with \pycs and using a free-knot spline interpolation. This allows us to unveil the variations due purely to microlensing over a very long time scale. When we are not able to measure the time delay, we subtract the curves pair-wise without shifting them in time.

\begin{figure*}[htbp!]
    \centering
    \begin{minipage}[c]{0.99\textwidth}
    \includegraphics[width=\textwidth]{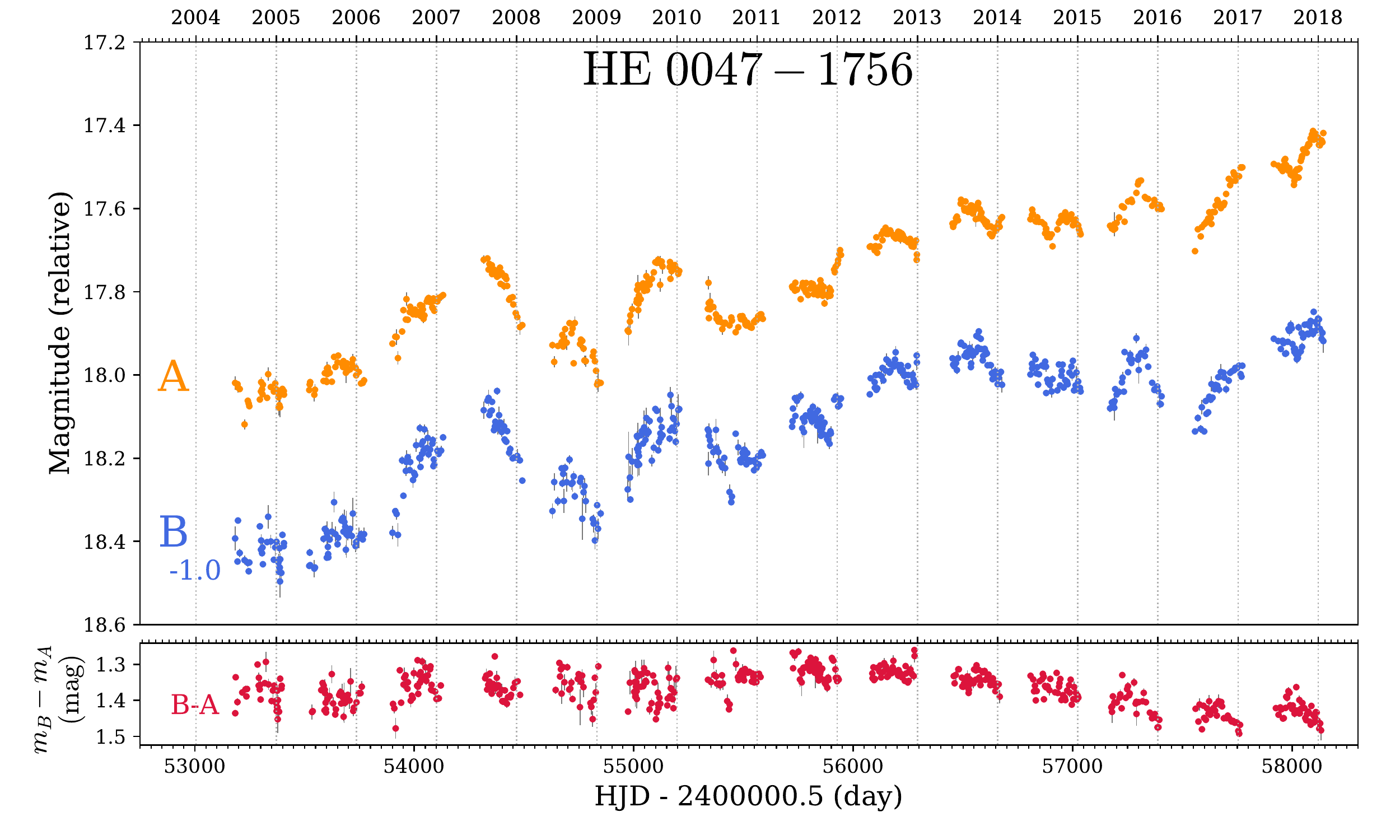}
    \end{minipage} 
    \begin{minipage}[c]{0.99\textwidth}
    \includegraphics[width=\textwidth]{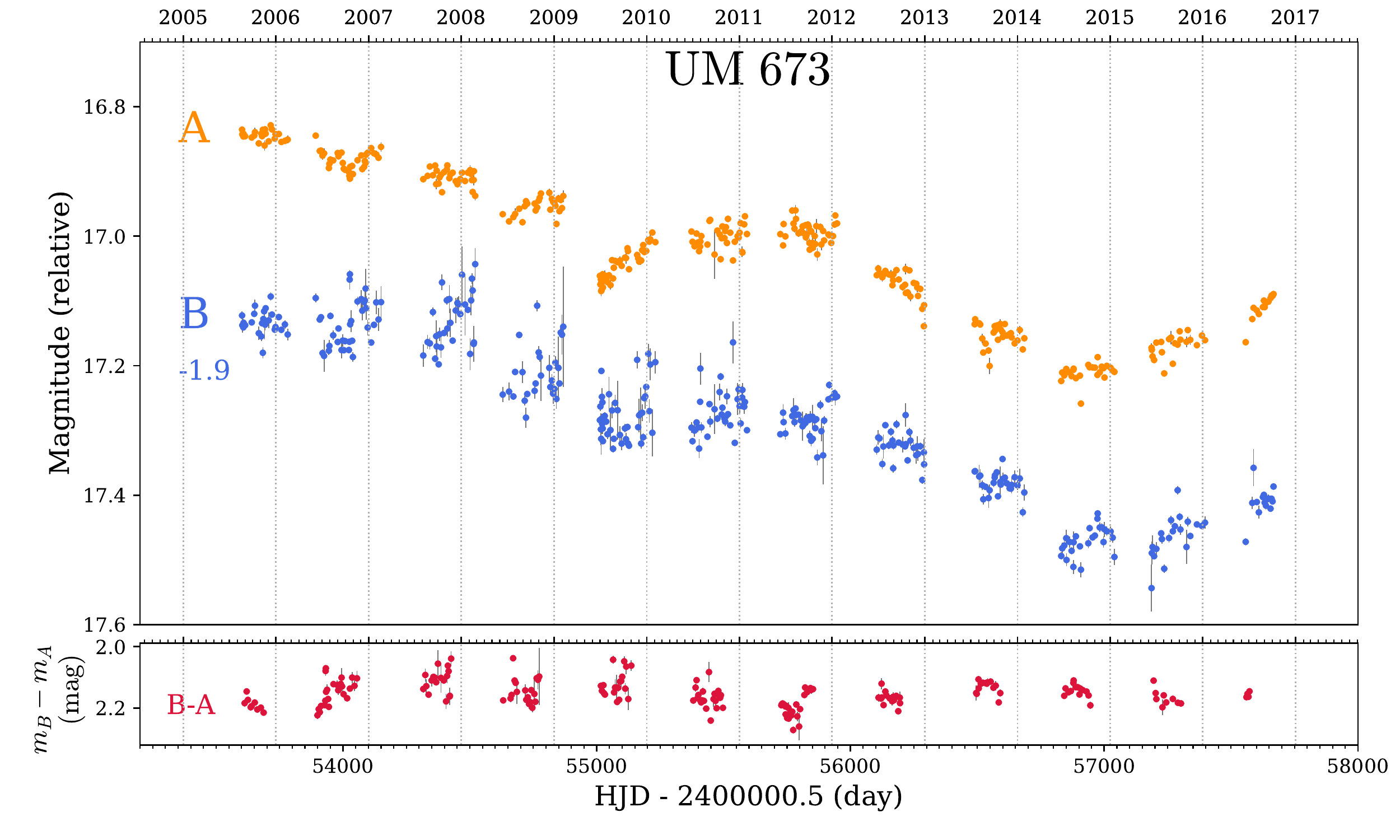}
    \end{minipage} 
    \caption{Light curves for lensed QSO \HEzerozero and \UMsix. Legend is the same as for Fig.\ref{fig:lcs_1131_1226}.}
    \label{fig:annex_lcs1}
\end{figure*}

\begin{figure*}[htbp!]
    \centering
    \begin{minipage}[c]{0.99\textwidth}
    \includegraphics[width=\textwidth]{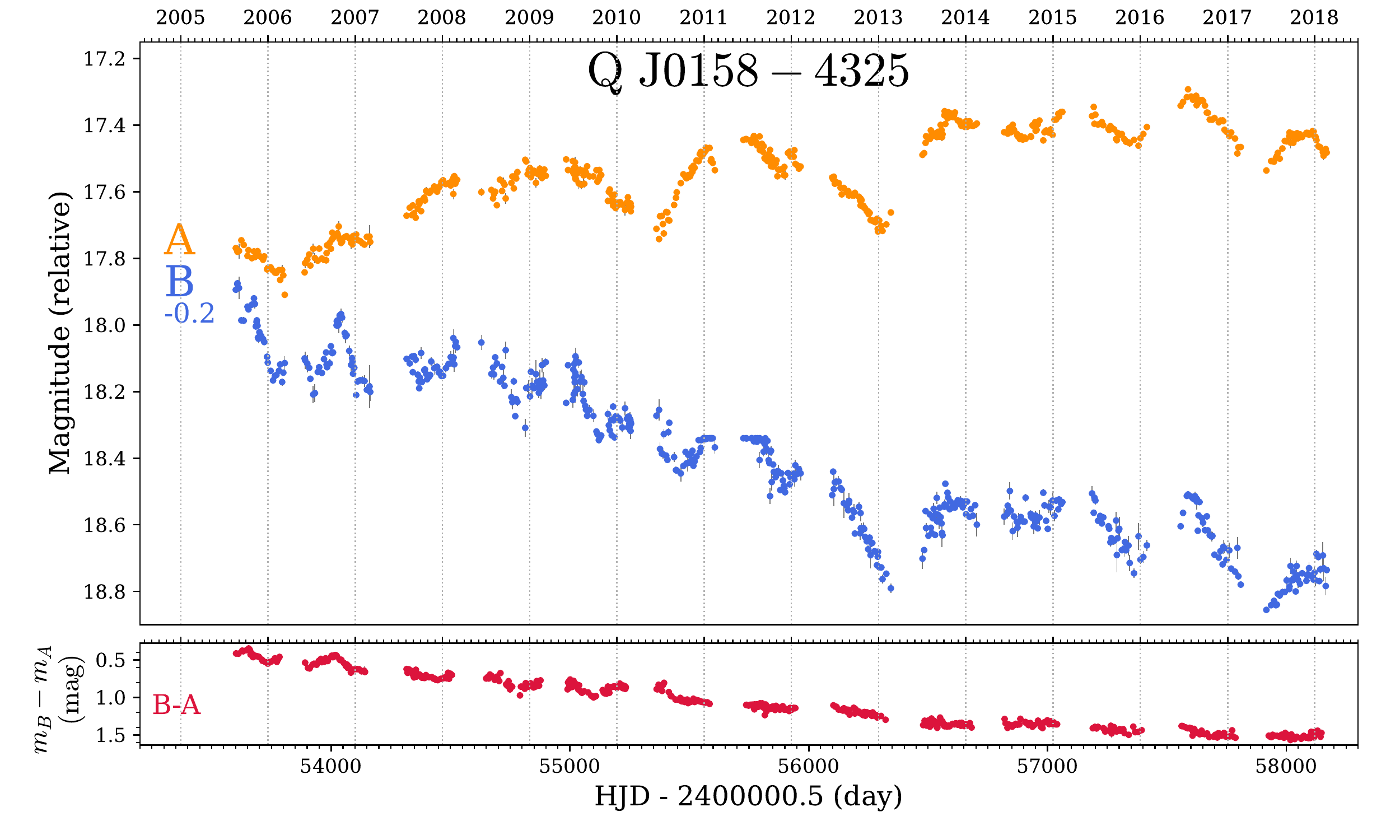}
    \end{minipage}
    \begin{minipage}[c]{0.99\textwidth}
    \includegraphics[width=\textwidth]{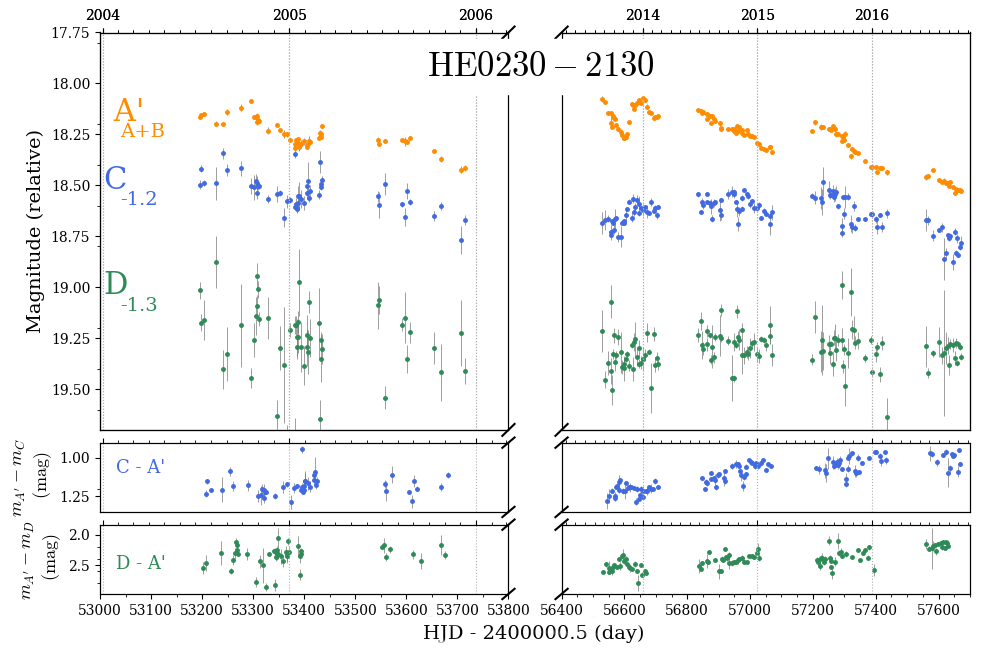}
    \end{minipage}
    \caption{Continuation of Fig.~\ref{fig:annex_lcs1}}
    \label{fig:annex_lcs2}
\end{figure*}
    
\begin{figure*}[htbp!]
    \centering
    \begin{minipage}[c]{0.99\textwidth}
    \includegraphics[width=\textwidth]{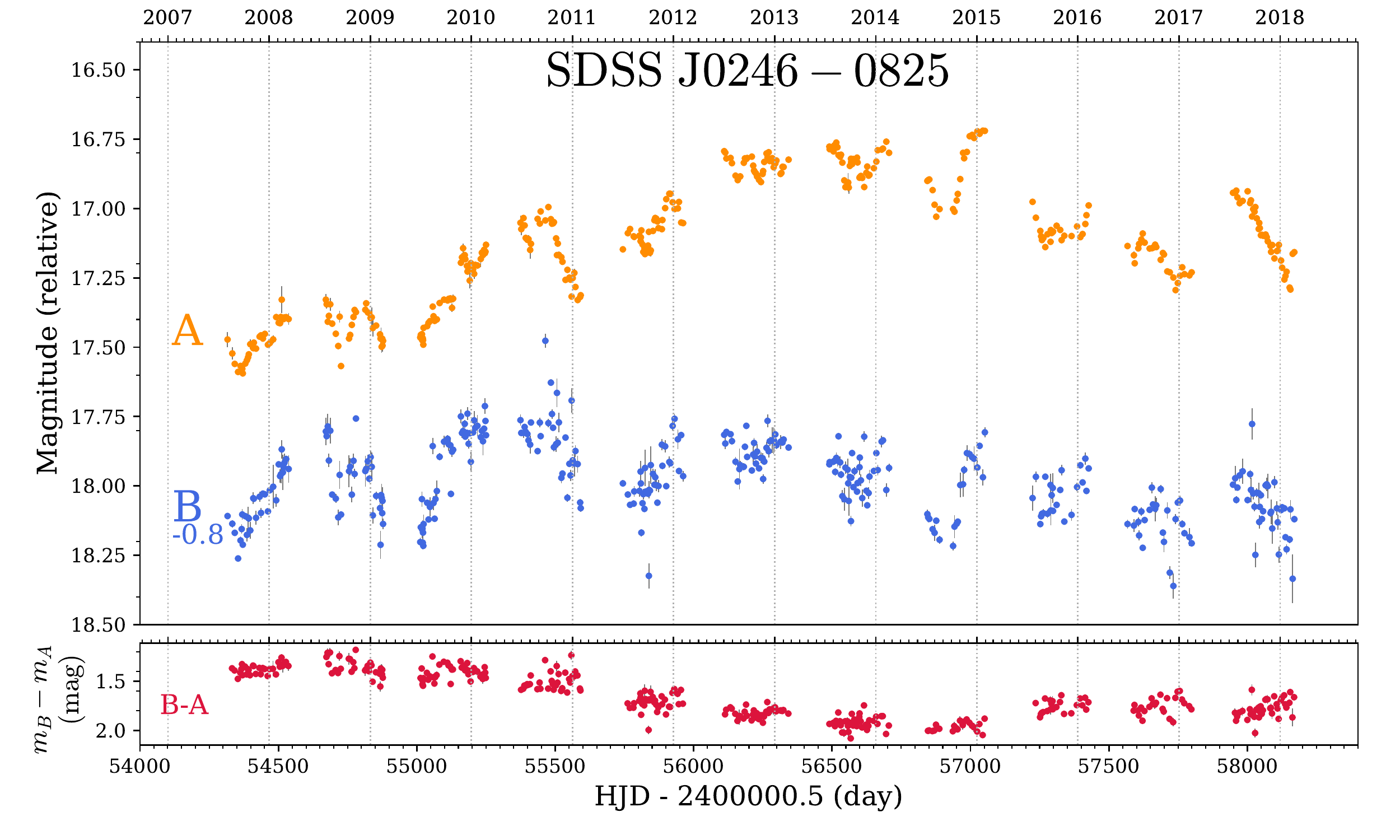}
    \end{minipage}    
    
    \begin{minipage}[c]{0.99\textwidth}
    \includegraphics[width=\textwidth]{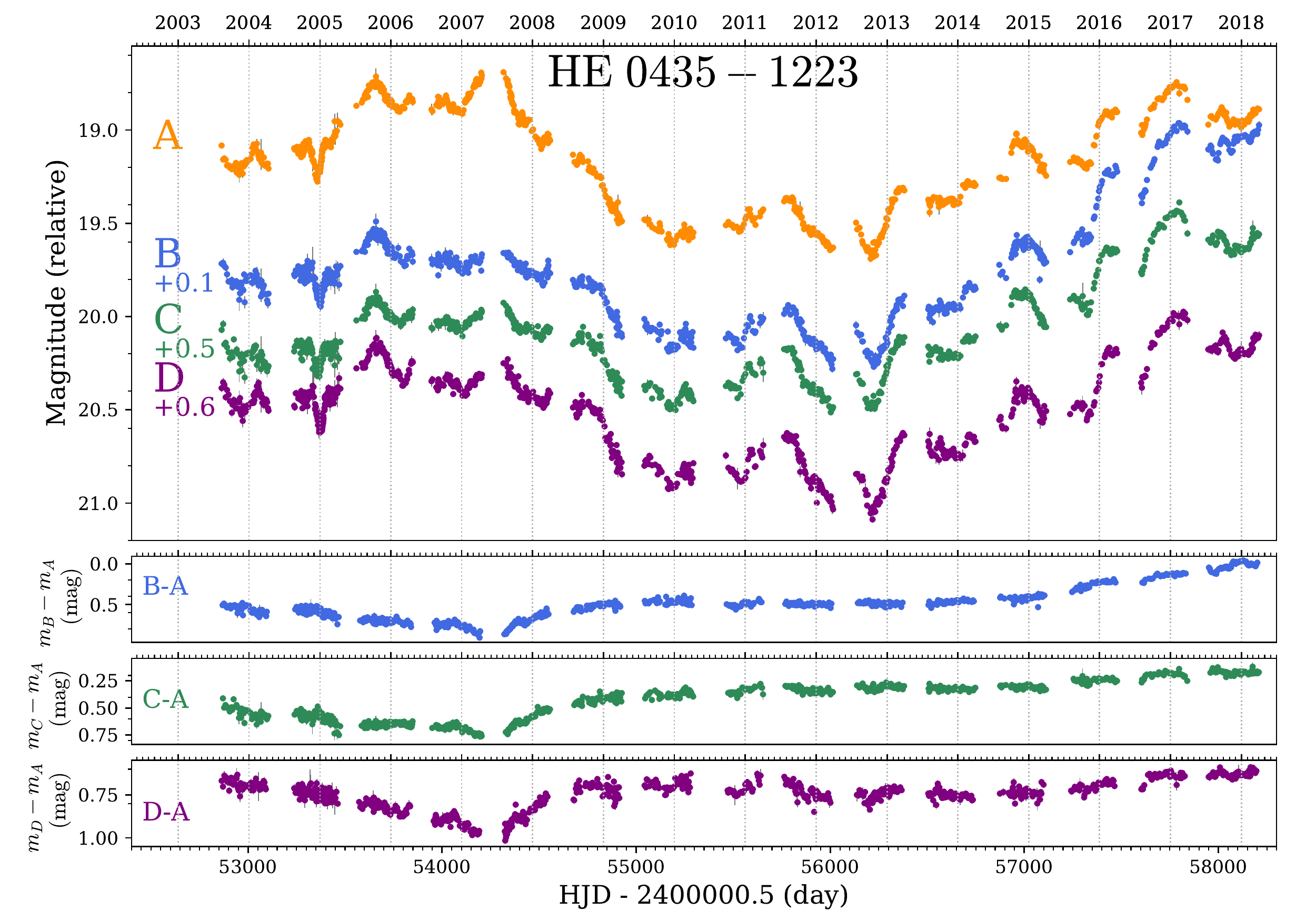}
    \end{minipage}
    \caption{Continuation of Fig.~\ref{fig:annex_lcs2}}
    \label{fig:annex_lcs3}
\end{figure*}

\begin{figure*}[hp!]
    \begin{minipage}[c]{0.99\textwidth}
    \includegraphics[width=\textwidth]{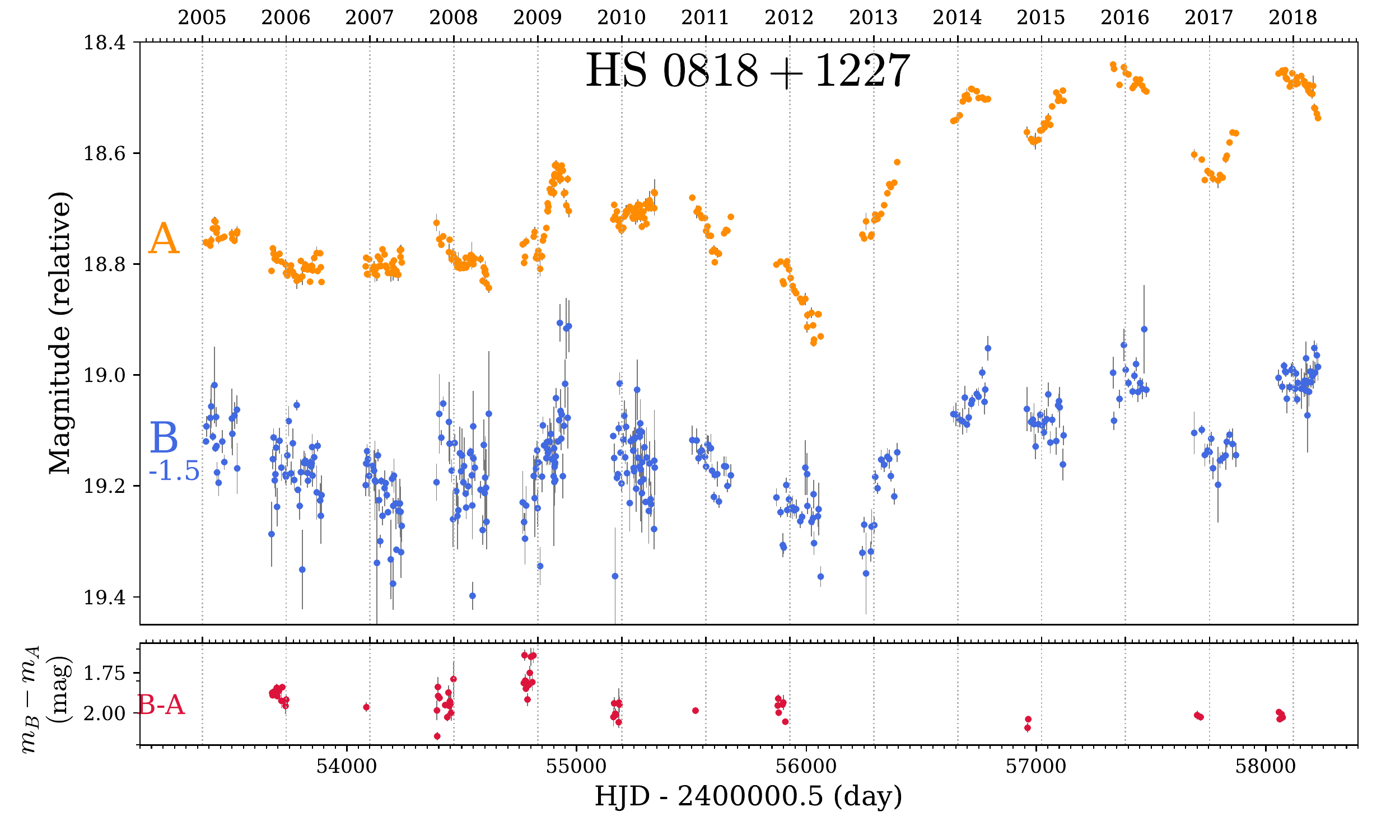}
    \end{minipage}
    \begin{minipage}[c]{0.99\textwidth}
    \includegraphics[width=\textwidth]{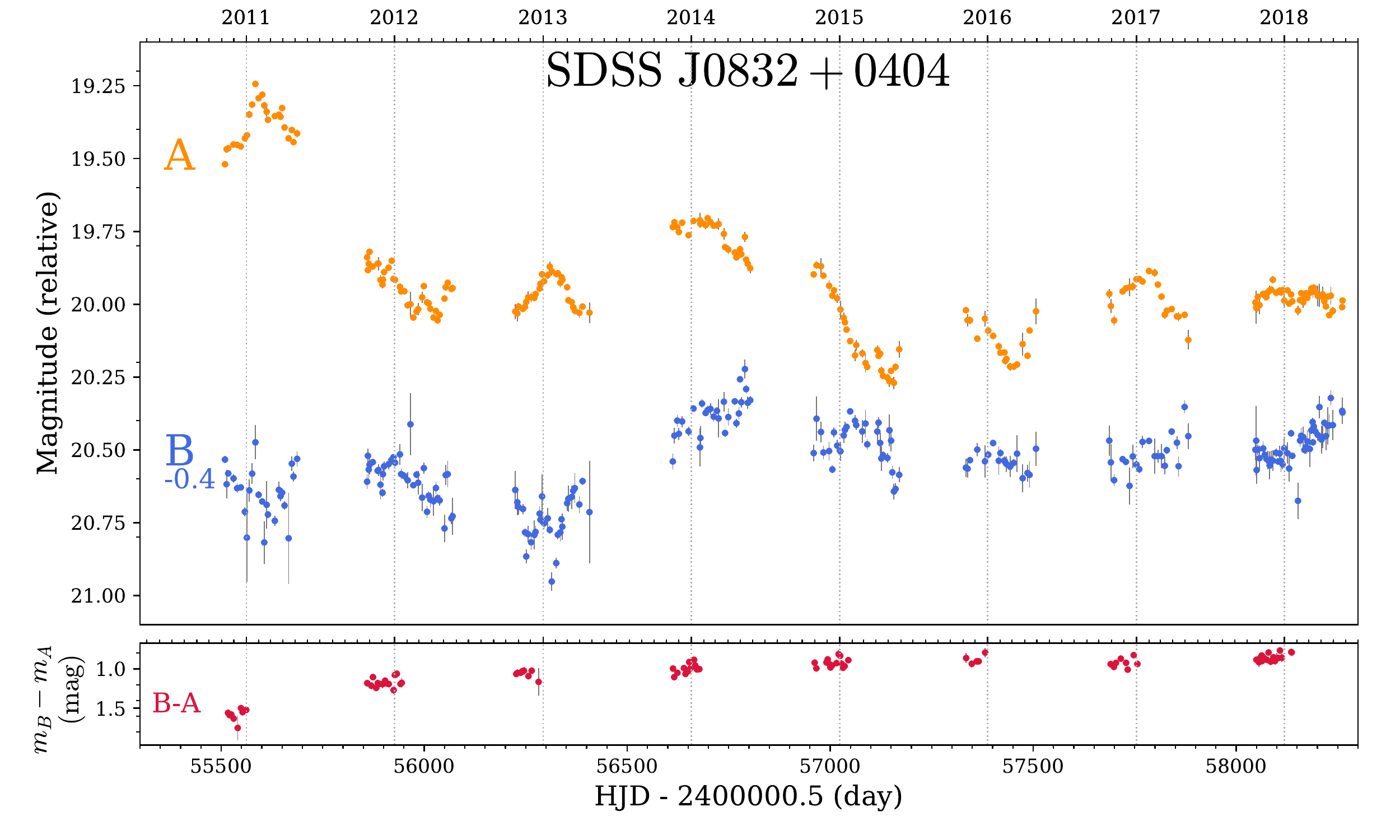}
    \end{minipage} 
    \caption{Continuation of Fig.~\ref{fig:annex_lcs3}}
    \label{fig:annex_lcs4}
\end{figure*}

\begin{figure*}[hp!]
    \centering
    \begin{minipage}[c]{0.99\textwidth}
    \includegraphics[width=\textwidth]{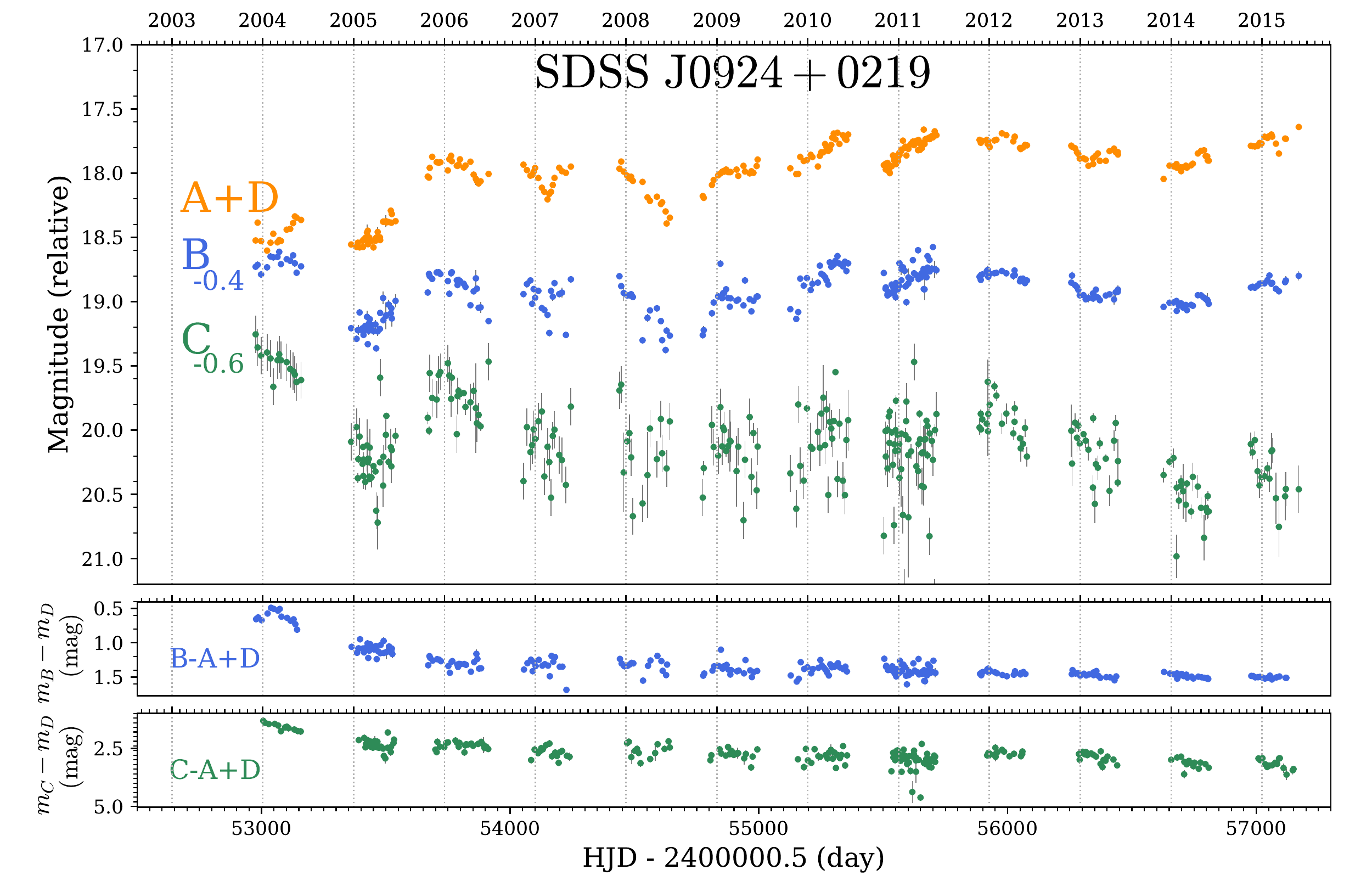}
    \end{minipage} 
    \begin{minipage}[c]{0.99\textwidth}
    \includegraphics[width=\textwidth]{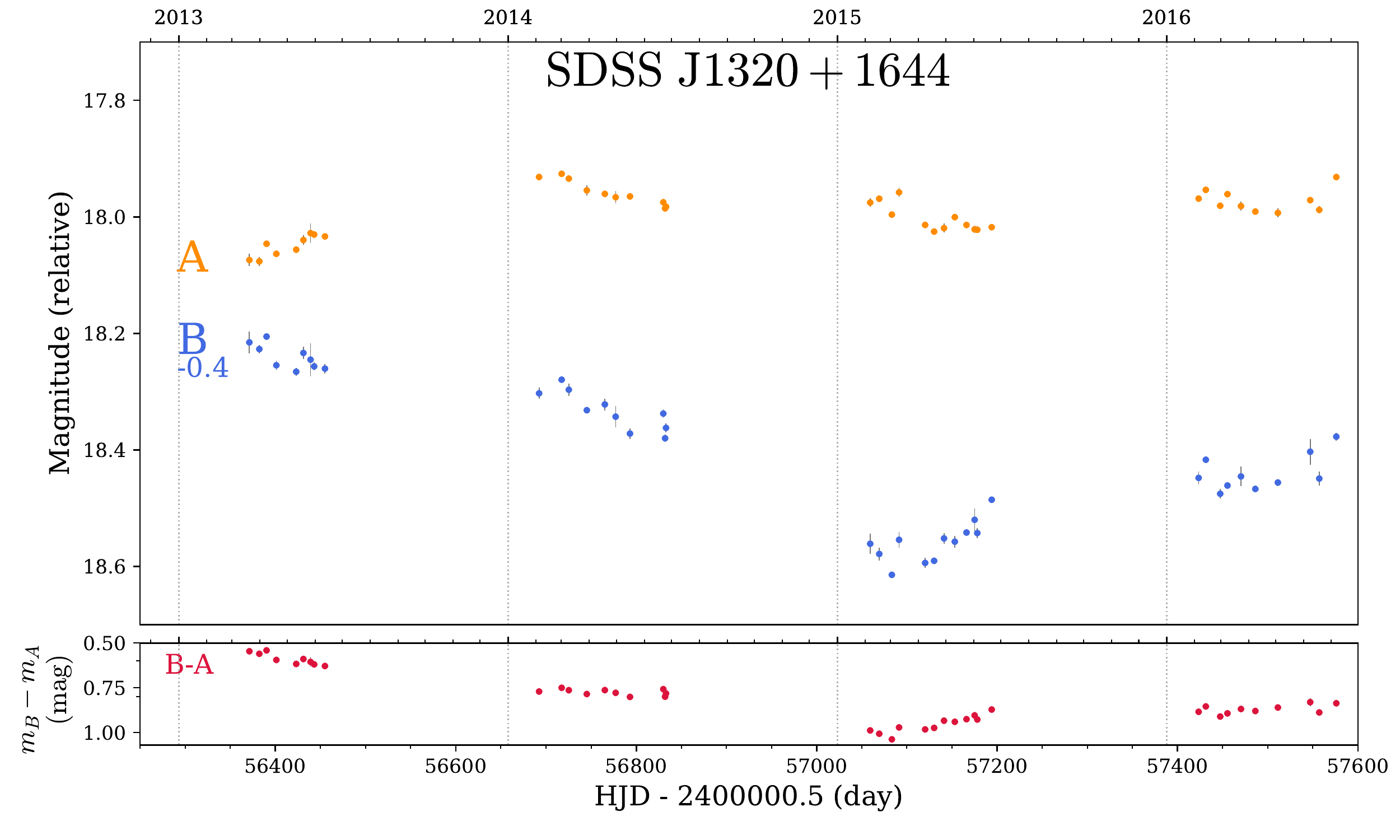}
    \end{minipage}
    \caption{Continuation of Fig.~\ref{fig:annex_lcs4}}
    \label{fig:annex_lcs5}
\end{figure*}   

\begin{figure*}[hp!]
    \centering
    \begin{minipage}[c]{0.99\textwidth}
    \includegraphics[width=\textwidth]{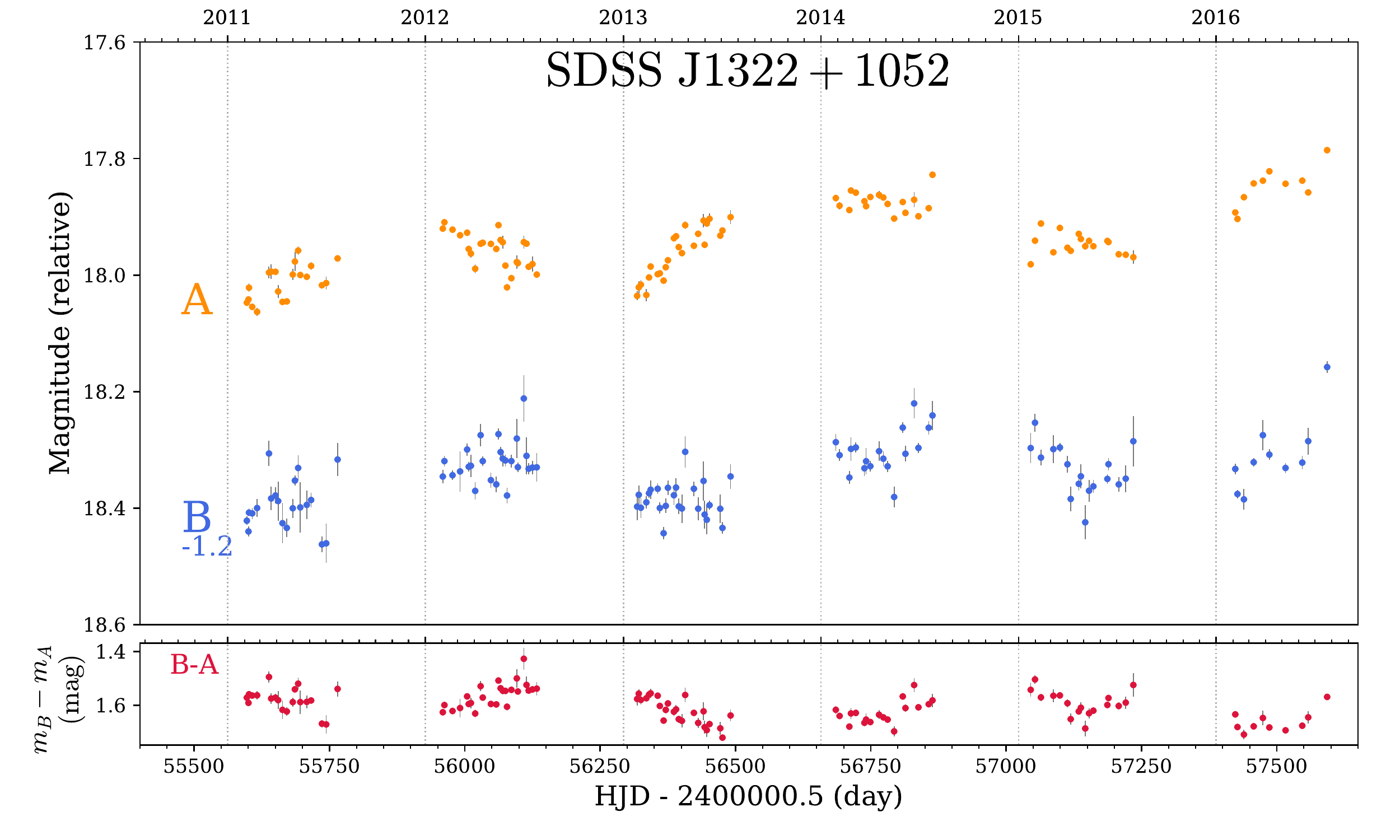}
    \end{minipage} 
    \begin{minipage}[c]{0.99\textwidth}
    \includegraphics[width=\textwidth]{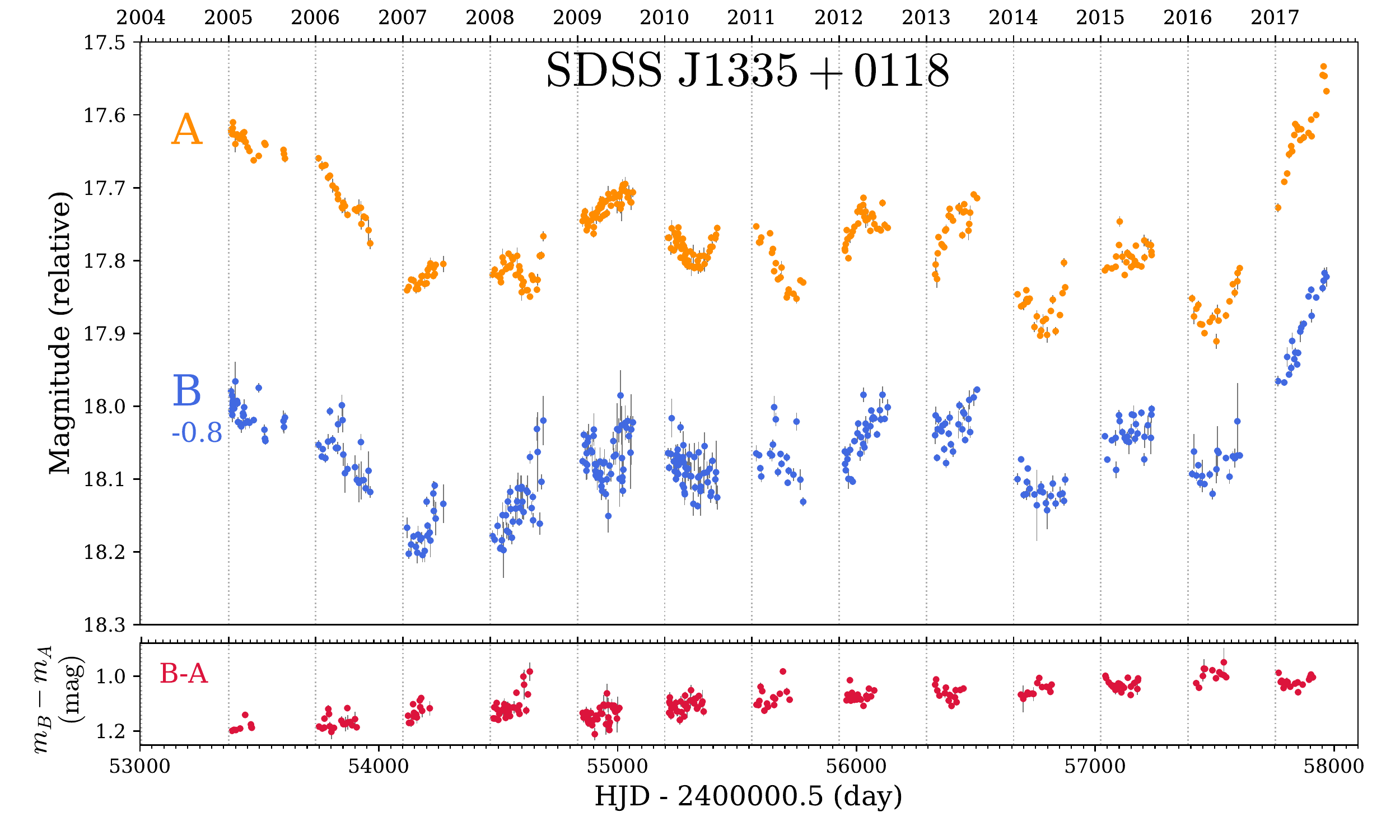}
    \end{minipage}
    \caption{Continuation of Fig.~\ref{fig:annex_lcs5}}
    \label{fig:annex_lcs6}
\end{figure*}

\begin{figure*}[hp!]
    \centering
    \begin{minipage}[c]{0.99\textwidth}
    \includegraphics[width=\textwidth]{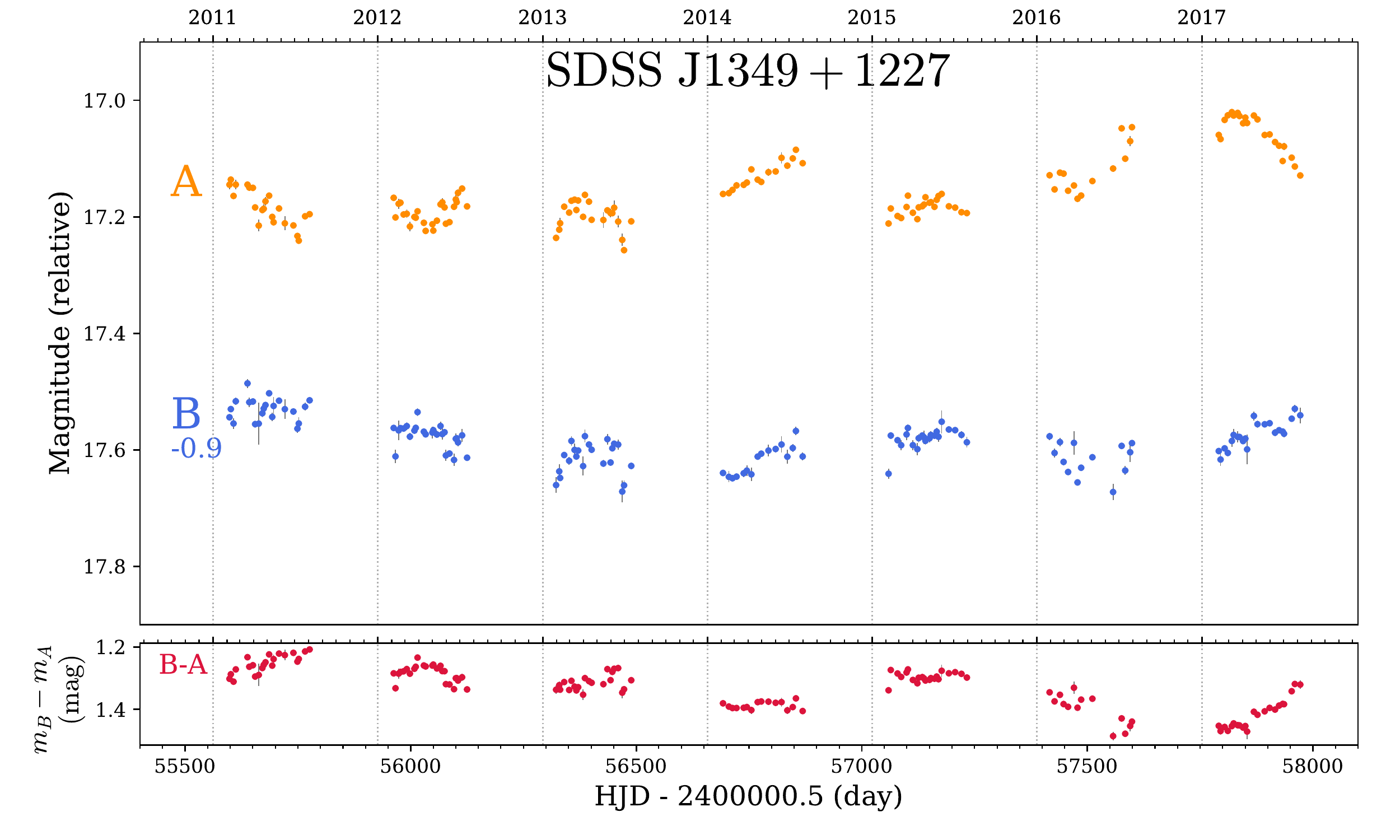}
    \end{minipage} 
    \begin{minipage}[c]{0.99\textwidth}
    \includegraphics[width=\textwidth]{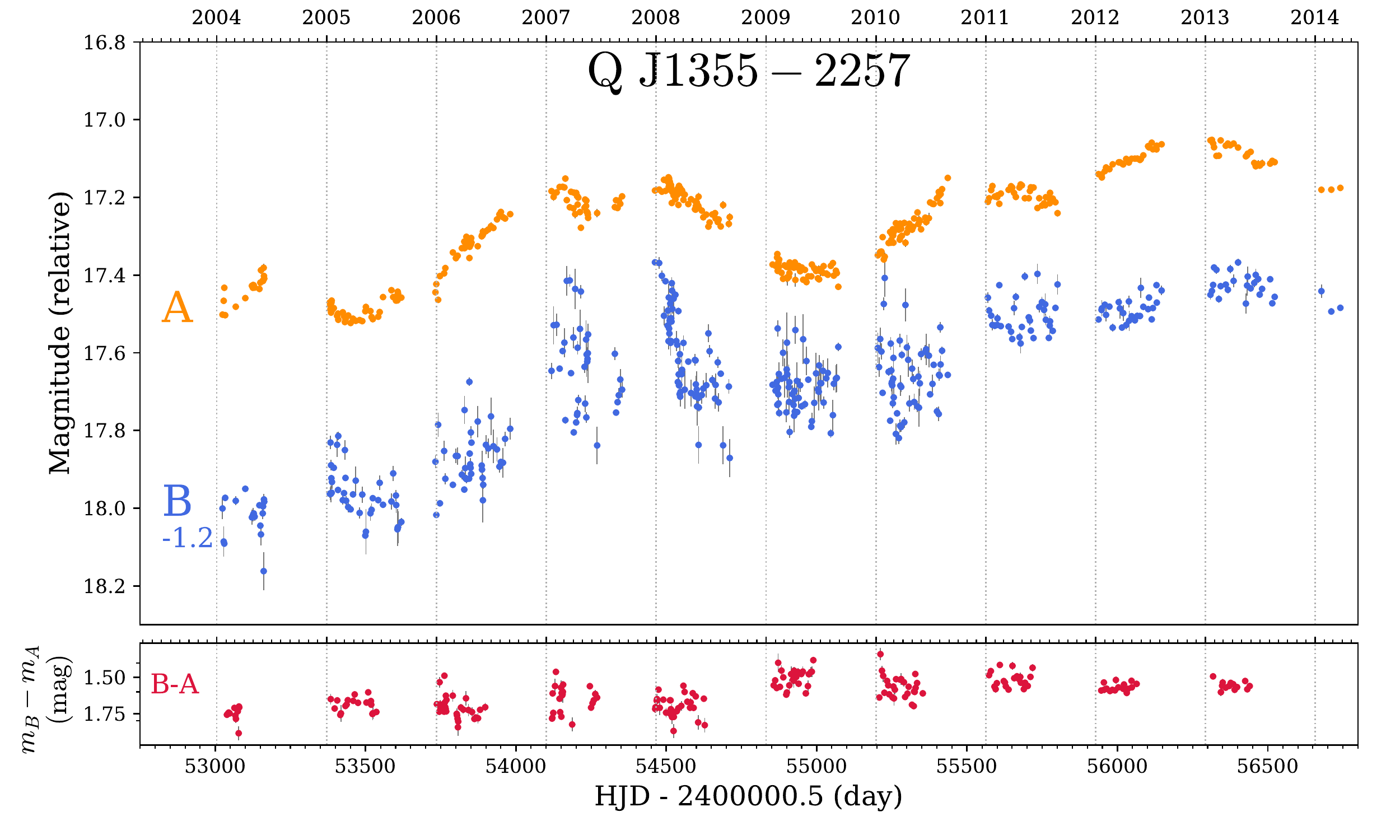}
    \end{minipage} 
    \caption{Continuation of Fig.~\ref{fig:annex_lcs6}}
    \label{fig:annex_lcs7}
\end{figure*}   

\begin{figure*}[hp!]
    \centering
    \begin{minipage}[c]{0.99\textwidth}
    \includegraphics[width=\textwidth]{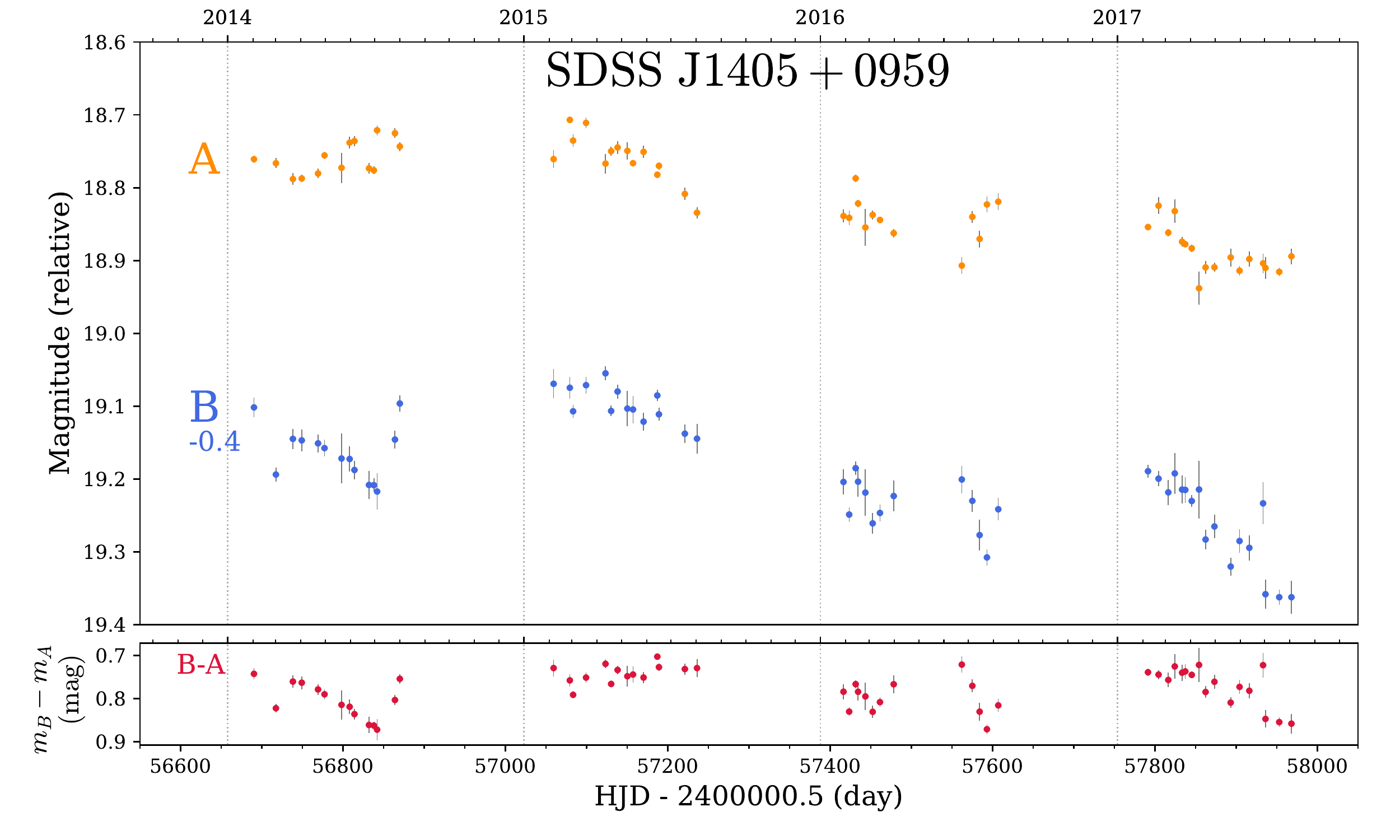}
    \end{minipage}
    \begin{minipage}[c]{0.99\textwidth}
\includegraphics[width=\textwidth]{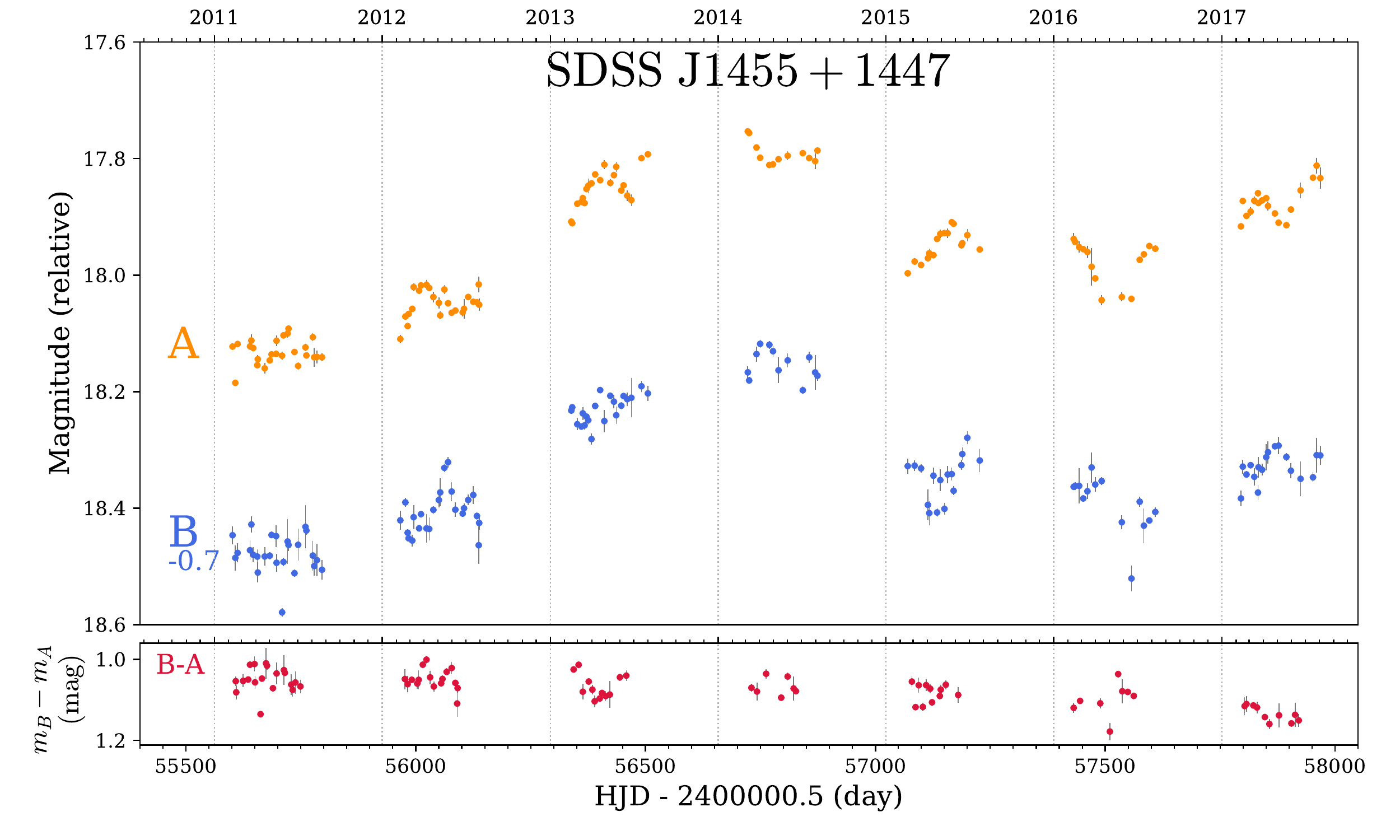}
    \end{minipage} 
    \caption{Continuation of Fig.~\ref{fig:annex_lcs7}}
    \label{fig:annex_lcs8}
\end{figure*}

\begin{figure*}[hp!]
    \centering
    
    \begin{minipage}[c]{0.99\textwidth}
    \includegraphics[width=\textwidth]{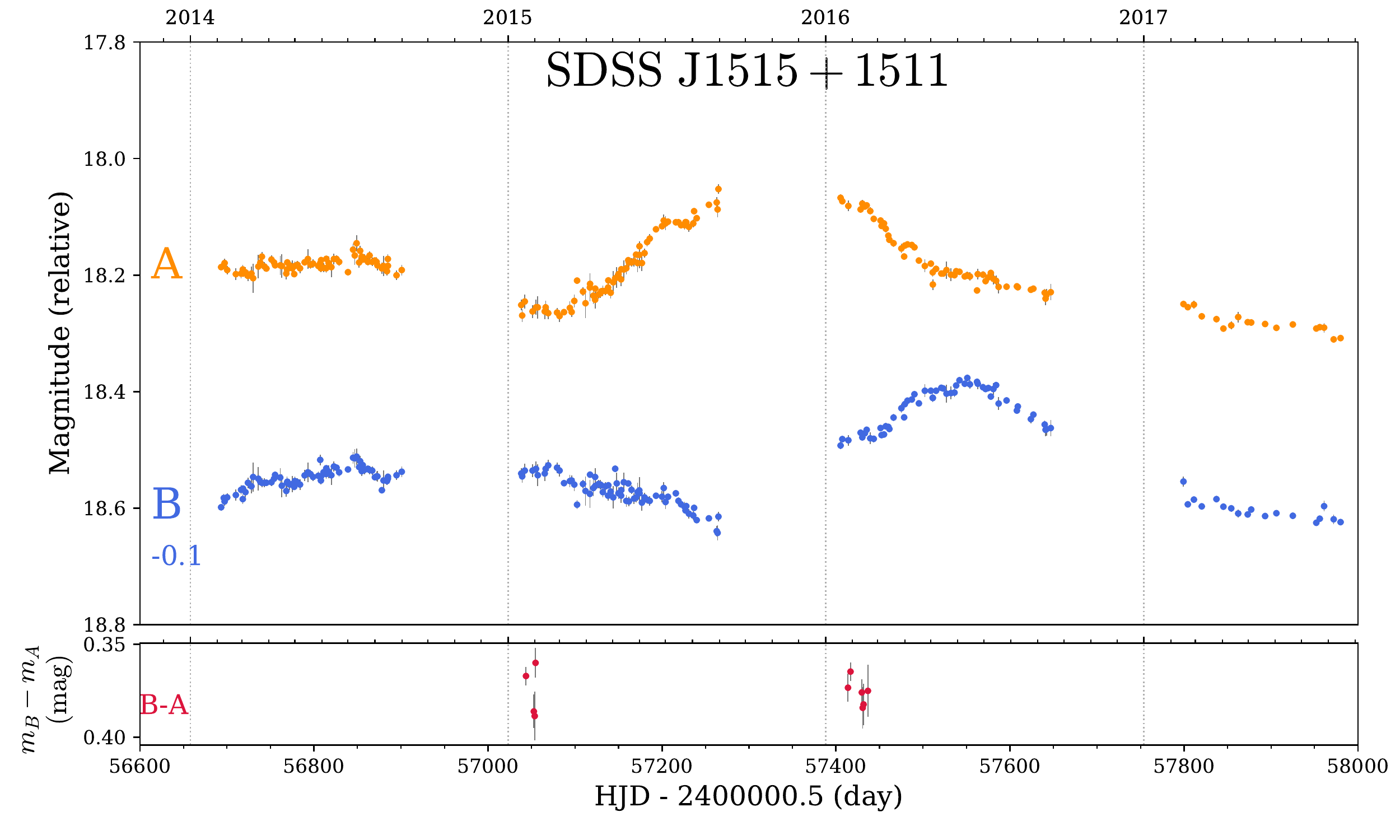}
    \end{minipage}
    
    \begin{minipage}[c]{0.99\textwidth}
    \includegraphics[width=\textwidth]{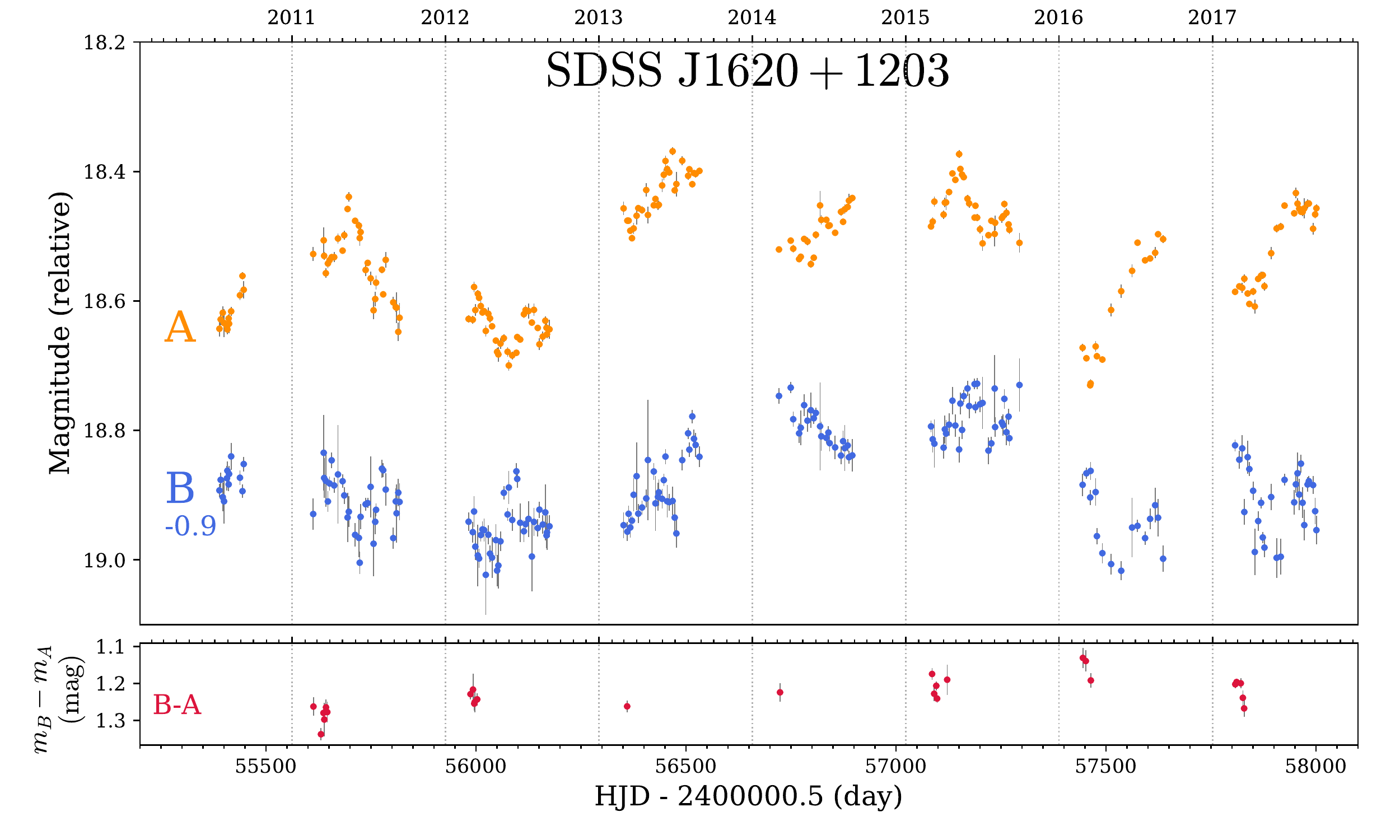}
    \end{minipage} 
    
    \caption{Continuation of Fig.~\ref{fig:annex_lcs8}}
    \label{fig:annex_lcs9}
\end{figure*}

\begin{figure*}[ht!]
    \centering
    \begin{minipage}[c]{0.99\textwidth}
    \includegraphics[width=\textwidth]{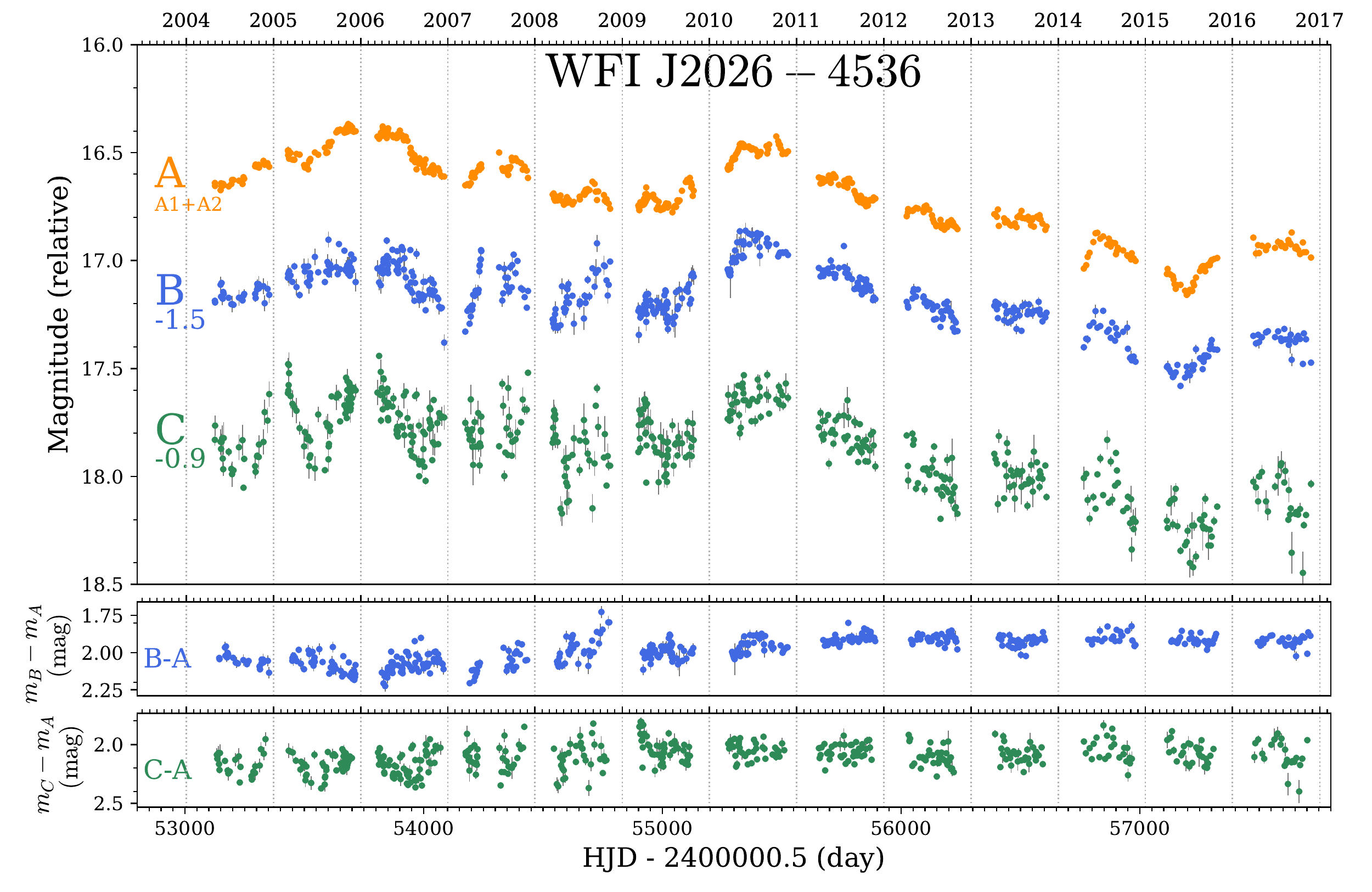}
    \end{minipage}
    \begin{minipage}[c]{0.99\textwidth}
    \includegraphics[width=\textwidth]{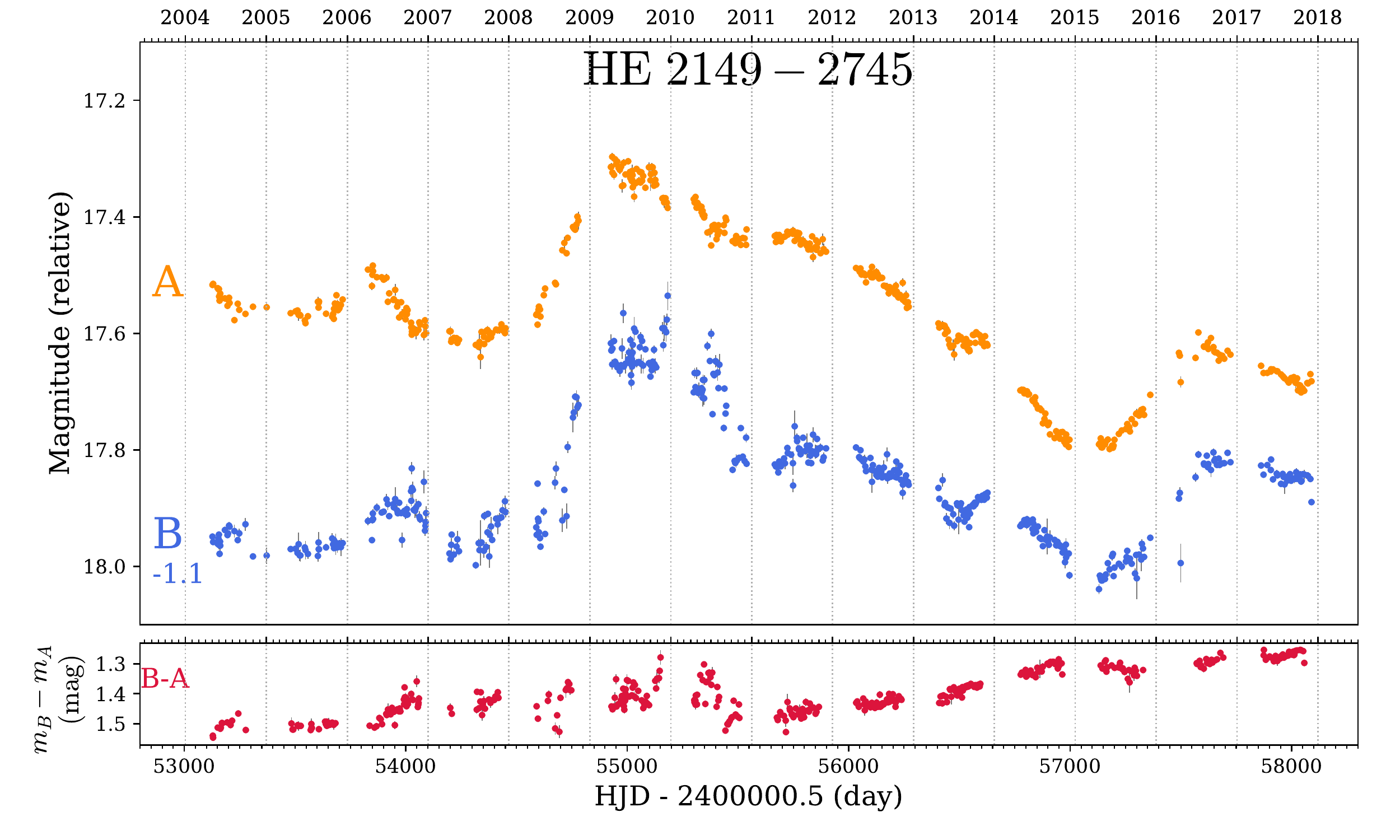}
    \end{minipage} 
    \caption{Continuation of Fig.~\ref{fig:annex_lcs9}}
    \label{fig:annex_lcs10}
\end{figure*}

\begin{figure*}[ht!]
    \centering
    \begin{minipage}[c]{0.99\textwidth}
    \includegraphics[width=\textwidth]{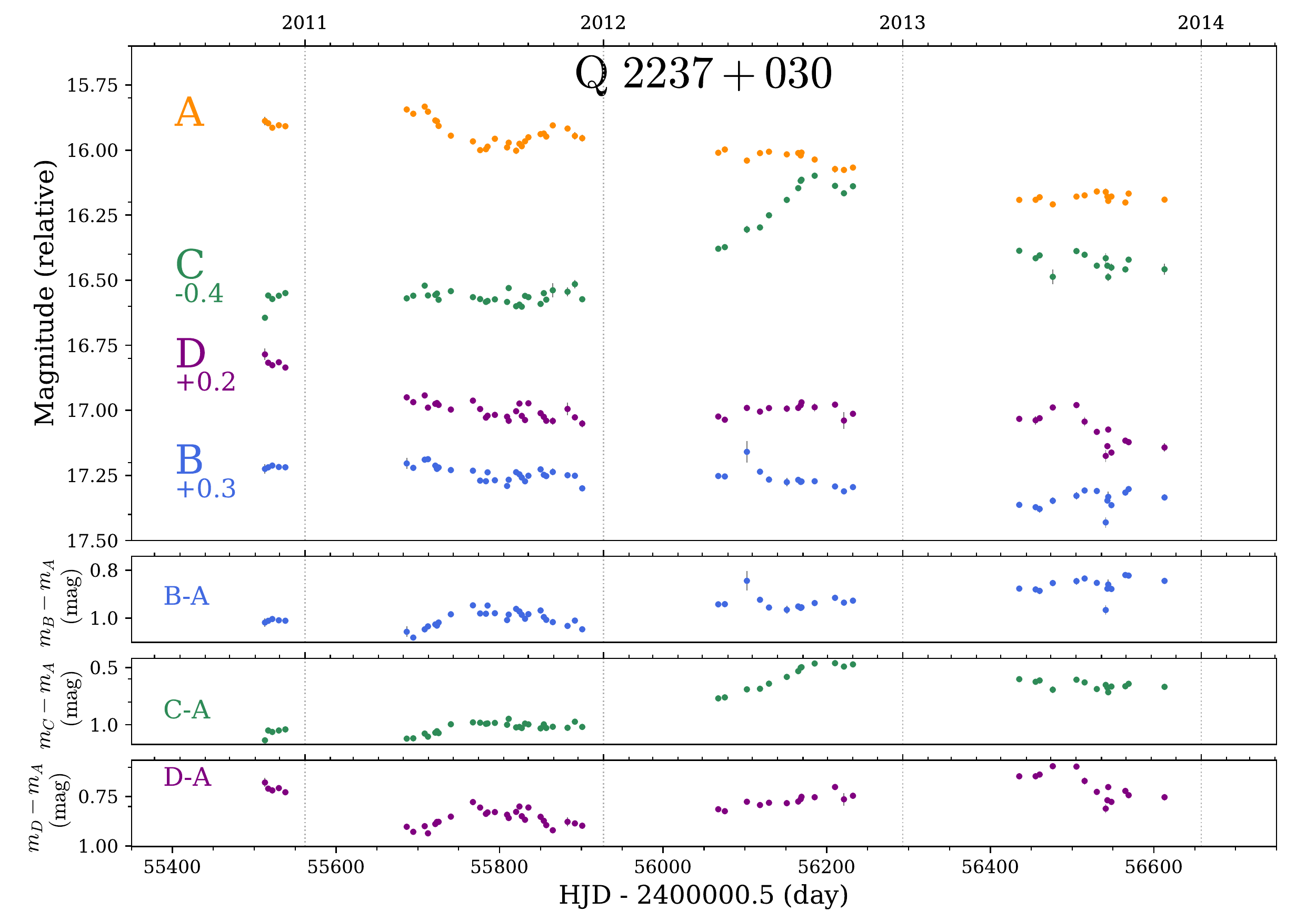}
    \end{minipage}
    \caption{Continuation of Fig.~\ref{fig:annex_lcs10}}
    \label{fig:annex_lcs11}
\end{figure*}

\end{document}